\documentclass{article}
\usepackage[letterpaper, margin=1.1in]{geometry}
\usepackage{libertine}
\usepackage[T1]{fontenc}
\usepackage[utf8]{inputenc}
\usepackage[english]{babel}
\usepackage{caption}
\usepackage{subcaption}
\usepackage[labelformat=simple]{subcaption}
\usepackage{multirow}
\usepackage{mathtools}
\usepackage{array, caption}
\usepackage{graphicx}
\usepackage{makecell}
\usepackage{setspace}
\usepackage{hyperref}
\usepackage[nameinlink,capitalise]{cleveref}

\newcommand{\lw}[1]{\smash{\lower2.0ex\hbox{#1}}}
\newcommand{\lwm}[1]{\smash{\lower1.5ex\hbox{#1}}}
\newcommand{\lwh}[1]{\smash{\lower1.0ex\hbox{#1}}}

\newcommand{\multisubref}[2]{\ref{#1}-\subref{#2}}
\usepackage{csquotes}
\usepackage{biblatex}
\usepackage{color}
\usepackage{xcolor}
\usepackage{colortbl}

\definecolor{hanpurple}{rgb}{0.32, 0.09, 0.98}
\definecolor{majorelleblue}{rgb}{0.38, 0.31, 0.86}
\definecolor{mediumpersianblue}{rgb}{0.0, 0.4, 0.65}
\definecolor{mediumseagreen}{rgb}{0.24, 0.7, 0.44}

\DeclareUnicodeCharacter{0301}{\'}

\begin{document} 

\begin{flushleft}  \bigskip
 {\Large
 \textbf{Self-supervised Deep Learning for Reading Activity Classification}
 }
 \newline
\\
 MD. RABIUL ISLAM, Osaka Prefecture University, Japan\\
 \smallskip
 SHUJI SAKAMOTO, Osaka Prefecture University, Japan\\
 \smallskip
 YOSHIHIRO YAMADA, Osaka Prefecture University, Japan\\
 \smallskip
 ANDREW VARGO, Osaka Prefecture University, Japan\\
 \smallskip
 MOTOI IWATA, Osaka Prefecture University, Japan\\
 \smallskip
 MASAKAZU IWAMURA, Osaka Prefecture University, Japan\\
 \smallskip
 KOICHI KISE, Osaka Prefecture University, Japan
 \end{flushleft}
{\small
Reading analysis can give important information about a user's confidence and habits and can be used to construct feedback to improve a user's reading behavior. A lack of labeled data inhibits the effective application of fully-supervised Deep Learning~(DL) for automatic reading analysis. In this paper, we propose a self-supervised DL method for reading analysis and evaluate it on two classification tasks. We first evaluate the proposed self-supervised DL method on a four-class classification task on reading detection using electrooculography~(EOG) glasses datasets, followed by an evaluation of a two-class classification task of confidence estimation on answers of multiple-choice questions (MCQs) using eye-tracking datasets.
Fully-supervised DL and support vector machines~(SVMs) are used to compare the performance of the proposed self-supervised DL method. The results show that the proposed self-supervised DL method is superior to the fully-supervised DL and SVM for both tasks, especially when training data is scarce. This result indicates that the proposed self-supervised DL method is the superior choice for reading analysis tasks. The results of this study are important for informing the design and implementation of automatic reading analysis platforms. 

\hfill \break
}
\section{Introduction}
Reading analysis is essential for understanding and advancing human learning strategies because it is possible to obtain a wide variety of information from reading activities~\cite{rayner_psychology_2012}. There are many different types of aspects of reading that can be analyzed and classified~\cite{cs_Pascual_10, cs_Okoso_11, cs_Cheng_12}. For example, one basic classification task is reading detection, where the objective is to detect whether the user is reading or not reading~\cite{rd_campbell_4, rd_landsmann_10}. Other research has tackled problems like identifying the type of text the user is reading, such as reading English text or Japanese text~\cite{rd_Ishimaru_13}. Finally, another reading activity classification task would be a problem-solving task such as confidence estimation in answering multiple-choice questions~(MCQs)~\cite{mcq_Yamada_2}, which requires significant cognitive activity from the reader. In this paper, we use the term reading activity to cover not only the activity of reading plain text but also problem-solving tasks completed via reading.

There are multiple ways in which to approach this endeavor. Traditional machine learning methods have achieved satisfactory results in laboratory settings where features are manually selected. This requires additional feature engineering expertise.
In addition, for the outside laboratory settings (in-the-wild) studies, these methods may not produce similar results~\cite{rd_Ishimaru_9} due to various reasons, such as noise which obfuscates important features that need to be extracted. Deep Learning~(DL), on the other hand, has been successfully used to solve a broad set of difficult problems in the areas of image recognition~\cite{imageDL_imagenet, imageDL_chan, ImageDL_rawat}, speech recognition~\cite{speech_graves, speech_Park_2019}, natural language processing~\cite{NLP_stanford, NLP_ruder}, human activity recognition~\cite{DLHAR_ronao, ubiDL_sheng_3, DLHAR_jobanputra} and eliminated the need for manual feature engineering.
The key to successful DL is to prepare enough labeled samples for successfully training the network.
In most fields, the difficulty of having enough amount of labeled samples is a serious issue~\cite{data_Roh}.

Lack of labeled data is also a problem for reading activity classification. Obtaining large and well-curated reading activity datasets is problematic because the annotation costs and the time it takes to generate a satisfactory dataset are prohibitive. In addition, the diversity of devices, types of embedded sensors, variations in specifications regarding sampling rates, and different deployment environments make dataset construction a challenge. For these reasons, it is very difficult to apply a fully-supervised DL method in this domain directly.

The self-supervised DL presents a potential solution to these problems. This method employs a ``pretext'' task to pre-train the network, before training for the task of interest (target task). Because labeled samples for the pretext task is generated without manual labeling,
the network can be trained with a much larger amount of data. In general, this helps to improve classification
accuracy.

In the field of human activity recognition, some researchers have attempted to employ self-supervised DL to solve the issue of the lack of labeled data and found it effective~\cite{SSL_saeed, SSL_Harish, HAR_SSL_Zaki}. For example, the method proposed by Saeed et al.~\cite{SSL_saeed} employs simple signal transformations such as flipping to produce the pretext task for sensor data. Because their task is to recognize human activities, both an accelerometer and a gyroscope are employed as sensors to distinguish human body movements that characterize each activity. However, we do not know whether similar approaches can be applied to cognitive activities
requiring fewer bodily movements such as the reading, which is typically captured by sensors like an eye-tracker.

This research aims to clarify how effective the self-supervised DL is at solving the lack of labeled data issue in reading activity classification. As a step toward this goal, we propose a  self-supervised DL method and evaluate it for two different but related reading activity classification tasks placed at two extreme points on the reading activity spectrum. The first one is reading detection, a physical-level reading activity, where periods of reading are identified throughout the day. The second one is confidence estimation in answering MCQs, which is an intensive cognitive level reading activity. This allows us to obtain a full picture of the effectiveness of the proposed self-supervised DL method across the reading activity spectrum. In the evaluation process, we recorded eye-movement using electrooculography~(EOG) glasses for reading detection and measured eye gaze using an eye-tracker for confidence estimation. We compared the effectiveness of the proposed self-supervised DL method by training and evaluating the network for a different number of training samples per class, starting from the availability of all samples per class to 10 samples per class. We used the fully-supervised DL approach as a comparative method along with support vector machines (SVMs), a traditional machine learning method, as a baseline. 

The results show that the proposed self-supervised DL method is superior compared to other methods at both tasks. 
Specifically, the proposed self-supervised DL method demonstrates better performance than the fully-supervised DL 
except at the largest number of training samples, where the proposed self-supervised DL method performs equally well.
Although the fully-supervised DL performs worse than SVM with a smaller number of training samples,
due to the impact of insufficient training samples, the proposed self-supervised DL method does not face this problem; it is always superior to SVM as well. The statistical analysis supports the above statements.

From the results, we conclude that the proposed self-supervised DL method is superior to other methods over a wide range of training samples on both tasks, and is comparable with the fully-supervised DL when the number of training samples available is high.
This indicates that we can recommend the self-supervised DL method for any size of  available training samples. This insight can help system designers and researchers more efficiently pursue reading activity classification. 

The remainder of the paper is organized as follows: 
\Cref{sec:relatedwork} presents related work on reading detection, confidence estimation, and self-supervised DL.
\Cref{sec:proposedmethod} presents the proposed self-supervised DL method in detail for both reading detection and confidence estimation.
\Cref{sec:datacollection} presents the details of datasets, and the data collection framework.
In \Cref{sec:resultsanddiscussion} we present the experimental conditions, results, and discussion about the results obtained.
Finally, \Cref{sec:conclusion} presents the conclusion and future work.

\section{Related Work} \label{sec:relatedwork}
Our work relates to several active research areas, including reading detection, confidence estimation, and  self-supervised DL. In this section, we describe how our work builds on these fields.
\subsection{Reading Detection}
Reading detection strategy varies depending on its purpose, and over the past years, researchers have proposed many methods for different kinds of automatic reading detection. For example, they have proposed methods for reading detection as a part of other human activities such as reading in transit~\cite{rd_Bulling_20, rd_Steil_14}, in office settings~\cite{rd_Bulling_12}, and with talking~\cite{rd_Ishimaru_16, rd_Ishimaru_8} by exploring eye movements in controlled settings using classical machine learning approaches. In another eye-based activity recognition study~\cite{rd_srivastava}, authors detected reading with desktop activities such as search and writing by using traditional machine learning methods.
Many existing studies attempt to explore reading activity as an individual activity accomplishing different modes, such as regular reading, detailed reading, skimming, and spell-checking~\cite{rd_strukelj}. Researchers used traditional machine learning methods to detect whether the user is reading or skimming~\cite{rd_Kelton_2, rd_Biedert}, reading or searching~\cite{rd_campbell_4}  and reading or not reading~\cite{rd_landsmann_10, rd_Ishimaru_9, rd_keat_7} in laboratory settings.
Furthermore, some methods attempt to explore the type of document read by the user. Towards this, Kunze et al.~\cite{rd_kunze_17} proposed a method for document type detection using eye movement features. The authors evaluated their approach based on Japanese document types such as novel, manga, magazine, newspaper, and textbook in laboratory settings and applying classical machine learning techniques. Recently, Ishimaru et al.~\cite{rd_Ishimaru_13} proposed a classical machine learning method to classify the language of text segments, English or Japanese, read by the user. They were able to  demonstrate an ability to differentiate the language of the text by analyzing eye movement data obtained through an in-the-wild study.

The literature on reading detection suggests a strong relationship between eye movements and reading activities. Moreover, the existing body of work provides evidence that eye movement plays a crucial role in reading detection, making it compelling to explore the development of eye-based reading detection systems. However, most of the proposed methods occur in laboratory settings use classical machine learning approaches except for some preliminary work that applies DL methods~\cite{rd_Ishimaru_9, emDL_copeland_2}. Although some proposed methods using classical machine learning approaches produce satisfactory results in laboratory settings, they may not do so in-the-wild~\cite{rd_Ishimaru_13}.

\subsection{Confidence Estimation in Answering Multiple-choice Questions}
MCQs are fundamental forms of assessing knowledge, ability, and user performance~\cite{MCQ_haladyna} and are the most popular since they are easy, quick, and offer more objective scoring. However, in this assessment, one critically common question arises: has the user answered correctly by chance or with confidence in their knowledge of the correct answer. 
A user may answer MCQs correctly and know the subject matter, but some other phenomena may also happen, such as when a user guesses the correct answer without any knowledge of understanding of the subject area~\cite{MCQ_lau, MCQ_mckenna}. Another possibility is that a user may be skilled in strategies to answer MCQs correctly~\cite{MCQ_melovitz, MCQ_sam}.
Moreover, the user may have correct or incorrect knowledge and may answer correctly even though the user may be confused by other options~\cite{MCQ_nehm}, or the user may be confident even though the answer is incorrect.
Correctness and incorrectness alone do not give the actual picture of the user's concept and understanding, which is a significant drawback of MCQ assessment~\cite{MCQ_siren} whereby the user's understanding of the subject material may be misunderstood~\cite{MCQ_chan}. 
Therefore, an assessment system must provide accurate information and give feedback. This feedback is essential for both users and instructors for developing further diagnoses~\cite{MCQ_brassil}. However, it is not possible to manually track each user in a group, so there is a need to develop a way to estimate confidence automatically.

As a step toward this automatic confidence estimation, researchers propose some methods. Tsai et al.~\cite{mcq_Tsai_1} analyzed user's visual attention spans when solving MCQs by using eye-tracking under laboratory settings and with the application of a traditional machine learning method. This gives a picture of the user's concept and understanding. The results show that successful problem solvers focus and spend more time examining relevant factors than irrelevant ones, while unsuccessful problem solvers spend more time decoding the problem, and have difficulty in recognizing the relevant factors. 
Yamada et al.~\cite{mcq_Yamada_2} proposed a method to classify whether the user is confident or not when answering MCQs through manually selected features from the eye gaze recorded with an eye-tracker in a controlled environment. All these proposed methods occurred in laboratory settings and employed traditional machine learning approaches. This means that a significant challenge remains for in-the-wild datasets. 

\subsection{Self-supervised Deep Learning}
In the past decade, the development and application of DL has successfully solved many problems in the field of ubiquitous computing~\cite{ubiDL_Hammerla_1, ubiDL_sheng_3, ubiDL_Chen_4, ubiDL_Ordez_2}, pervasive intelligence~\cite{perveDL_Sarhan_1, perveDL_Yao_2}, health~\cite{DLhealth_nonaka_1, DLhealth_Hannun_2, DLhealth_Ji_3}, and well-being~\cite{DLwellbeing_Saeed_1, DLwellbeing_Liu_2}. Most of the methods use fully-supervised DL approaches that need large and carefully labeled data that is feasible for use in domains such as computer vision~\cite{CV_toshev,CV_moschoglou} but unfeasible in others, particularly because of the tremendous cost and amount of effort to manually label data. 
To overcome the innate limitations of the fully-supervised DL approaches, researchers introduced several unsupervised methods. 

Recently, researchers proposed a DL technique called self-supervised DL~\cite{SSL_amis,SSL_Liu, SSL_agrawal_1, SSL_Lan} that encourages a network for representation learning by generating pretext tasks to utilize the massive amount of unlabeled data.
Self-supervised DL is now an active research approach in various domains such as computer vision and robotics~\cite{ SSL_imagevision_gidaris_4,  SSL_imagevision_lee_8, SSL_imagevision_owens_12, SSL_imagevision_pathak_13} and its achievements show that it is effective. When it comes to human activity recognition tasks, the same issue of the lack of labeled data occurs. Researchers attempted applying self-supervised DL for learning representations by utilizing some simple signal transformations such as flipping and adding noise~\cite{SSL_saeed}. Their findings show that a self-supervised DL approach is effective in this domain. 

Inspired by the recent success of applying self-supervised DL to address the issue of insufficient labeled data, we set out a research agenda to explore the generalize efficacy of self-supervised DL for eye movement sensory data for physical and cognitive intensive reading tasks.
To the best of our knowledge, the work presented here is the first attempt at applying the self-supervised DL method in reading activity classification. The proposed self-supervised DL method is evaluated for two eye movement datasets that are collected in real-life settings without any major restrictions. We applied the proposed self-supervised DL method for in-the-wild datasets.

\section{Proposed Method} \label{sec:proposedmethod}
\begin{figure}[tb]
      \begin{minipage}[b]{\textwidth}
      \centering
        \includegraphics[width=.71\hsize]{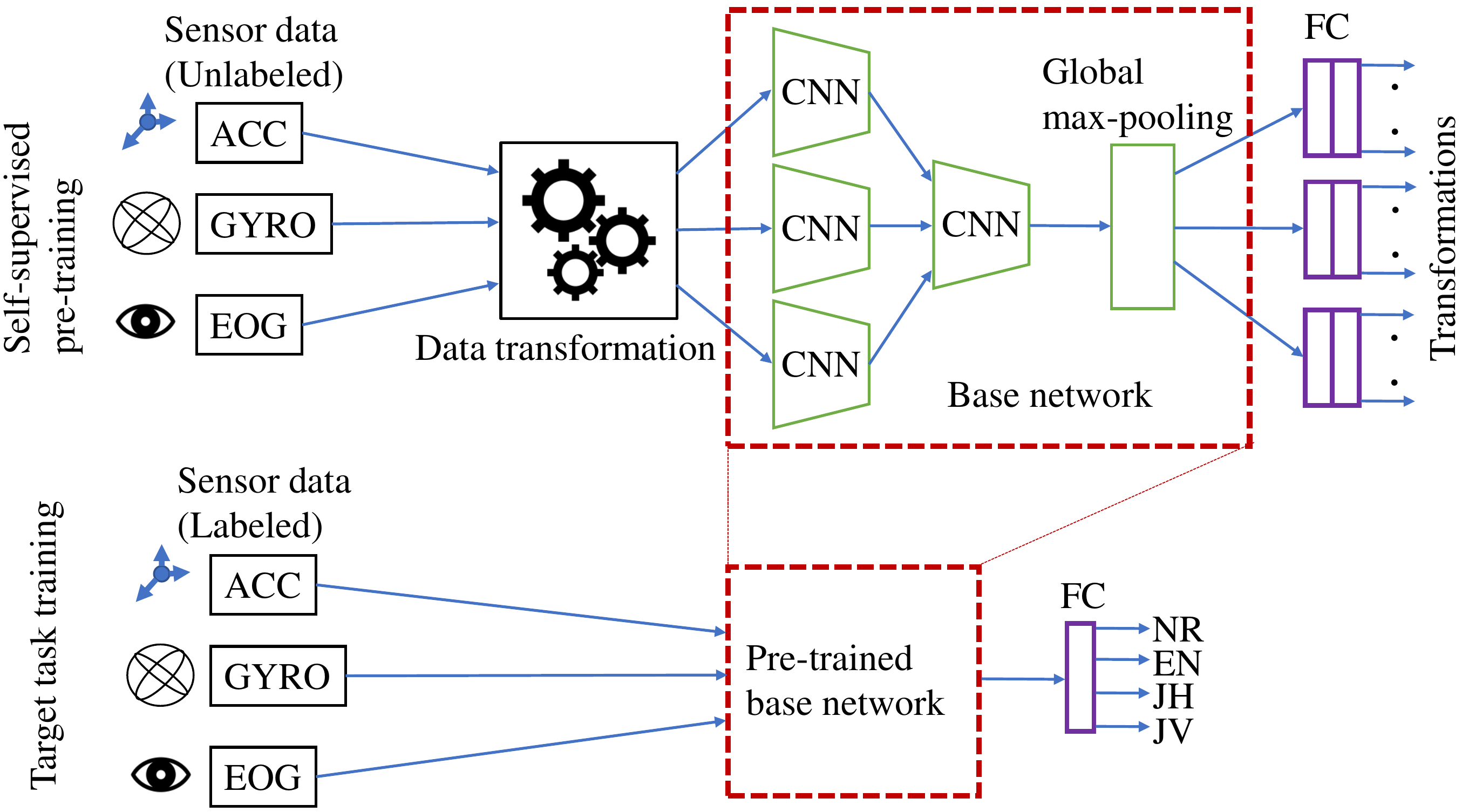}
        \subcaption{Reading detection}
        \label{fig:proposed_rd}
      \end{minipage}\\

      \begin{minipage}[b]{\textwidth}
        \centering
        \includegraphics[width=.71\hsize]{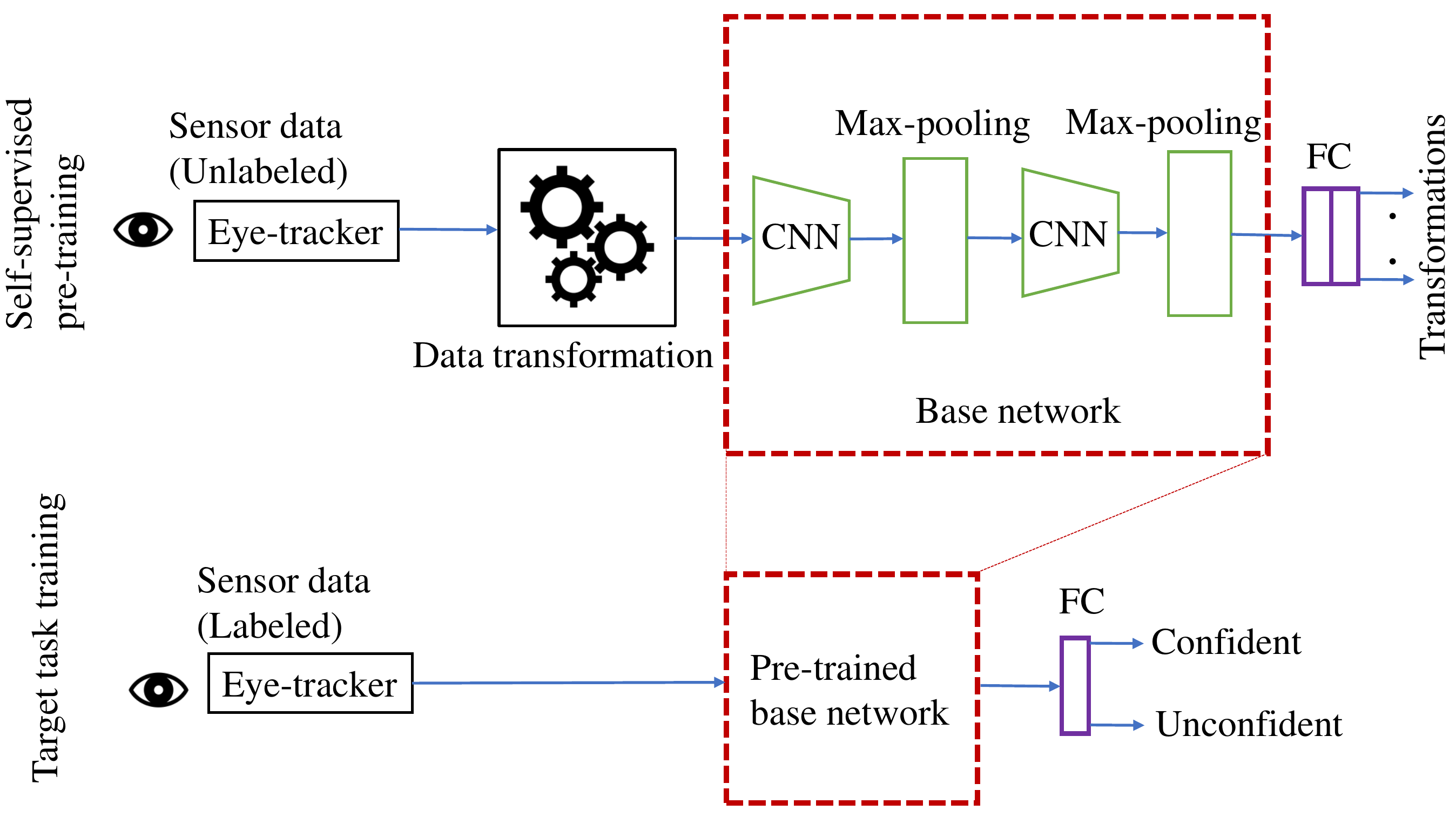}
        \subcaption{Confidence estimation}
        \label{fig:proposed_ce}
      \end{minipage}
    \caption{Proposed self-supervised DL method for reading activity classification.}
    \label{fig:proposed}
  \end{figure}

We propose a self-supervised DL method for reading activity classification using sensor data, as shown in \autoref{fig:proposed} that consists of two stages. The first stage shown in the upper parts of Figures~\ref{fig:proposed_rd} and \subref{fig:proposed_ce} 
is self-supervised pre-training consisting of solving the pretext task, automatically applied to a large collection of unlabeled sensor data. 
The second stage shown in the lower parts of Figures~\ref{fig:proposed_rd} and \subref{fig:proposed_ce} is target task training, i.e., training of a reading activity classification network by fine-tuning the pre-trained base network using labeled sensor data. 

To evaluate our method, we implemented the proposed self-supervised DL method on two very different reading activity tasks: 
reading detection and confidence estimation in answering MCQs. The reasons for this is that reading activities are distributed on a wide spectrum. For example, some activities are merely related to quantity of reading, such as reading periods or the number of read words. On the other hand, other activities involve the quality of reading such as understanding and confidence. To show the applicability of the proposed self-supervised DL method,
we apply it to tasks which belong to these respective categories. We selected reading detection that segments reading periods from all activities and confidence estimation in answering MCQs. 
Another reason for selecting these two activities is that they are recorded by using two different devices: EOG glasses for reading detection, and an eye-tracker for confidence estimation.
We consider that the proposed self-supervised DL method is general and effective enough if it works for both tasks using different devices.
In the following, we explain details of the proposed self-supervised DL method for each task.

\subsection{Reading Detection}
Reading detection aims to identify periods of reading from all other activities. This is implemented as a classification task; the user activities are divided into short segments and then classified into different segments of activity.

The devices employed to measure reading detection are EOG glasses that generate EOG data of eye movements, and accelerometer (ACC) and gyroscope (GYRO) data from the movement of the EOG glasses themselves.
From the EOG signals, we obtained data of horizontal and vertical eye movements (EOG\_H and EOG\_V). ACC and GYRO data consist of $x$, $y$ and $z$ components (ACC\_X, ACC\_Y, ACC\_Z, GYRO\_X, GYRO\_Y and GYRO\_Z). In total we collected eight different kinds of data.
The details of data recording are described in \Cref{sec:RD_dataset}.

As shown in \autoref{fig:proposed_rd}, the proposed self-supervised DL method is divided into two phases: self-supervised pre-training and training of the target task.

\subsubsection{Self-supervised Pre-training}
\begin{figure}[tb]
    \centering
    
      \begin{minipage}[b]{0.32\hsize}
        \centering
        \includegraphics[scale=.7]{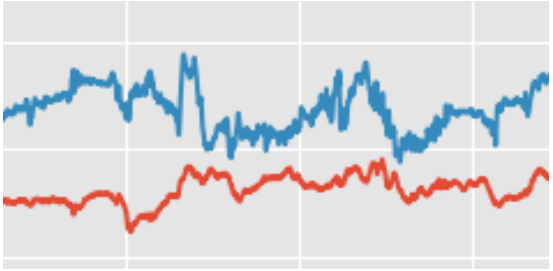}
        \subcaption{No transformation}
    \end{minipage}
    \hfill
      \begin{minipage}[b]{0.32\hsize}
        \centering
        \includegraphics[scale=.7]{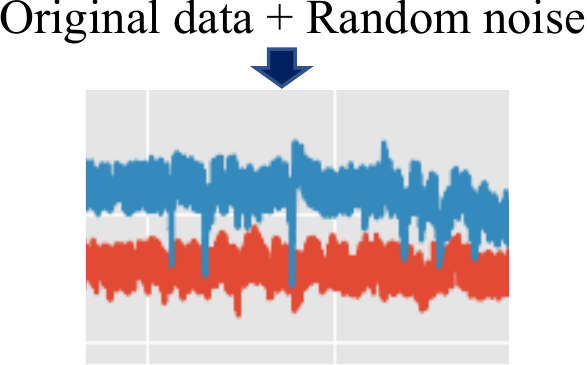}
        \subcaption{Random noise}
      \end{minipage}
    \hfill
      \begin{minipage}[b]{0.32\hsize}
        \centering
        \includegraphics[scale=.7]{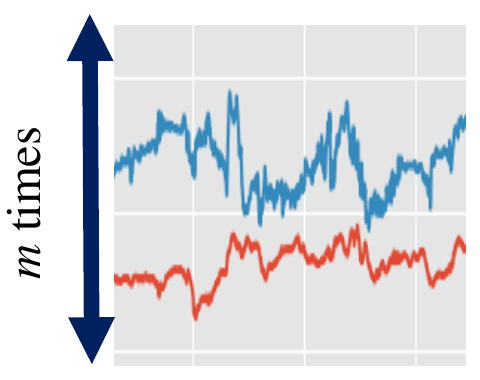}
        \subcaption{Scale}
      \end{minipage}

  \vspace*{\intextsep}

      \begin{minipage}[b]{0.32\hsize}
        \centering
        \includegraphics[scale=.7]{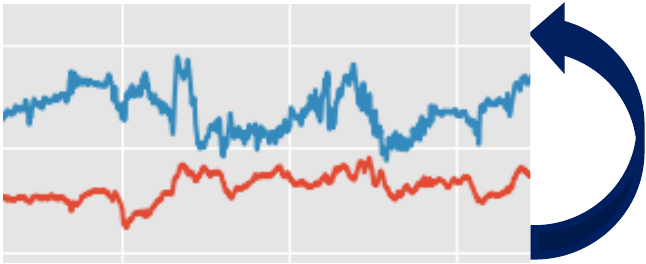}
        \vspace{5.67mm}
        \subcaption{Vertical flip}
      \end{minipage} 
    \hfill
      \begin{minipage}[b]{0.32\hsize}
        \centering
        \includegraphics[scale=.7]{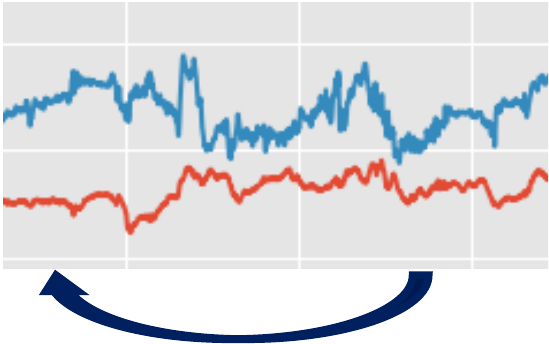}
        \subcaption{Horizontal flip}
      \end{minipage} 
    \hfill
      \begin{minipage}[b]{0.32\hsize}
        \centering
        \includegraphics[scale=.7]{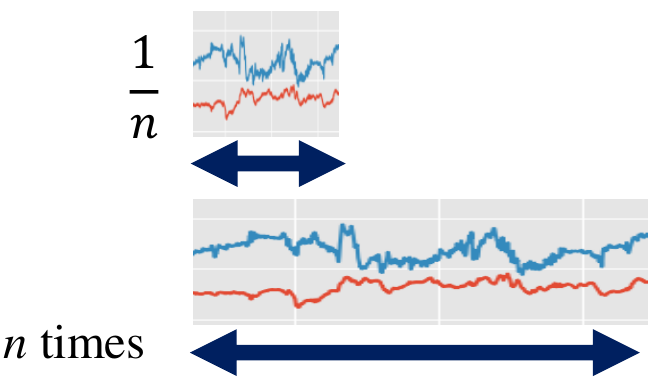}
        \subcaption{Time warp}
      \end{minipage}
      
        \vspace*{\intextsep}

      \begin{minipage}[b]{0.32\hsize}
        \centering
        \includegraphics[scale=.7]{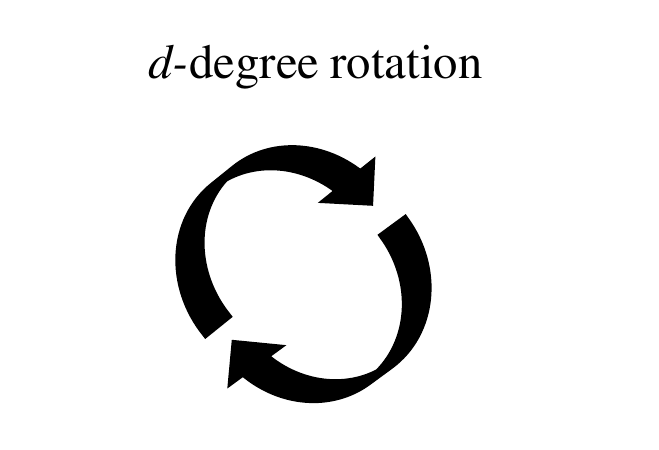}
        \subcaption{Rotation}
      \end{minipage} 
    \hfill
      \begin{minipage}[b]{0.32\hsize}
        \centering
        \includegraphics[scale=.7]{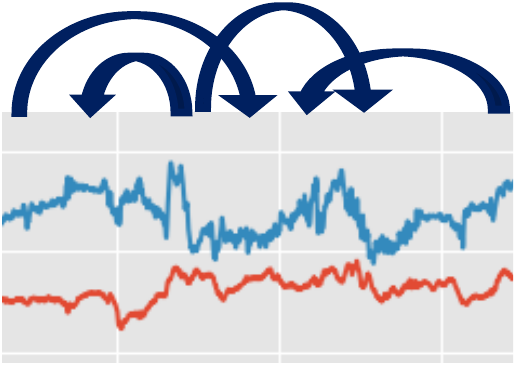}
        \subcaption{Permutation}
      \end{minipage} 
    \hfill
      \begin{minipage}[b]{0.32\hsize}
        \centering
        \includegraphics[scale=.7]{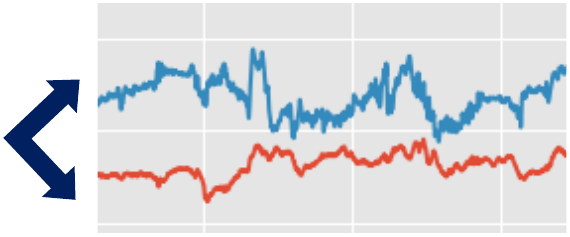}
        \subcaption{Channel shuffle}
      \end{minipage}

    \caption{The pretext tasks for reading detection proposed by Saeed et al.~\cite{SSL_saeed}.}
    \label{fig:transform_exp}
\end{figure}

\begin{figure}[tb]
      \begin{minipage}[b]{0.32\hsize}
        \centering
        \includegraphics[width=\hsize]{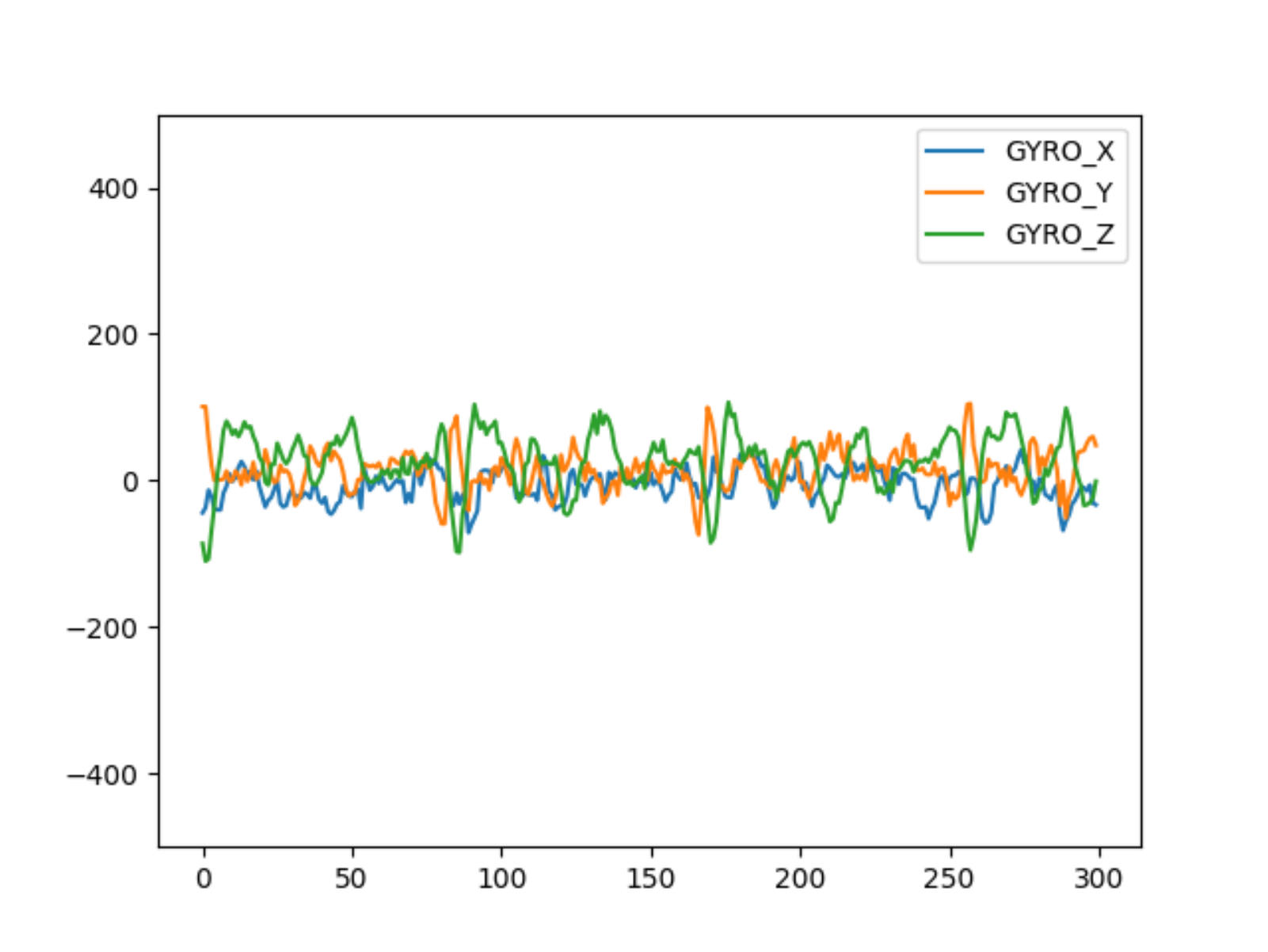}
        \subcaption{No transformation}
      \end{minipage}
    \hfill
      \begin{minipage}[b]{0.32\hsize}
        \centering
        \includegraphics[width=\hsize]{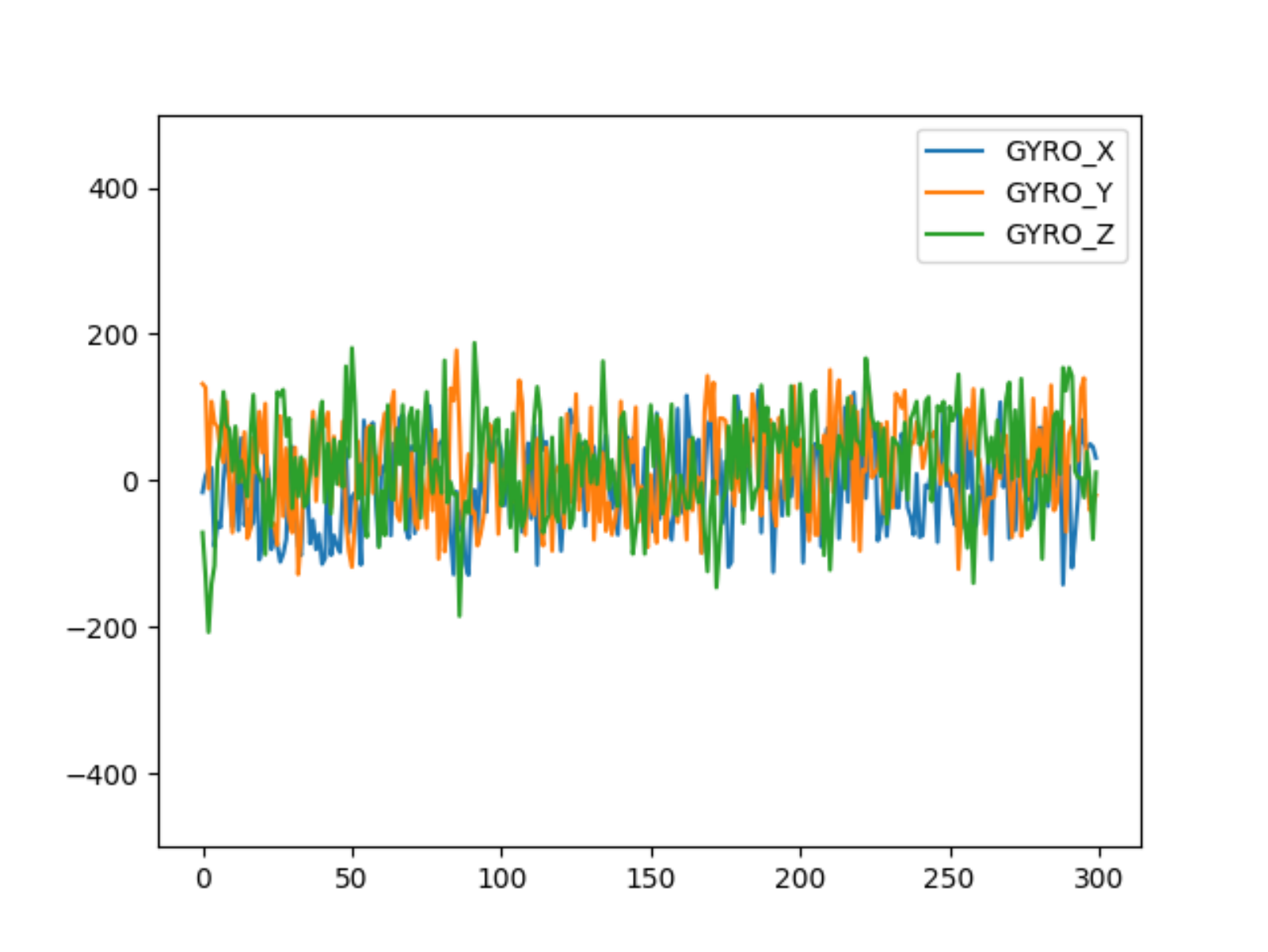}
        \subcaption{Random noise}
      \end{minipage}
    \hfill
       \begin{minipage}[b]{0.32\hsize}
       \centering
        \includegraphics[width=\hsize]{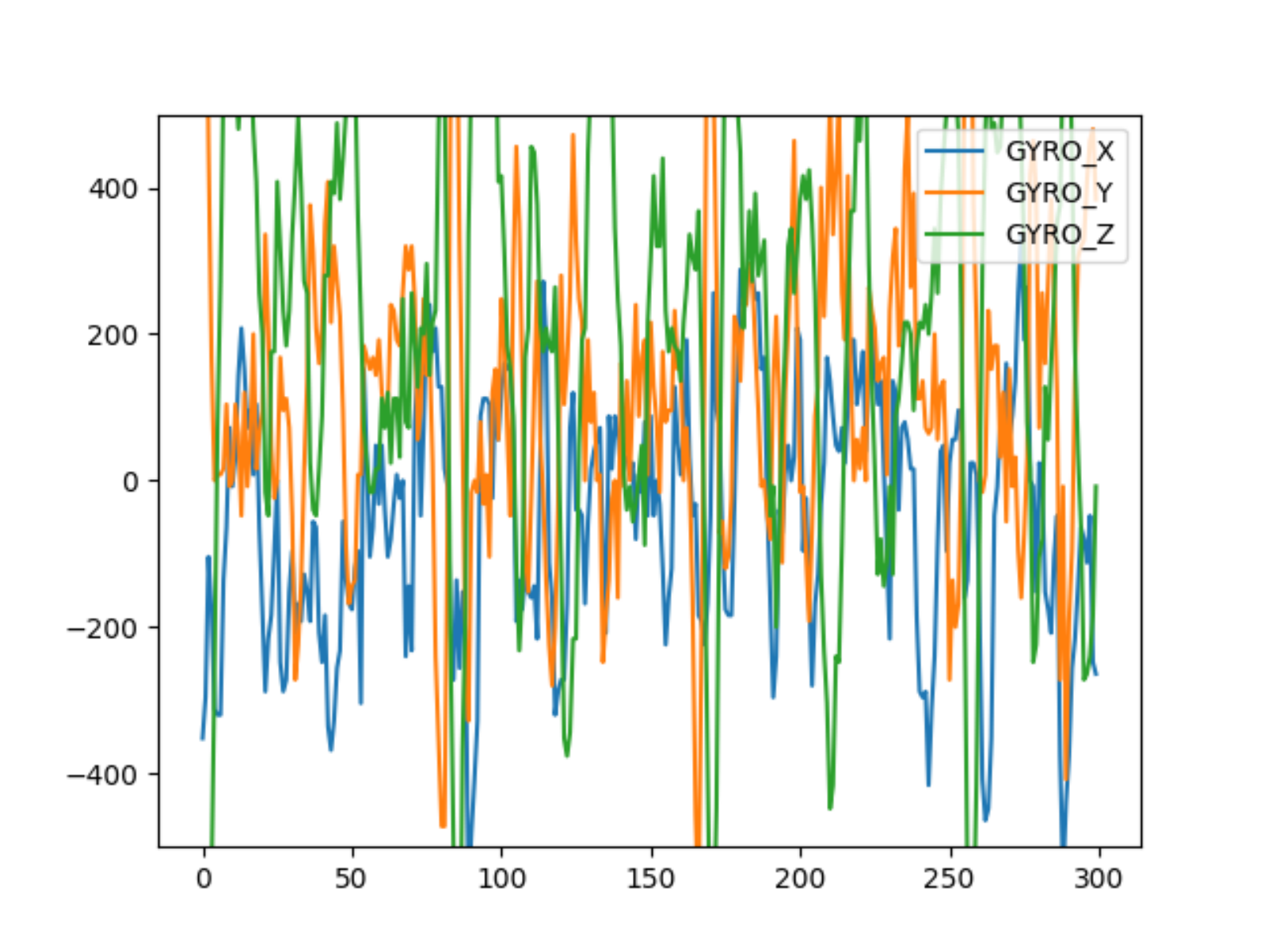}
        \subcaption{Scale}
      \end{minipage}
      
        \vspace*{\intextsep}
      
       \begin{minipage}[b]{0.32\hsize}
       \centering
        \includegraphics[width=\hsize]{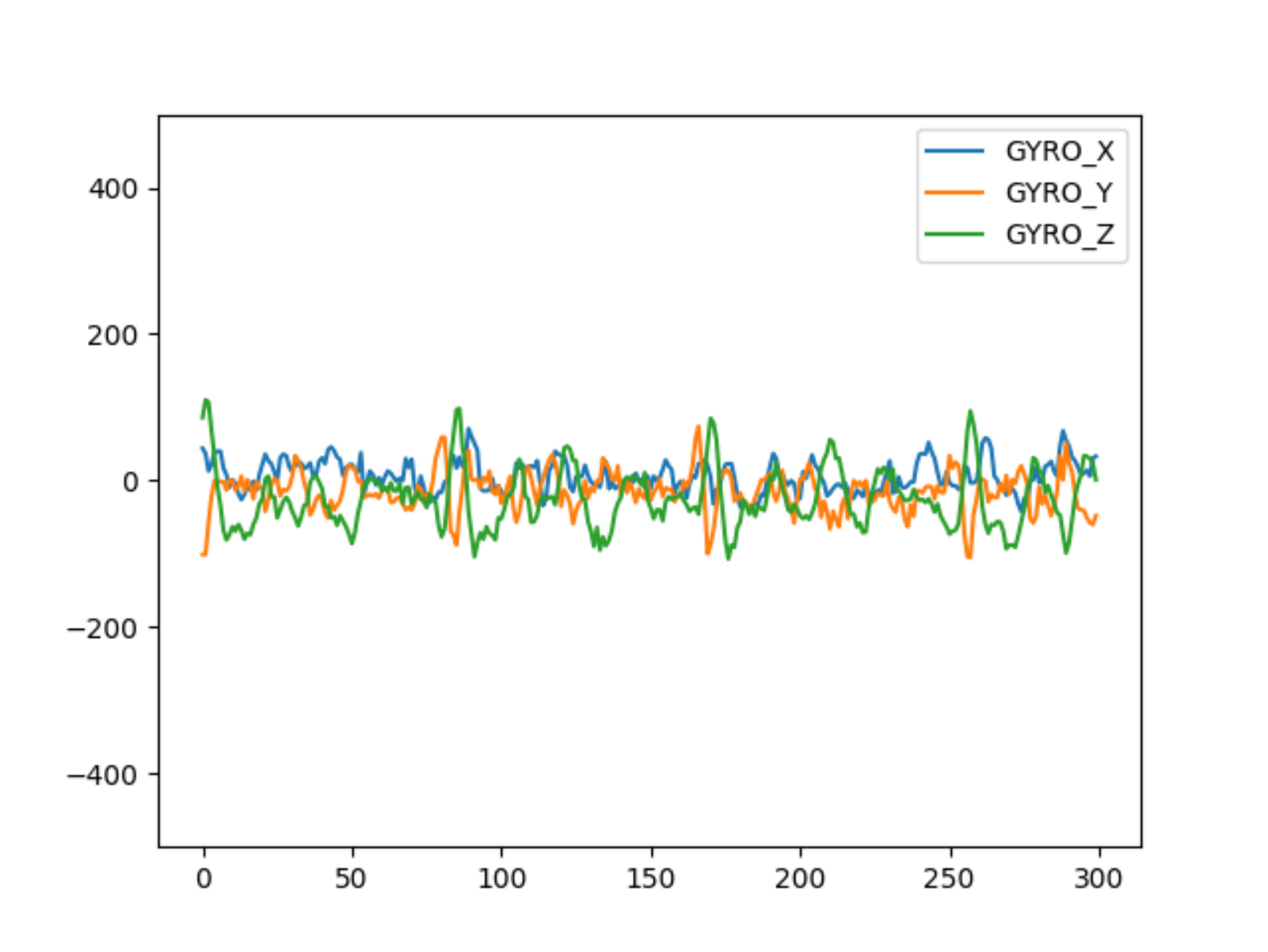}
        \subcaption{Vertical flip}
      \end{minipage}
    \hfill
       \begin{minipage}[b]{0.32\hsize}
       \centering
        \includegraphics[width=\hsize]{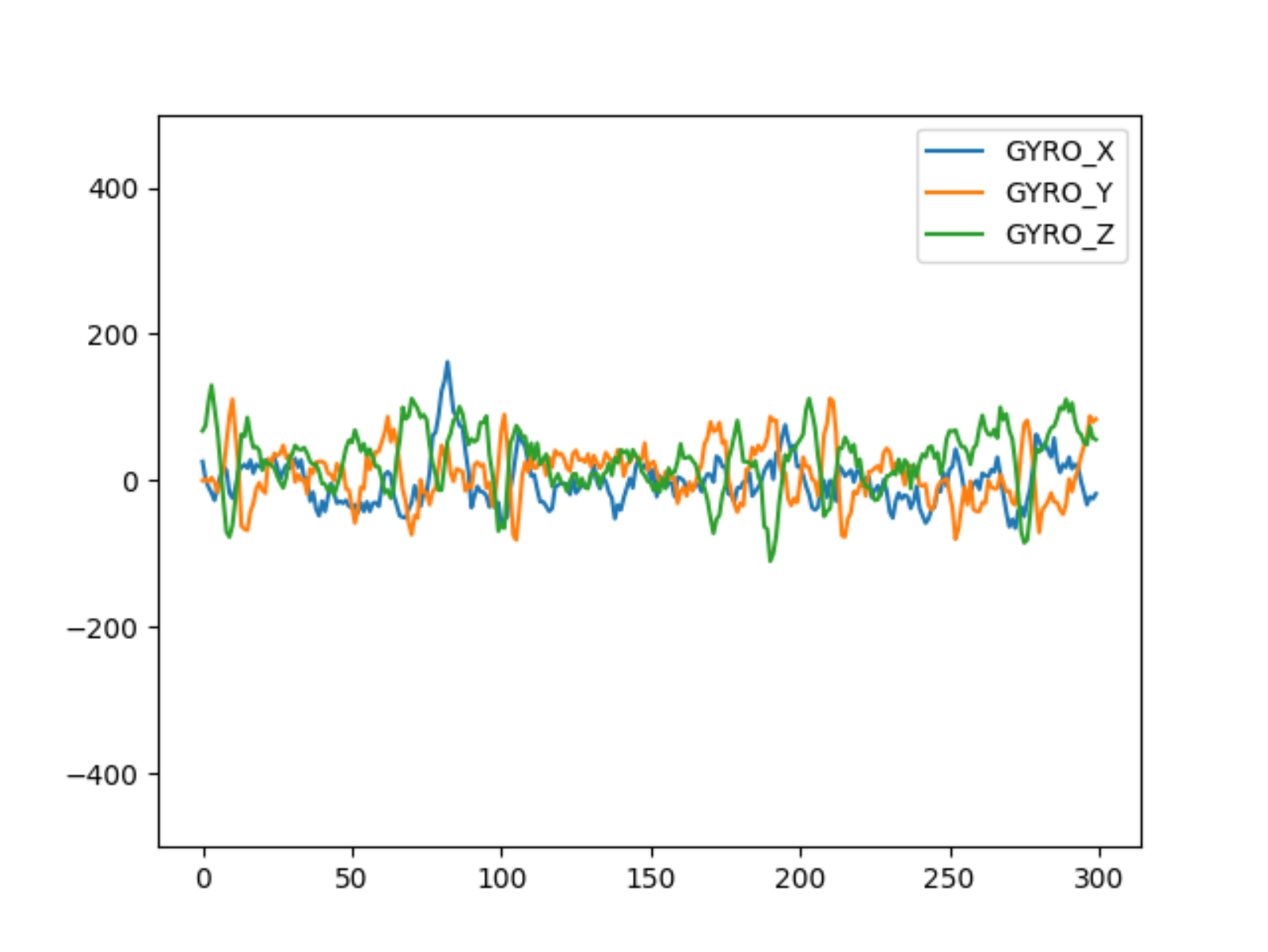}
        \subcaption{Horizontal flip}
      \end{minipage} 
    \hfill
       \begin{minipage}[b]{0.32\hsize}
       \centering
        \includegraphics[width=\hsize]{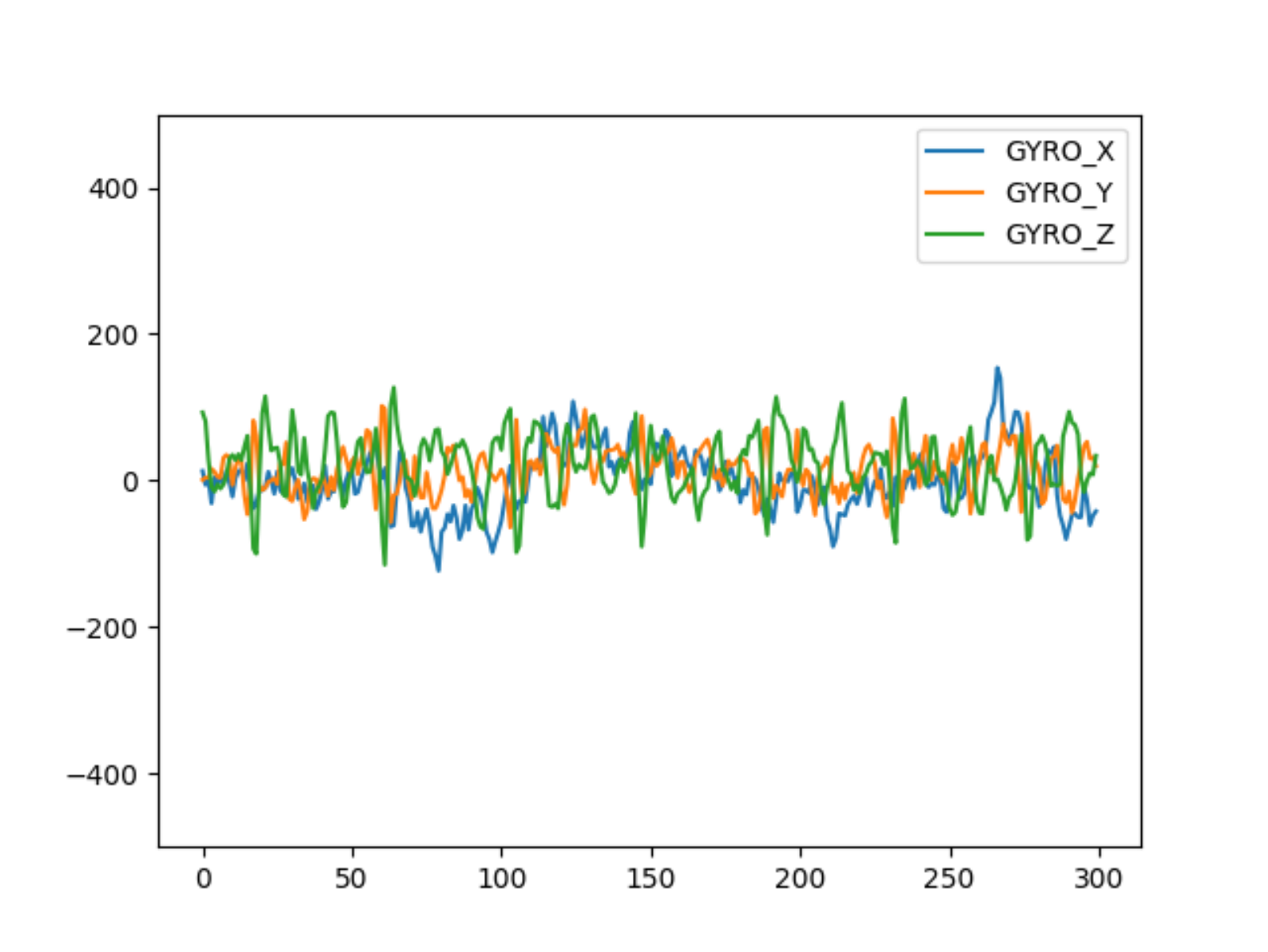}
        \subcaption{Time warp}
      \end{minipage}
      
        \vspace*{\intextsep}

       \begin{minipage}[b]{0.32\hsize}
       \centering
        \includegraphics[width=\hsize]{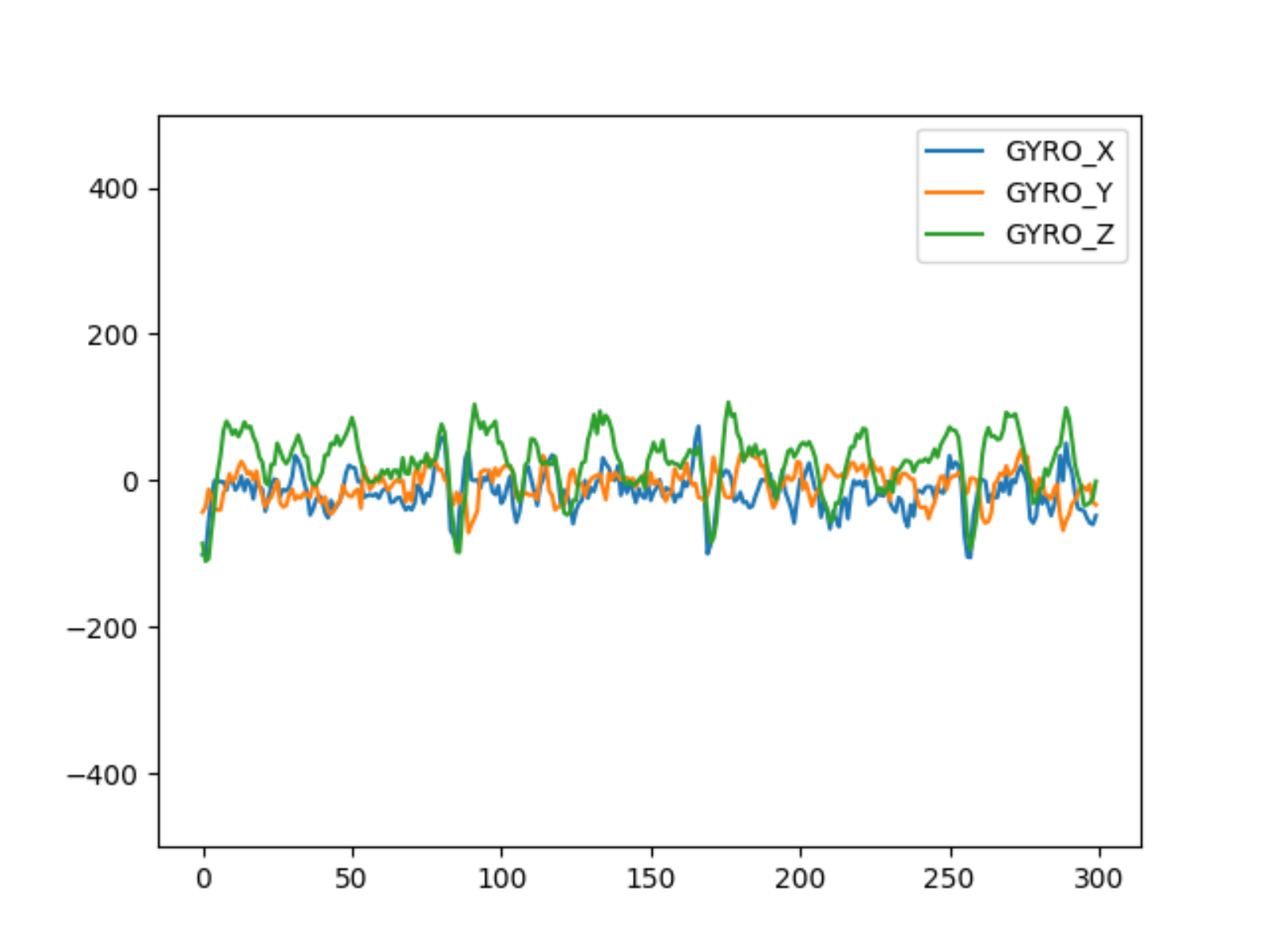}
        \subcaption{Rotation}
      \end{minipage} 
    \hfill
       \begin{minipage}[b]{0.32\hsize}
       \centering
        \includegraphics[width=\hsize]{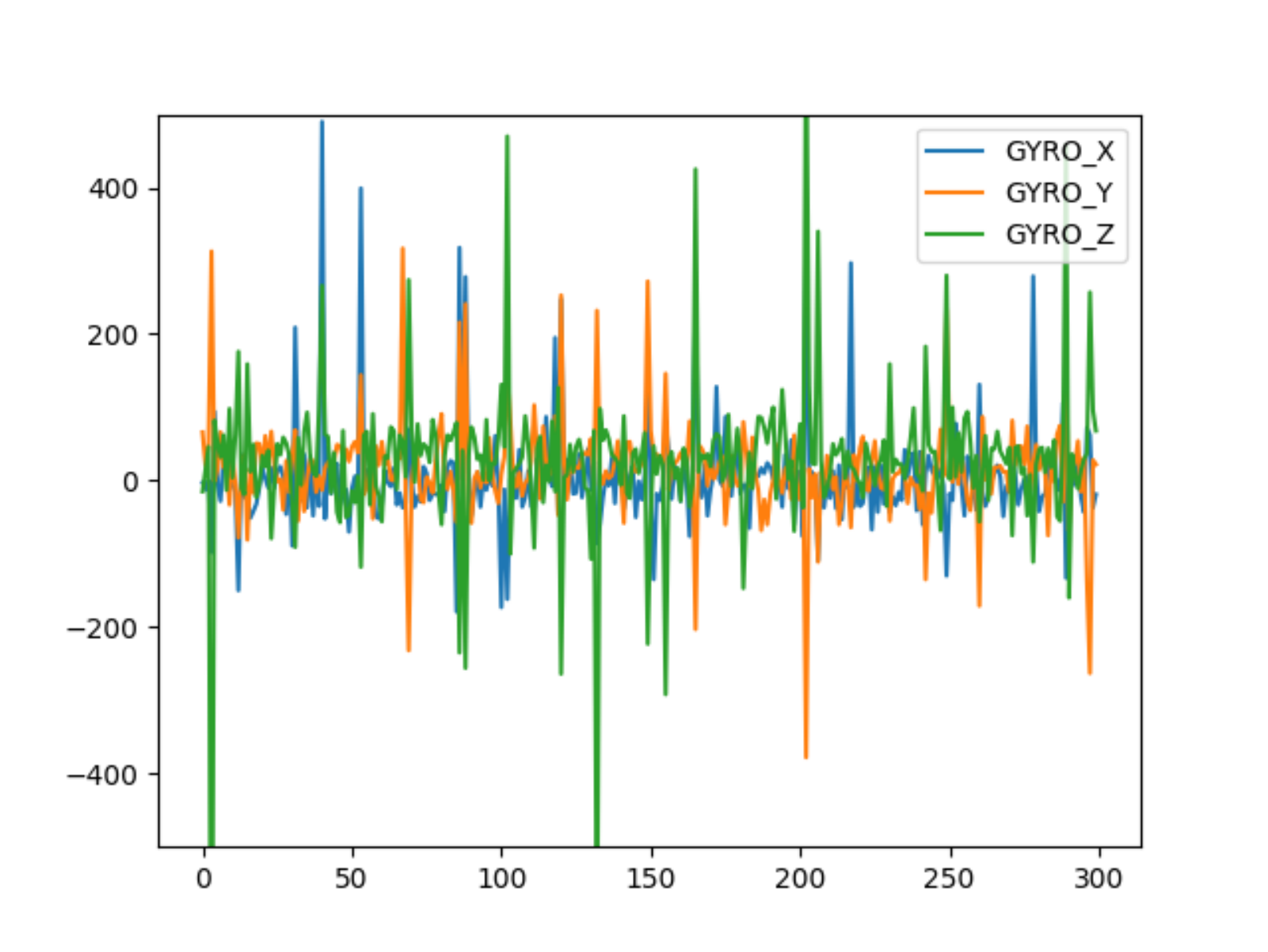}
        \subcaption{Permutation}
      \end{minipage}
     \hfill     
       \begin{minipage}[b]{0.32\hsize}
        \centering
        \includegraphics[width=\hsize]{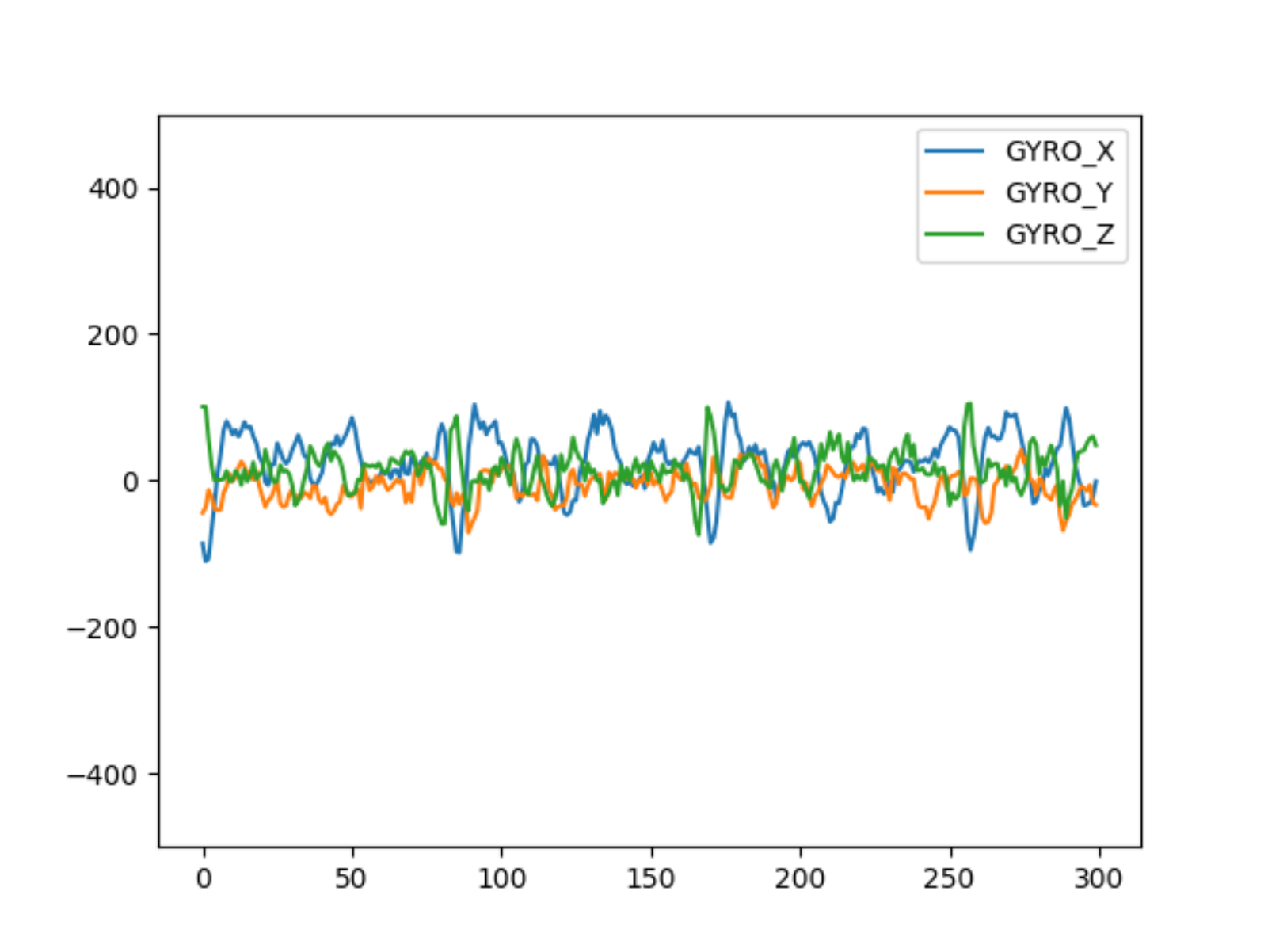}
        \subcaption{Channel shuffle}
      \end{minipage}

    \caption{Examples of GYRO data after applying transformations of \autoref{fig:transform_exp}.}
    \label{fig:transform}
  \end{figure}

Self-supervised pre-training involves learning the representation of signal data by using a pretext task. For the pretext task, we employed the task proposed by Saeed et al.~\cite{SSL_saeed}; the pretext task is to recognize the transformation applied to an input signal. 
For ACC and GYRO data, we employ eight transformations as shown in \autoref{fig:transform_exp}. Because rotation is meaningless for EOG data, seven transformations excluding rotation are applied instead.
Examples of transformed data for GYRO are shown in \autoref{fig:transform}. In the self-supervised pre-training, we solved the eight-class classification (seven transformations $+$ not transformed) problem for EOG data, and nine-class classification (eight transformations $+$ not transformed) problem for ACC and GYRO data.

The red-dashed rectangle in the upper part of \autoref{fig:proposed_rd} illustrates the base network trained by the pretext task.
It consists of three Convolutional Neural Network (CNN) blocks for EOG, ACC and GYRO data, a CNN block that concatenates three CNN layers, and a global max-pooling layer.
Each CNN block consists of three 1D CNN layers. The numbers of units in the CNN layers are 32, 64, and 96, respectively, and the kernel sizes are 24, 16, and 8, respectively. We applied batch normalization after each CNN layer, and a dropout layer after the global max-pooling layer.
Finally, we added three classifiers at the end of the base network for EOG, ACC, and GYRO data, respectively. Each classifier consists of two fully connected (FC) layers, and the numbers of units in the FC layers are 256 and 512, respectively.
We use Rectified Linear Unit (ReLU)~\cite{Nair_ICML2010} as the activation function for all CNN and FC layers, the softmax function as the output layer, and Adam~\cite{Kingma_ICLR2015} as the optimizer.

\subsubsection{Target Task Training}
The final step is the target task training. 
For the reading detection, we have four target classes: not reading (NR), reading English text (EN), reading Japanese horizontal text (JH), and reading Japanese vertical text (JV).
Japanese scripts can be written horizontally or vertically. 
Japanese horizontal writing is similar to English writing except there are no spaces between words, which causing different eye movements when reading the text. In the vertical writing system, characters are read from top to bottom, going right to left~\cite{japanese_h_v}. 

We use the pre-trained base network to create a reading detection network by fine-tuning the pre-trained base network using labeled sensor data with a supervised approach, as shown in the lower part of \autoref{fig:proposed_rd}.
The FC layer in the target task training has 1024 units.
We use the same activation function, optimizer, and output layer as used in the self-supervised pre-training of reading detection.

\subsection{Confidence Estimation}
\begin{figure}[tb]
     \centering
     \includegraphics[width=0.7\hsize]{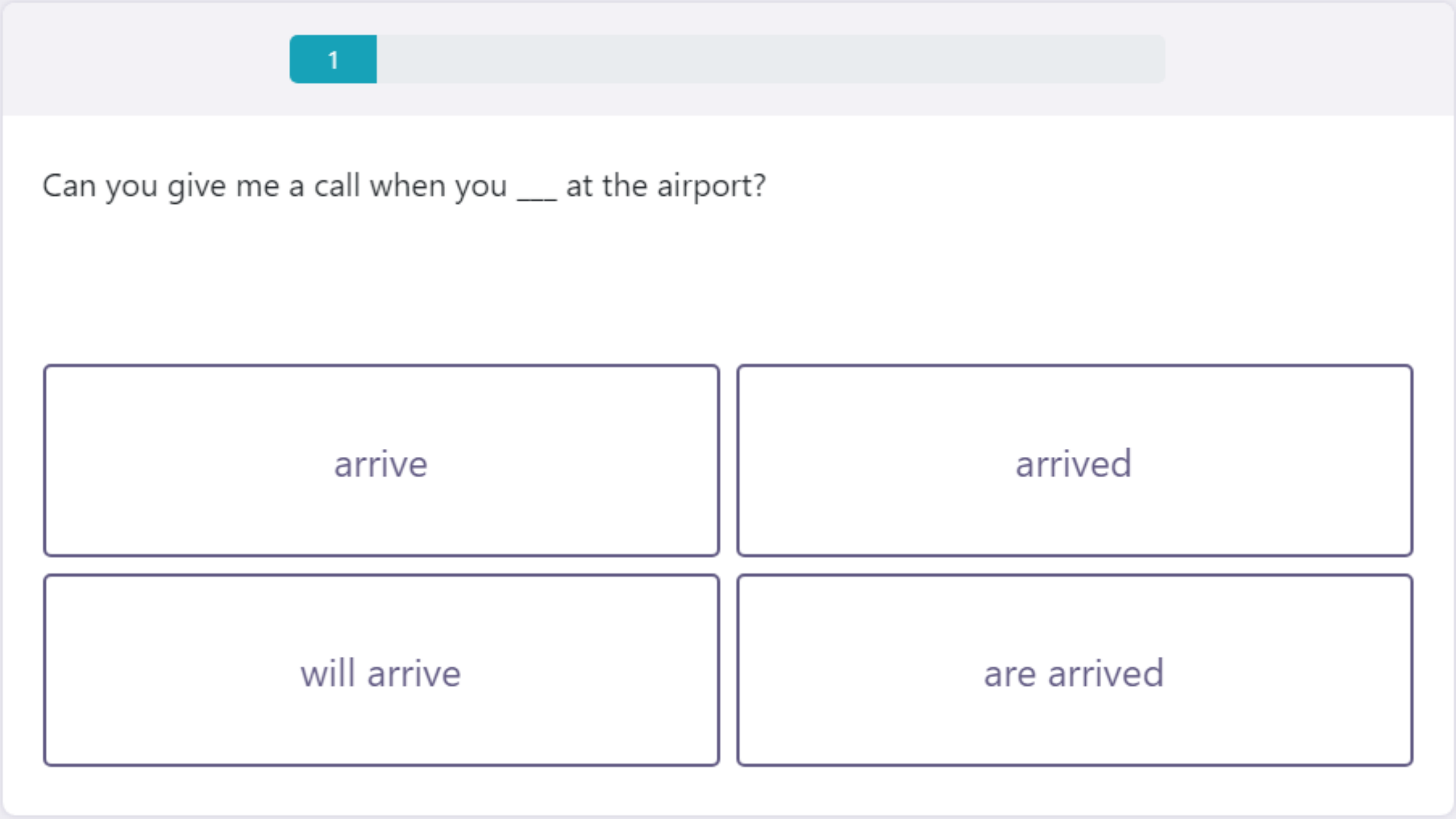}
     \caption{MCQ format used for confidence estimation.
     The user is asked to select one choice to fill in the blank.}
     \label{fig:mcq}
 \end{figure}

\begin{figure}[tb]
      \begin{minipage}[b]{0.24\hsize}
        \centering
        \includegraphics[scale=.12]{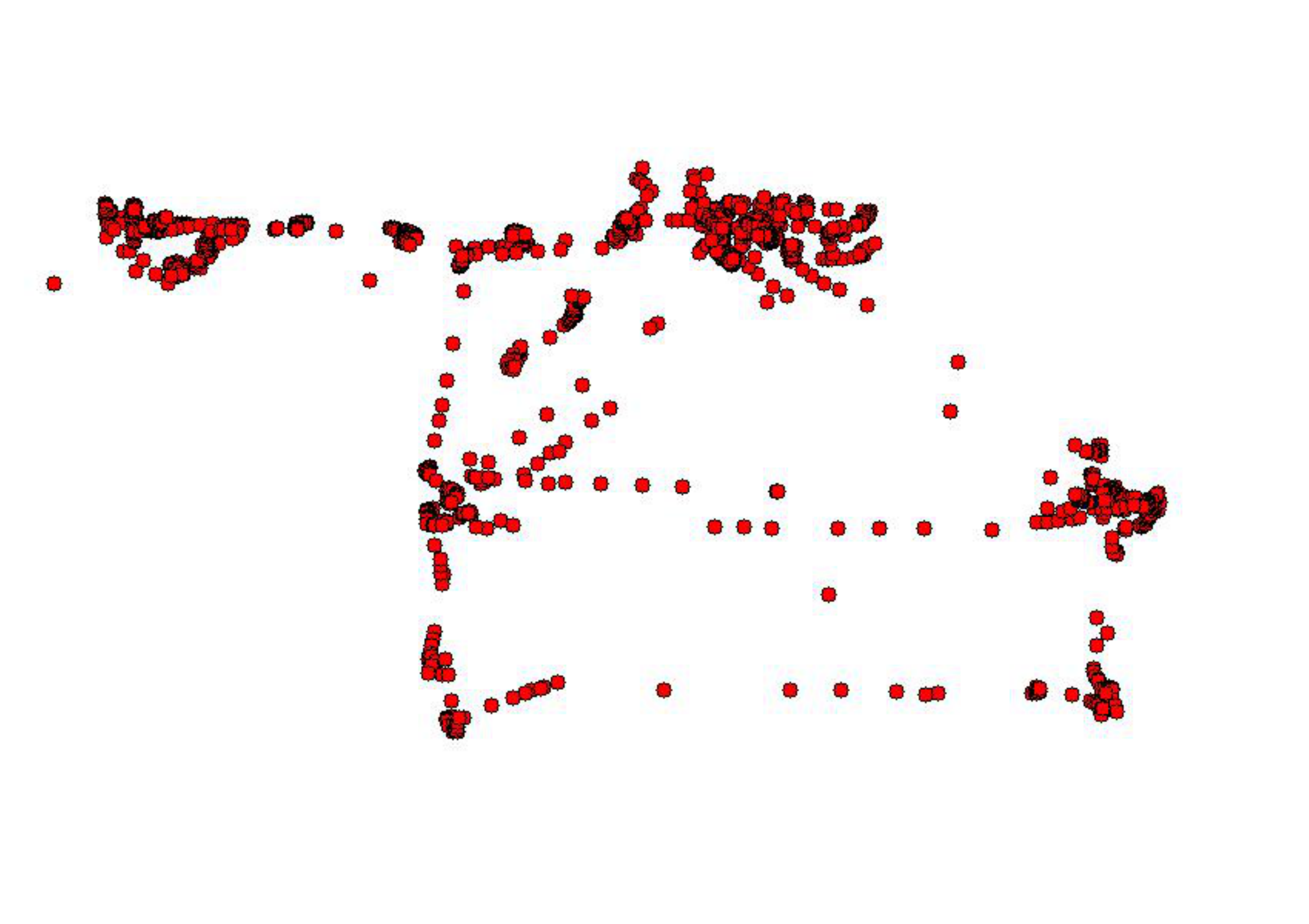}
        \subcaption{No transformation}
        \label{fig:not_transformed_ce}
      \end{minipage}
      \hfill
      \begin{minipage}[b]{0.24\hsize}
        \centering
        \includegraphics[scale=.12]{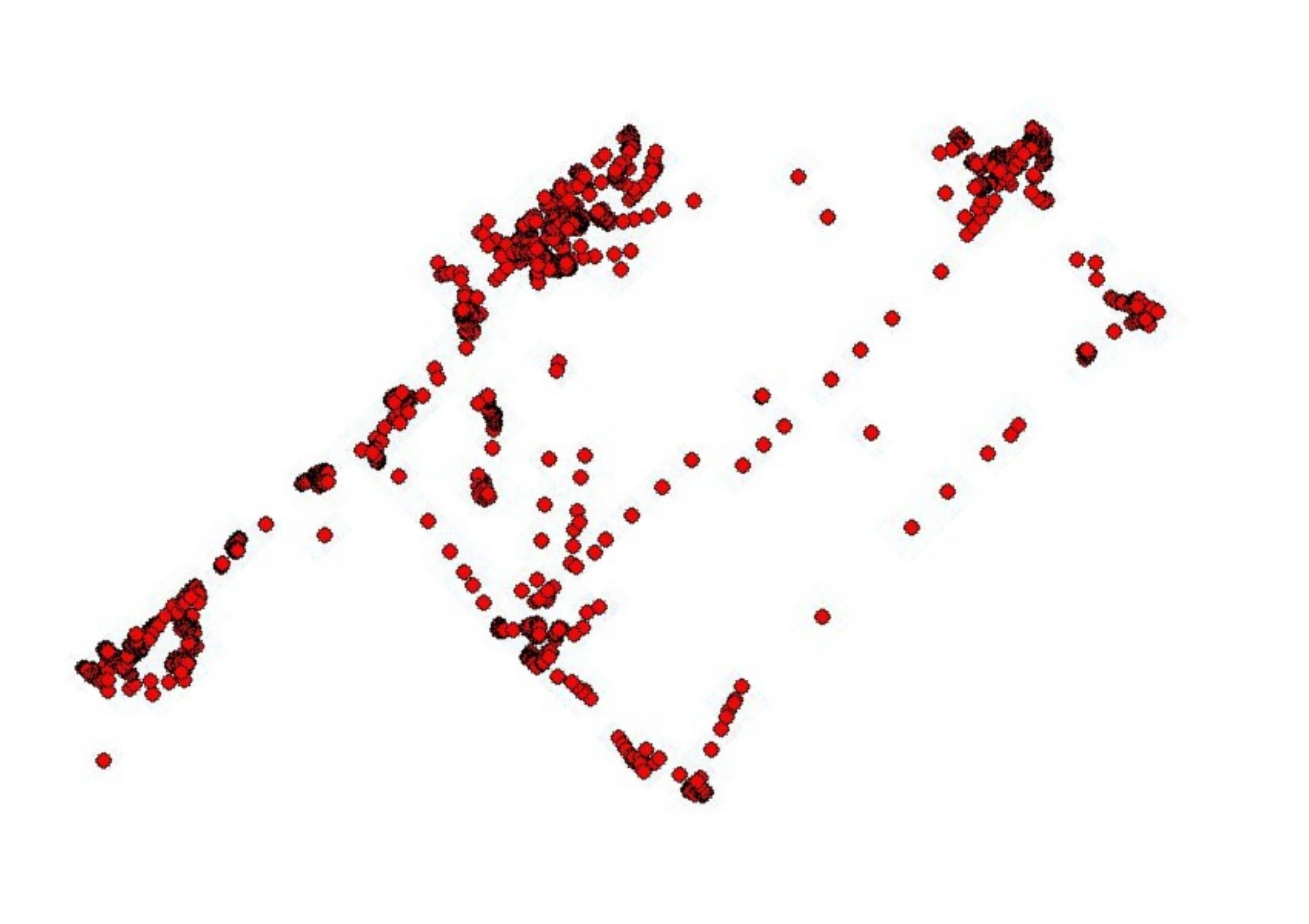}
        \subcaption{Rotation}
        \label{fig:rotated_ce}
      \end{minipage} 
      \hfill
      \begin{minipage}[b]{0.24\hsize}
        \centering
        \includegraphics[scale=.12]{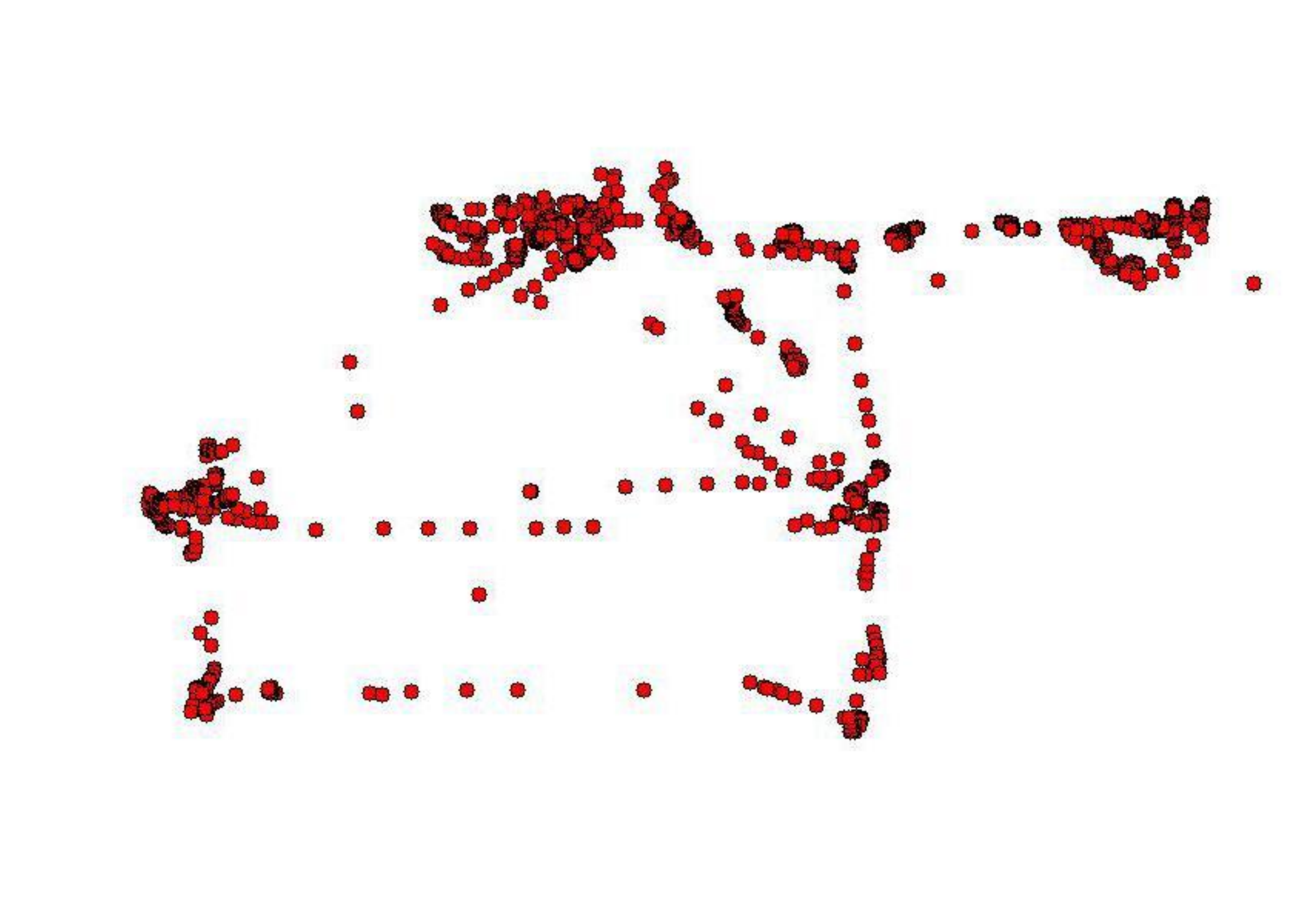}
        \subcaption{Reflection about $y$-axis}
        \label{fig:Mirror}
      \end{minipage} 
      \hfill
       \begin{minipage}[b]{0.24\hsize}
        \centering
        \includegraphics[scale=.12]{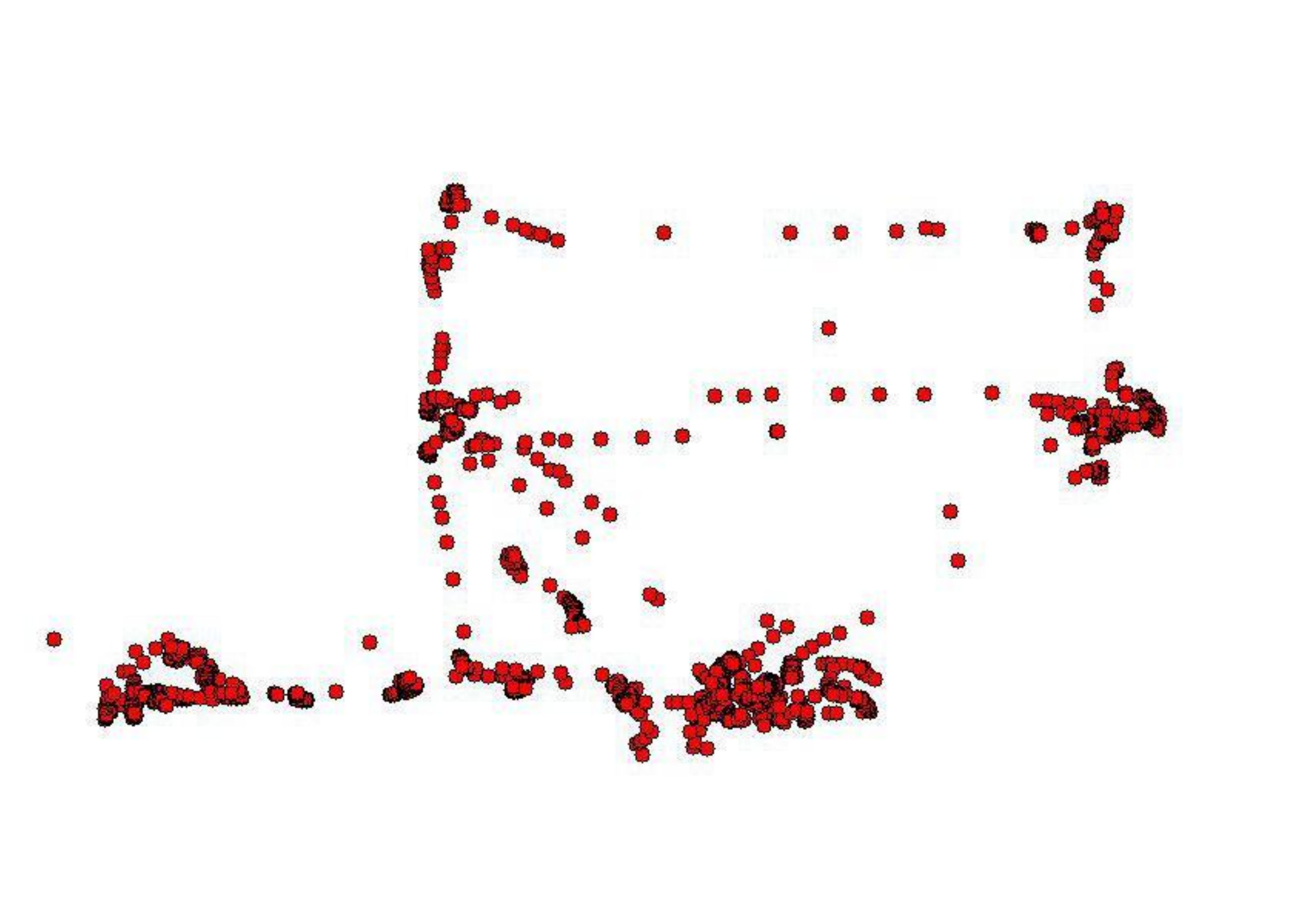}
        \subcaption{Reflection about $x$-axis}
        \label{fig:Flipped}
      \end{minipage}
    \caption{Examples of transformed eye gaze image data in confidence estimation.}
    \label{fig:transformation_ce}
  \end{figure}

Confidence estimation in answering MCQs involves classifying whether the answer is produced 
with confidence or not.
The format for how we handled MCQs in this paper is shown in \autoref{fig:mcq};
the user is asked to select one choice to fill in the blank.
We employed an eye-tracker to describe the user's behaviors.
Unlike the classification of fixed length segments in the reading detection activity, 
the amount of sensor data varies in this task.
To cope with such an issue, researchers of human activity recognition transformed time-series data into images 
to solve the classification task using CNN~\cite{timeseriestoimage_wang, timeseriestoimage_wu, timeseriestoimage_Jiang,timeseriestoimage_hatami,timeseriestoimage_chen}. 
We also convert the eye-tracking data by plotting eye gaze graphically as shown in \autoref{fig:not_transformed_ce}.

\subsubsection{Self-supervised Pre-training}
In the pretext task for confidence estimation, we consider three image transformations as shown in Figures~\multisubref{fig:rotated_ce}{fig:Flipped}. 
The rotation is to apply 45$^\circ$ anti-clockwise rotation to the original image.
Reflection about $x$ and $y$ axes means the transformation of each pixel at $(x,y)$ to $(x,-y)$ and $(-x,y)$, respectively.
In the self-supervised pre-training, we solve the four-class classification (three transformations $+$ not transformed) problem.

The red-dashed box in the upper part of \autoref{fig:proposed_ce} shows the base network that 
consists of two CNN blocks and a max-pooling layer after each CNN block. Besides, we add a dropout layer after the second max-pooling layer followed by a flatten layer.
Each CNN block consists of two 2D CNN layers. The numbers of units of CNN layers are 8 for the first CNN block and 16 for the second CNN block, respectively. The kernel size is $3\times3$ for all four CNN layers. We added a batch normalization after each CNN layer.
At the end of the base network, we add a classifier consisting of two FC layers to identify the type of transformations applied, and the number of units of both FC layers is 36. 
We use ReLU as the activation function for all CNN and FC layers, the softmax function as the output layer, and the stochastic gradient descent as the optimizer. The input image size is $64\times64\times3$.

\subsubsection{Traget Task Training}
After the pre-training, the target task training is performed by replacing the FC layers of the pre-trained network and fine-tuning the pre-trained base network using labeled eye gaze image data.
We designed the confidence estimation target task as a binary classification: confident or unconfident.
For the target task training, the number of units in the final FC layer is 64. 
We used the same input image size, activation function, optimizer, and output layer used in the self-supervised pre-training task for confidence estimation.

\section{Data Collection} \label{sec:datacollection}
\subsection{Reading Detection Datasets} \label{sec:RD_dataset}
\begin{figure}[tb]
\centering
    \begin{minipage}[b]{0.33\hsize}
        \centering
        \includegraphics[width=\hsize]{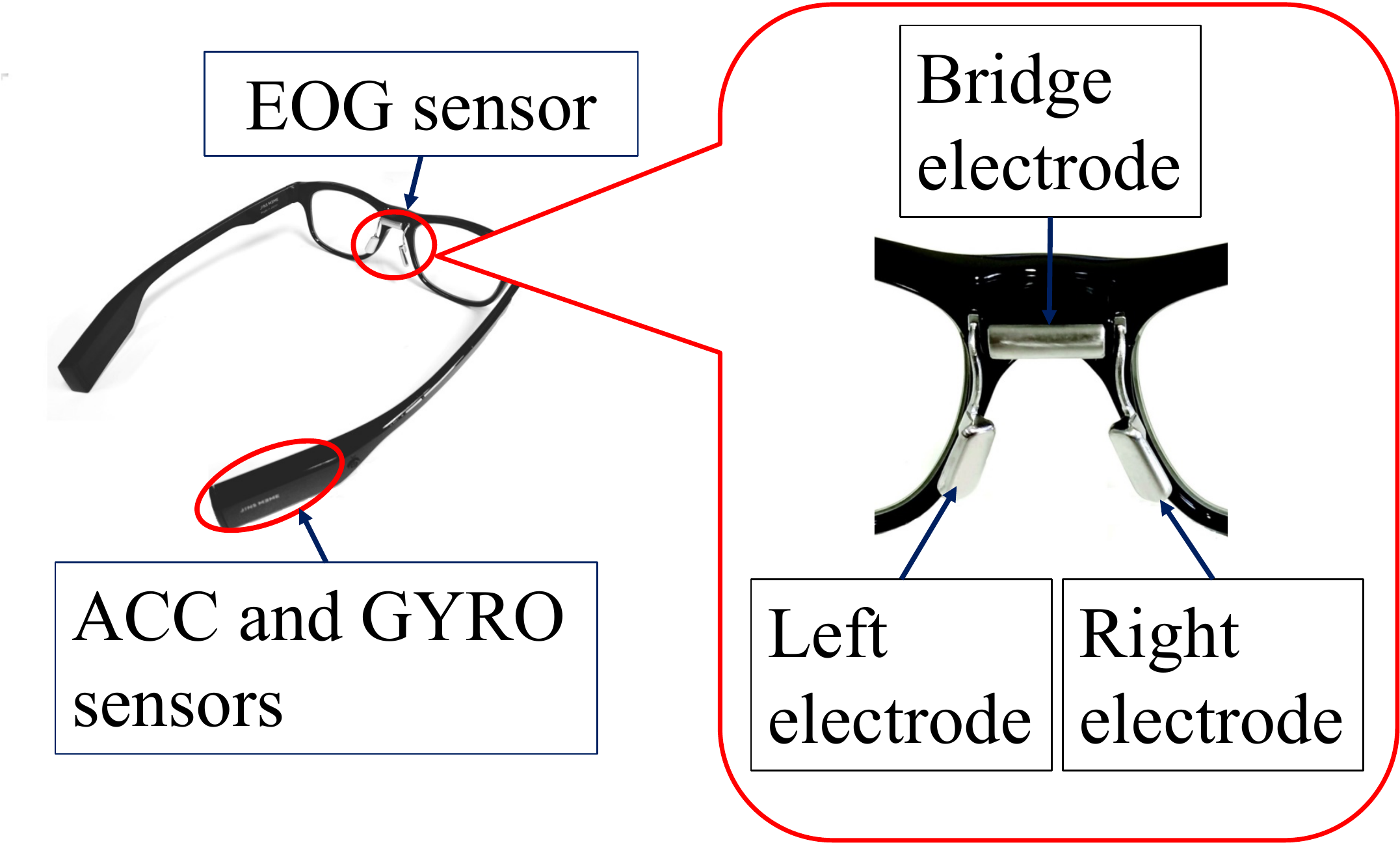}
        \subcaption{}
        \label{fig:jins}
    \end{minipage}
    \hfill
     \begin{minipage}[b]{0.32\hsize}
        \centering
        \includegraphics[width=\hsize]{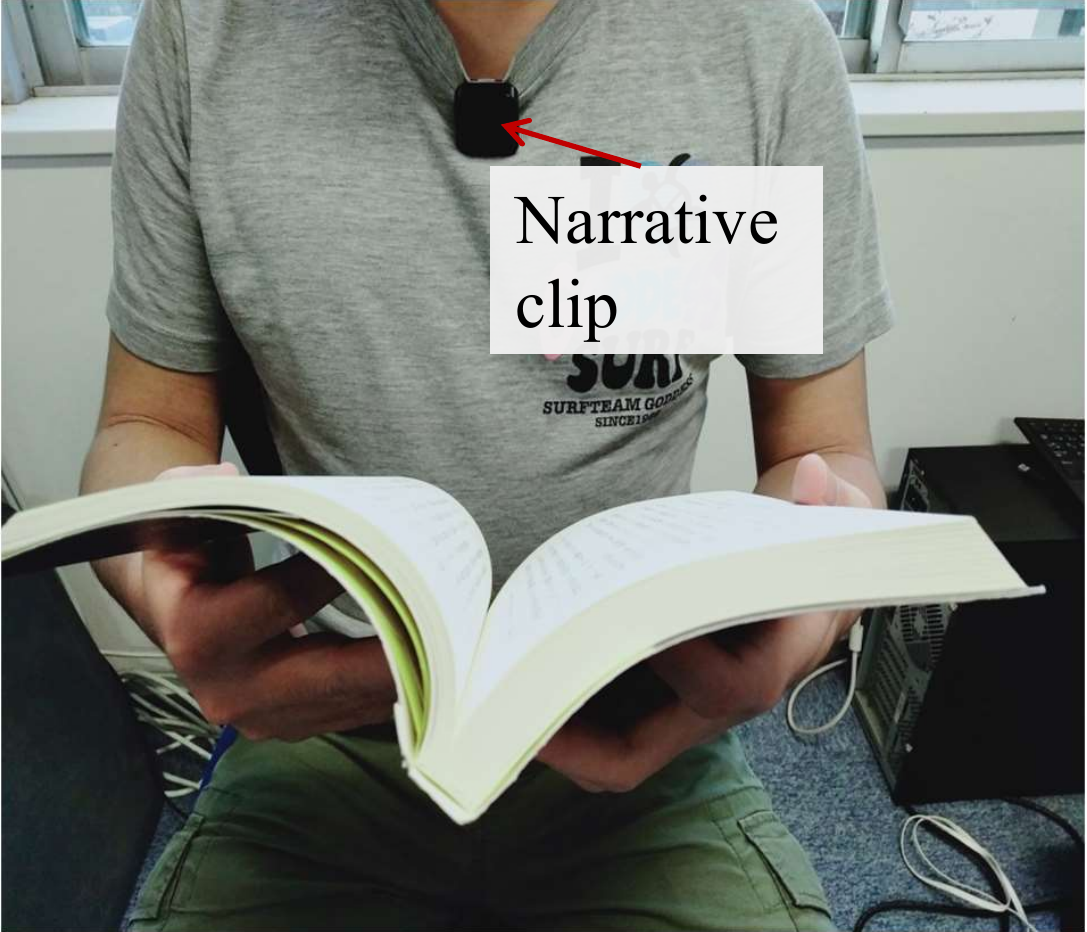}
        \subcaption{}
        \label{fig:jins_read1}
    \end{minipage}
    \hfill
    \begin{minipage}[b]{0.32\hsize}
        \centering
        \includegraphics[width=\hsize]{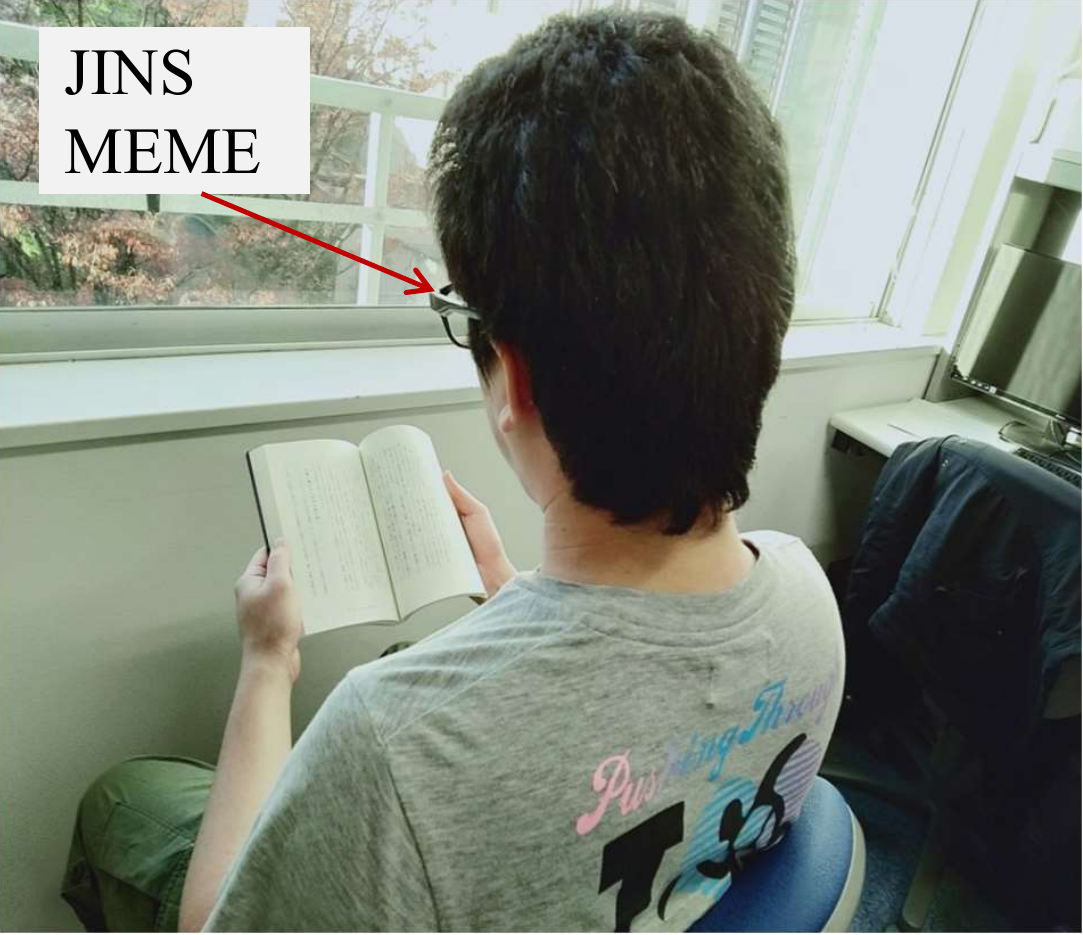}
        \subcaption{}
        \label{fig:jins_read2}
    \end{minipage}
\caption{Data recording for reading detection; (a) JINS MEME EOG glasses, and (b) and (c) a user reading text documents wearing narrative clip and JINS MEME EOG glasses.}
\label{fig:sensor}
\end{figure}

We used two datasets for reading detection: a labeled dataset and an unlabeled dataset. We recorded data for both datasets using JINS MEME EOG glasses~\cite{jins_meme}.
This is an eye-wear device developed by JINS, as shown in \autoref{fig:jins}, which equips EOG, ACC, and GYRO sensors.
EOG is a technique that measures the corneo-retinal standing potential between the front and the back of the human eye.
The EOG sensor consisting of left, right, and bridge electrodes, as shown in \autoref{fig:jins}, records the potential change due to eye movement in the horizontal and vertical directions. 
Let $L$, $R$, and $B$ be the potentials of left, right, and bridge, respectively. The horizontal EOG and vertical EOG are calculated as $L-R$ and  $B- \frac {L + R} {2}$, respectively.
The JINS MEME is also equipped with a three-axis ACC sensor and a three-axis GYRO sensor, as shown in \autoref{fig:jins}, and those measure head movements. The sampling rate of EOG, ACC, and GYRO sensors is 100 Hz.
We describe labeled and unlabeled datasets for reading detection in detail in the following two subsections.

\subsubsection{Labeled Dataset for Reading Detection}

\begin{figure}[tb]
       \begin{minipage}[t]{0.24\hsize}
        \centering
        \includegraphics[width=\hsize]{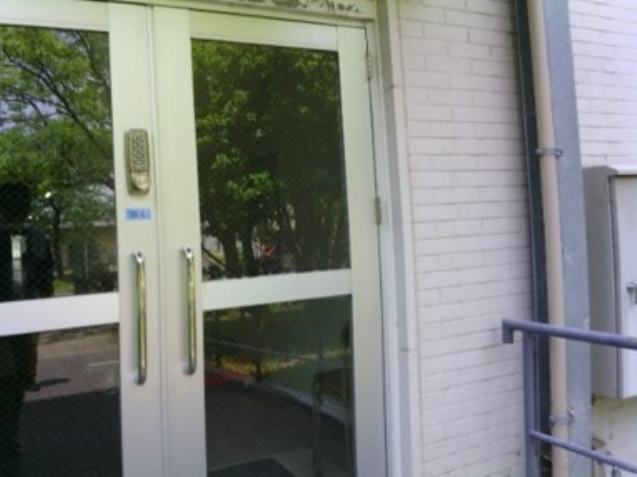}
        \subcaption{Not reading (NR)}
        \label{fig:n_reading}
      \end{minipage}
      \hfill
      \begin{minipage}[t]{0.24\hsize}
        \centering
        \includegraphics[width=\hsize]{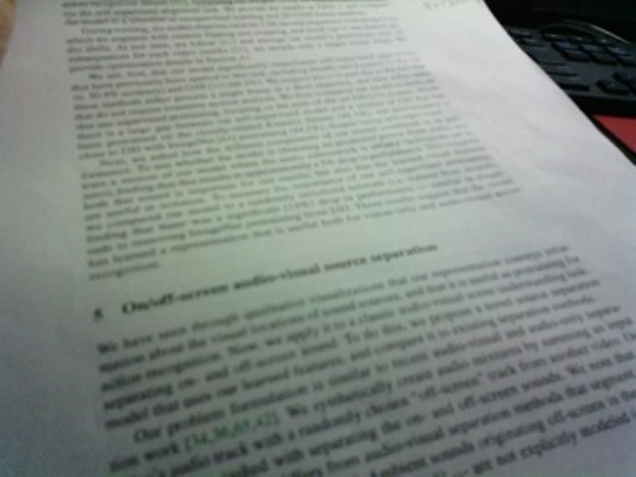}
        \subcaption{Reading English document (EN)}
        \label{fig:e_reading}
      \end{minipage}
      \hfill
      \begin{minipage}[t]{0.24\hsize}
        \centering
        \includegraphics[width=\hsize]{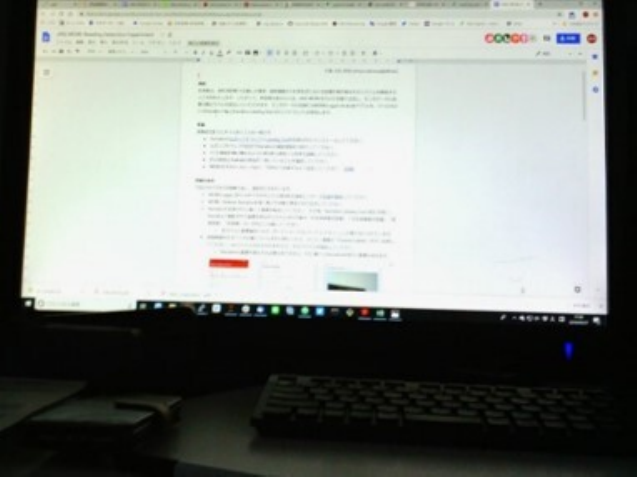}
        \subcaption{Reading Japanese horizontal document (JH)}
        \label{fig:h_reading}
      \end{minipage} 
      \hfill
      \begin{minipage}[t]{0.24\hsize}
        \centering
        \includegraphics[width=\hsize]{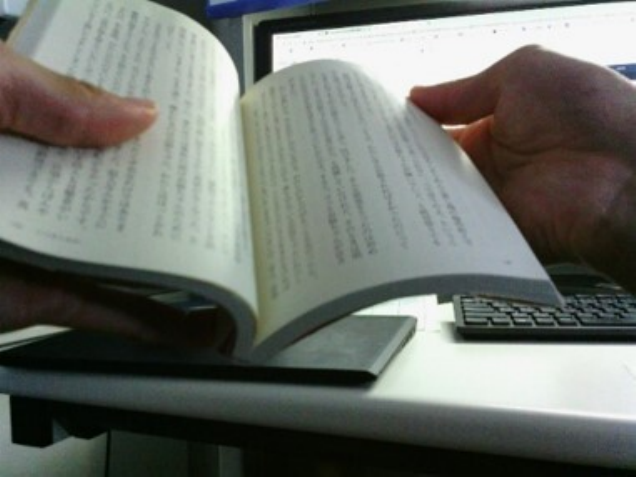}
        \subcaption{Reading Japanese vertical document (JV)}
        \label{fig:v_reading}
      \end{minipage}
    \caption{Images taken with the narrative clip for each type of the user's activity in reading detection.
    }
    \label{fig:narrative_clip}
  \end{figure}

For the labeled dataset, we employed OPU\_RD dataset, which was introduced by Ishimaru et al.~\cite{rd_Ishimaru_13}. 
Ten participants were recruited for data collection. 
Each participant wore the JINS MEME glasses as shown in \autoref{fig:jins_read2} for about 12 hours a day for two days and was asked to read English documents, vertically written Japanese documents, and horizontally written Japanese documents for about 1 hour for each in a day.
Participants also had a small camera called narrative clip on their clothing to take frontal images every 30 seconds as shown in \autoref{fig:jins_read1}. Some images taken with the narrative clip are shown in \autoref{fig:narrative_clip}. 

\begin{figure}[tb]
    \centering
    \begin{minipage}[b]{0.31\hsize}
        \centering
        \includegraphics[width=\hsize]{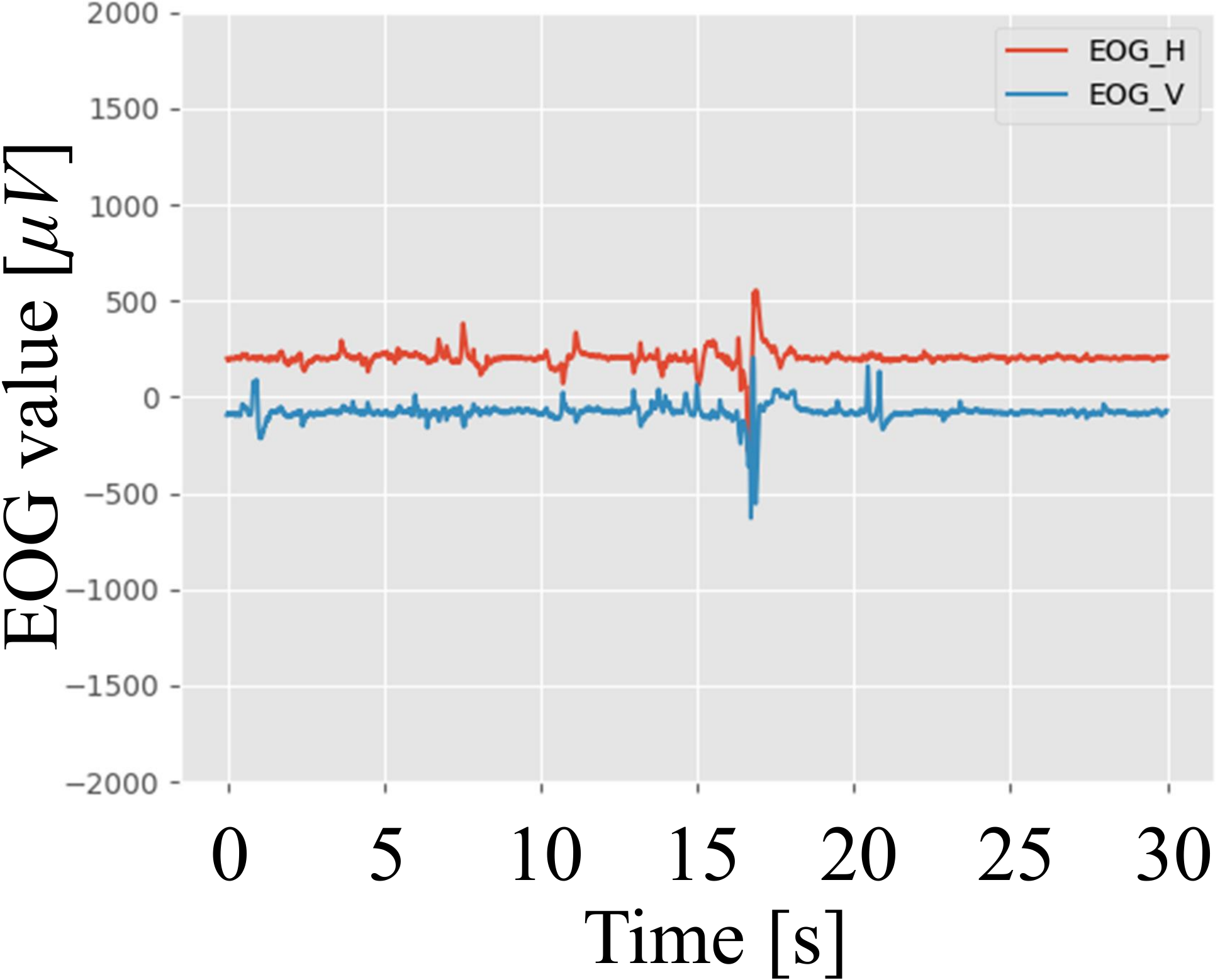}
        \subcaption{EOG}
      \end{minipage}
      \hfill
      \begin{minipage}[b]{0.31\hsize}
        \centering
        \includegraphics[width=\hsize]{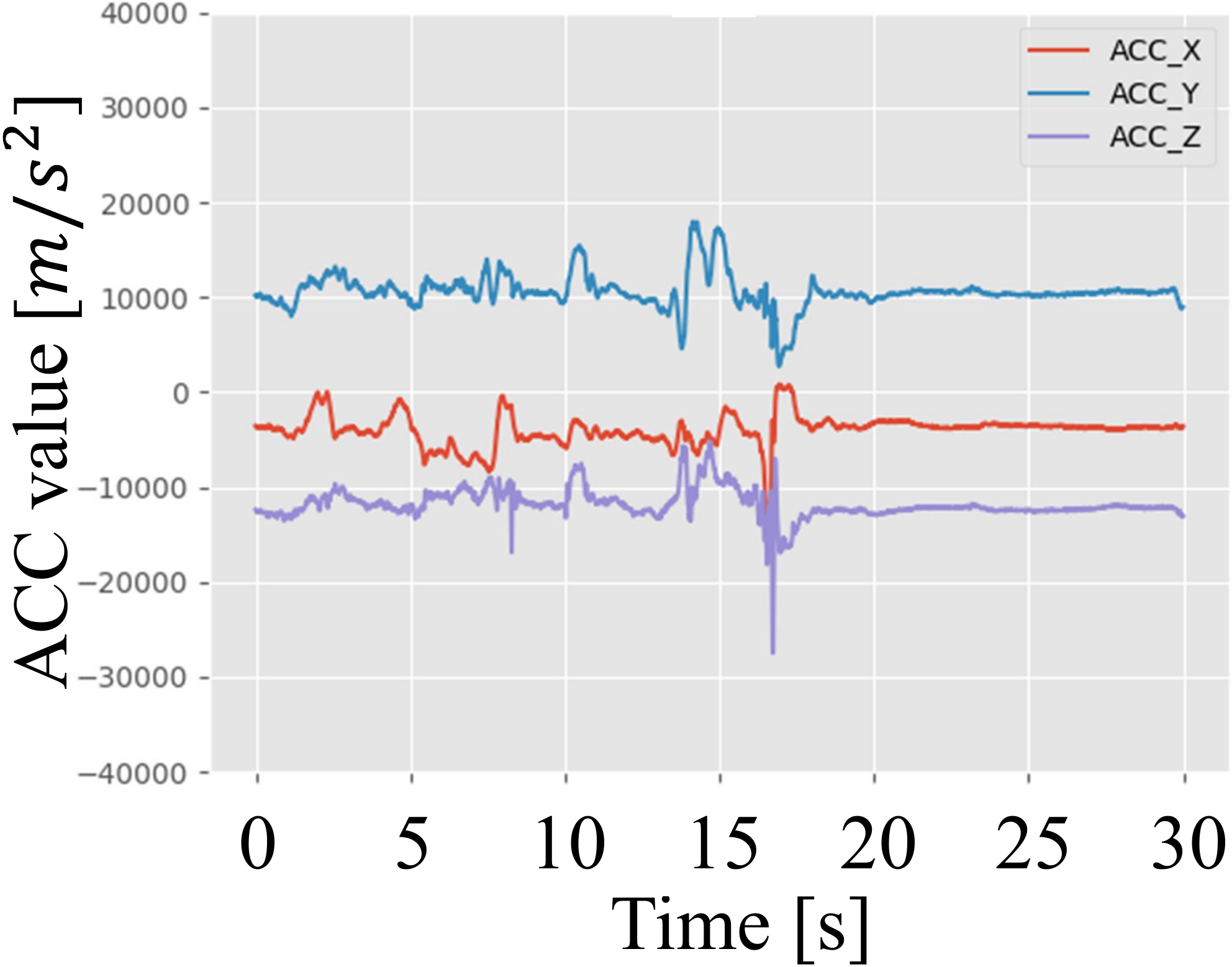}
        \subcaption{ACC} 
      \end{minipage} 
      \hfill
      \begin{minipage}[b]{0.31\hsize}
        \centering
        \includegraphics[width=\hsize]{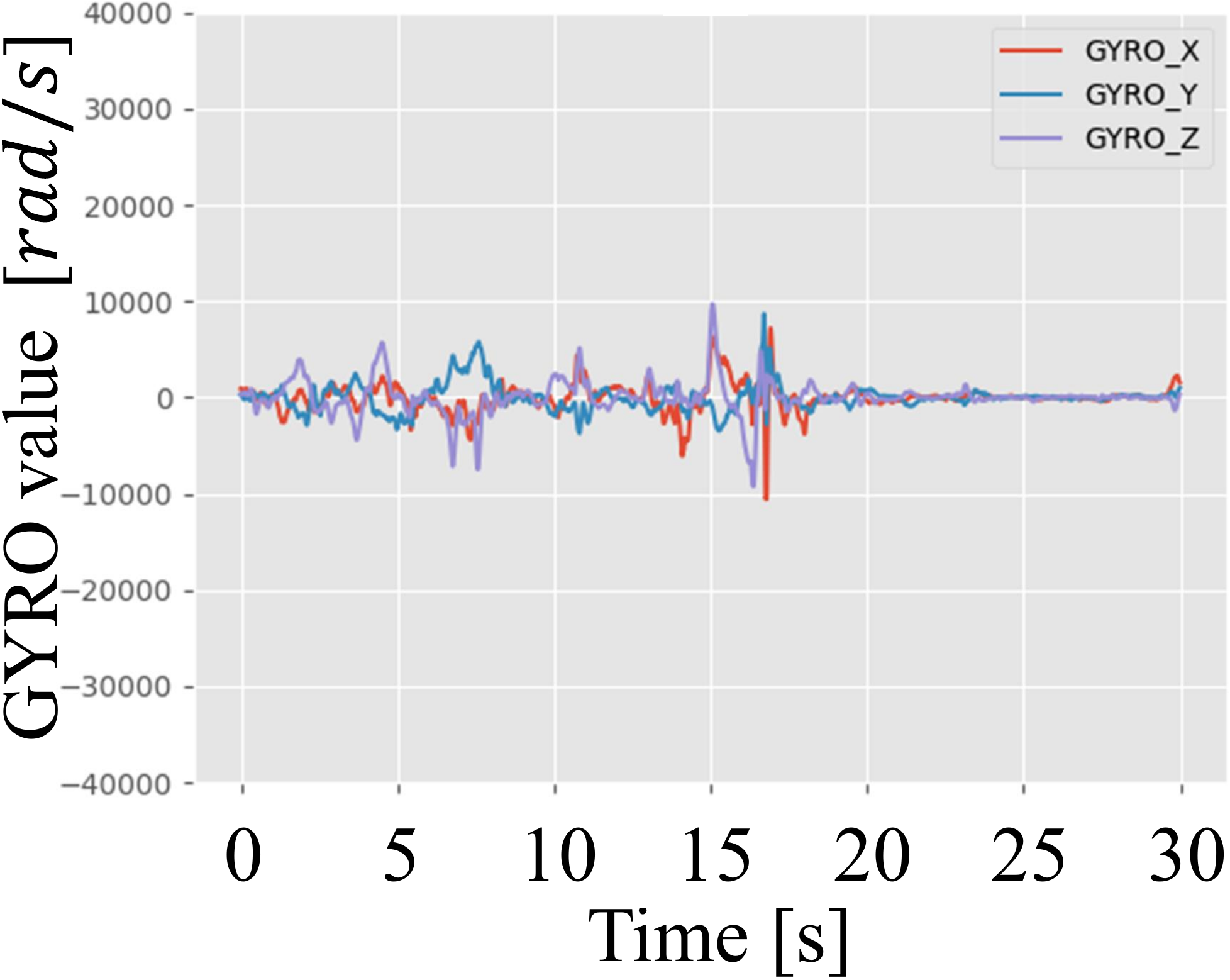}
        \subcaption{GYRO}
      \end{minipage} 
    \caption{Data samples of 30 seconds segments recorded with the JINS MEME EOG glasses in reading detection.}
    \label{fig:data_example_rd1}
\end{figure}

\begin{figure}[tb]
\centering
    \begin{minipage}[b]{0.48\hsize}
    \centering
    \includegraphics[scale=.4]{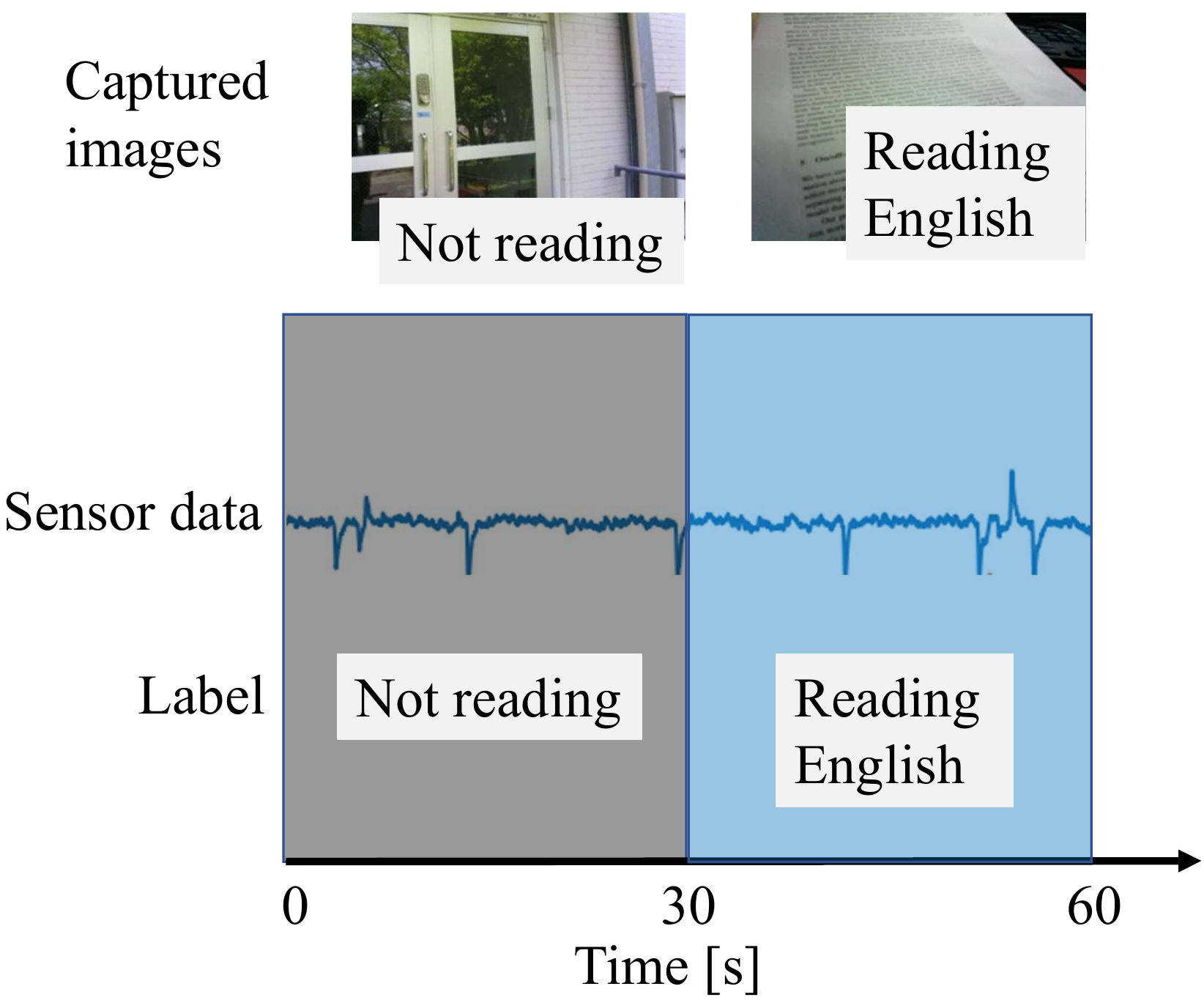}
    \subcaption{Labeling a 30 seconds segment with no label overlap}
    \label{fig:label1}
    \end{minipage}
    \hfill
    \begin{minipage}[b]{0.48\hsize}
    \centering
    \includegraphics[scale=.4]{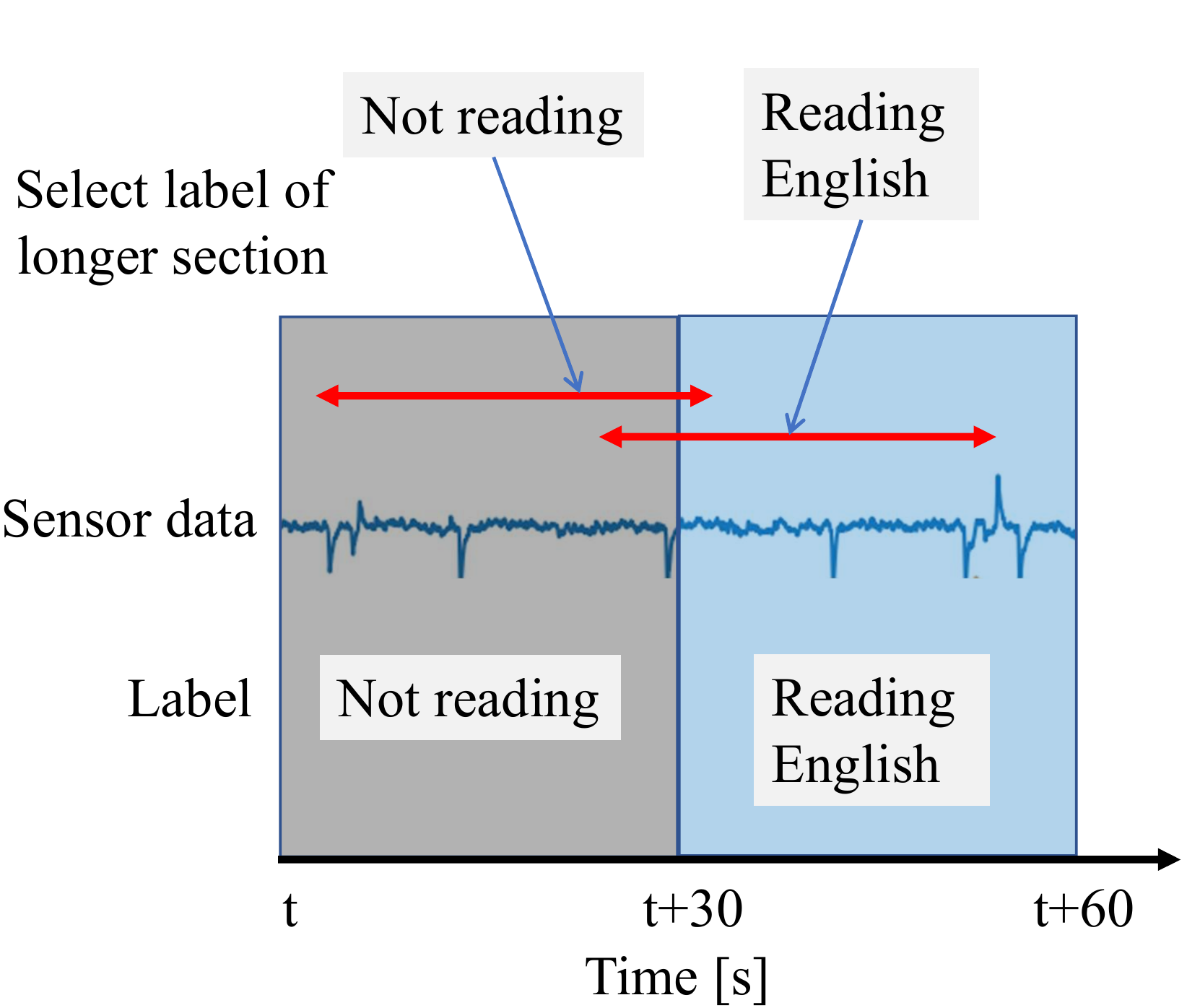}
    \subcaption{Labeling a 30 seconds segment with label overlap}
    \label{fig:label2}
    \end{minipage}
    \caption{Data labeling in reading detection.}
    \label{fig:label}
\end{figure}

The narrative clip was allowed to be removed in places where recording was inappropriate.
Except for the above-mentioned conditions, no restrictions were imposed during data recording.
Thus, the dataset can be regarded as ``in-the-wild.''
At the end of each day, participants labeled recorded data based on the images taken by the narrative clip.
The summary of the participant's activities on the first and second days is shown in \autoref{tbl:recording}.
\begin{table}[tb]
\centering
\caption{Recording time duration [minute] of labeled data for reading detection.}
\begin{minipage}[b]{\hsize}
\centering
\subcaption{Day 1}
\begin{tabular}{c|cccccccccc}
    \hline
    Type of activity & p1 & p2 & p3 & p4 & p5 & p6 & p7 & p8 & p9 & p10 \\
    \hline
    NR & 474 & 447 & 490 & 499 & 476 & 379 & 204 & 538 & 476 & 494 \\
    \hline
    EN & 54 & 73 & 74 & 67 & 63 & 82 & 91 & 62 & 59 & 49 \\
    \hline
    JH & 97 & 70 & 101 & 72 & 73 & 86 & 127 & 63 & 75 & 90 \\
    \hline
    JV & 74 & 101 & 71 & 60 & 83 & 73 & 65 & 66 & 113 & 65 \\
    \hline
\end{tabular}
\label{tbl:day1}
\end{minipage}

\vspace*{\intextsep}

\begin{minipage}[b]{\hsize}
\centering
\subcaption{Day 2}
    \begin{tabular}{c|cccccccccc}
    \hline
    Type of activity& p1 & p2 & p3 & p4 & p5 & p6 & p7 & p8 & p9 & p10 \\
    \hline
     NR & 466 & 499 & 429 & 567 & 283 & 409 & 358 & 496 & 307 & 509 \\
    \hline
    EN & 100 & 96 & 61 & 61 & 74 & 115 & 67 & 55 & 58 & 62 \\
    \hline
    JH & 71 & 74 & 117 & 53 & 64 & 60 & 53 & 58 & 58 & 68 \\
    \hline
    JV & 89 & 75 & 67 & 60 & 116 & 72 & 74 & 73 & 60 & 75 \\
    \hline
\end{tabular}
\label{tbl:day2}
\end{minipage}
\label{tbl:recording}
\end{table}
The data for classification were prepared as follows.
First, the EOG, ACC, and GYRO data were split into segments by using a segment of size 30 seconds slid by 15 seconds. Thus segments overlap with each other.
\autoref{fig:data_example_rd1} shows examples of segments.
After that, each segment is labeled. Since the length of a segment and the frame rate of the narrative clip are the same, one label is assigned to one segment as shown in \autoref{fig:label1}.
However, there are exceptional cases where two labels may be applied to one segment. 
This is because the frame rate of the narrative clip fluctuates. In these cases, the most overlapping label is selected as shown in \autoref{fig:label2}.

\begin{figure}[tb]
\centering
    \includegraphics[scale=.49]{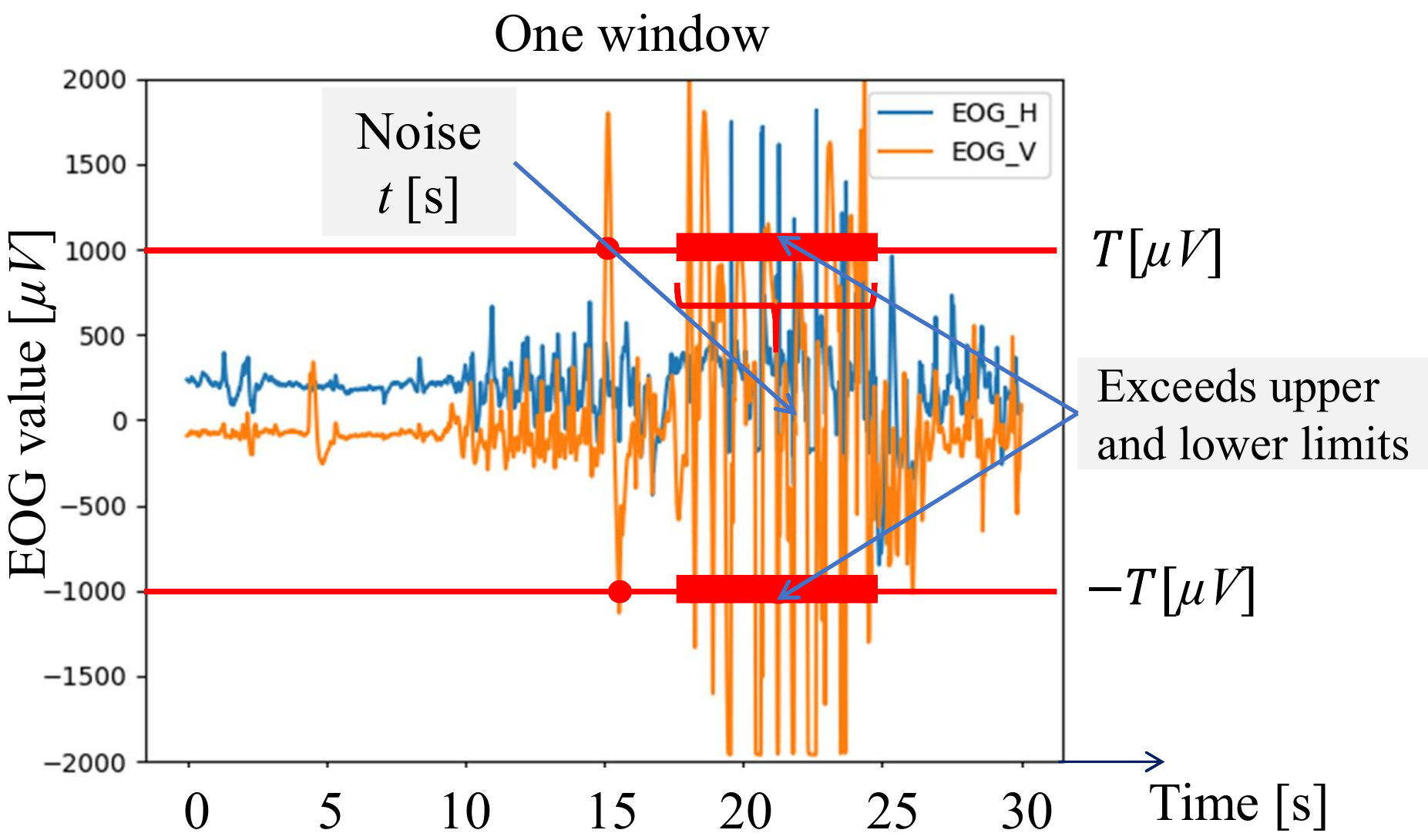}
    \caption{EOG data segment with burst noise and noise judgment criteria.}
    \label{fig:noise_example}
\end{figure}

EOG data sometimes suffers from bursts of noise of about several seconds as shown in \autoref{fig:noise_example} due to poor contact of the EOG electrodes to the skin. 
We found such EOG segments and discarded them using the following criterion:
We set the noise judgment criteria as EOG values above $T$ [$\mu V$] or below $-T$ [$\mu V$]  and exist for continuous $t$ seconds to be considered as noise as shown in \autoref{fig:noise_example}.
In this study, we set $T = 1,000$ [$\mu V$] and $t = 2$ seconds for the labeled dataset for noise determination. 
Also, the EOG, ACC, and GYRO data values deviate from the reference values due to the differences in the mounting sensors. 
Therefore, we corrected the reference value by subtracting the average value of segments from each segment of data. 
The number of segments of each class included in the labeled dataset is shown in \autoref{tbl:rd-labeled-samples}.
\begin{table}[tb]
\centering
\caption{Number of segments in the labeled dataset for reading detection.}
    \begin{tabular}{c|cc}
            \hline
            \multirow{2}{*}{Type of activity} & \multicolumn{2}{c}{Number of segments} \\
            \cline{2-3}
             & Before noise removal & After noise removal\\
            \hline
            NR & 35,228 & 32,708\\
            \hline
            EN & 5,724 & 5,340\\
            \hline
            JH & 6,166 & 5,792\\
            \hline
            JV & 6,161 & 5,798\\
            \hline
    \end{tabular}
    \label{tbl:rd-labeled-samples}
\end{table}

\subsubsection{Unlabeled Dataset for Reading Detection}

We recruited 13 Japanese university students. Each participant wore a JINS MEME device for three to eight days and read English document, horizontally written Japanese documents, and vertically written Japanese documents, or did not read anything at all. The measurement time is about 20 to 60 hours per person, and the total recorded time from all participants is 676 hours.
In addition, we employed an unlabeled dataset that recorded EOG, ACC, and GYRO data when 39 experimental participants attended presentations at a conference. Because there is no restriction,
these unlabeled datasets are considered ``in-the-wild.''
The unlabeled data totals about 1,359 hours. We also divided the unlabeled data into 30 seconds segments with 15 seconds overlap in the same manner as described for the labeled dataset. 
We discarded noisy segments using the noise judgment criteria
with the parameters $T = 1,000$ [$\mu V$] and $t = 0.01$ seconds and corrected the reference value. After noise removal, the total number of segments in the unlabeled dataset is 177,921.

\subsection{Confidence Estimation Datasets}
\begin{figure}[tb]
      \begin{minipage}[t]{0.32\hsize}
        \centering
        \includegraphics[scale=.28]{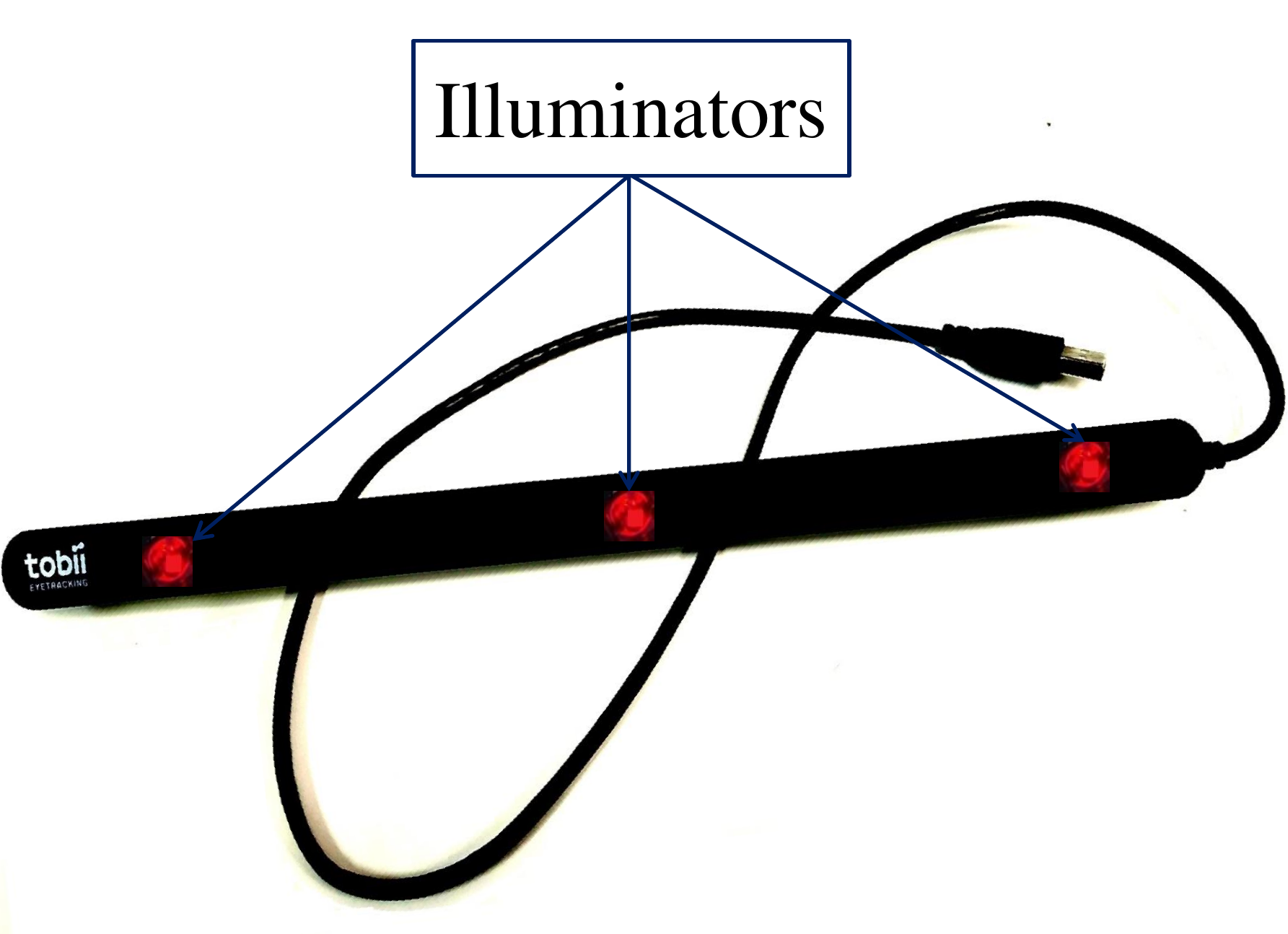}
        \subcaption{Tobii 4C eye-tracker}
        \label{fig:tobii}
    \end{minipage}
    \hfill
      \begin{minipage}[t]{0.31\hsize}
        \centering
        \includegraphics[width=\hsize]{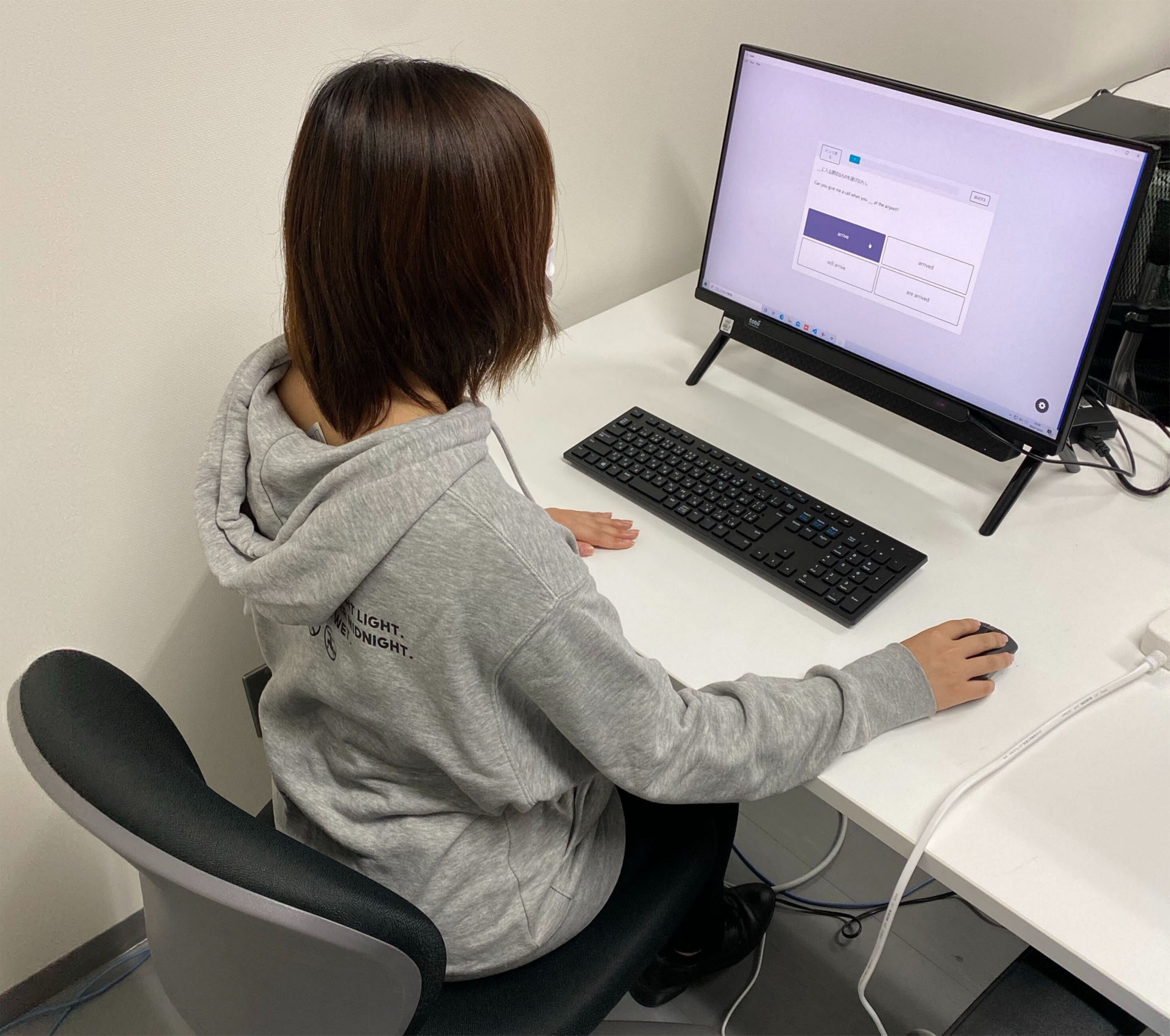}
        \subcaption{User answering MCQs}
        \label{fig:answering}
      \end{minipage}
        \hfill
      \begin{minipage}[t]{0.31\hsize}
        \centering
        \includegraphics[width=\hsize]{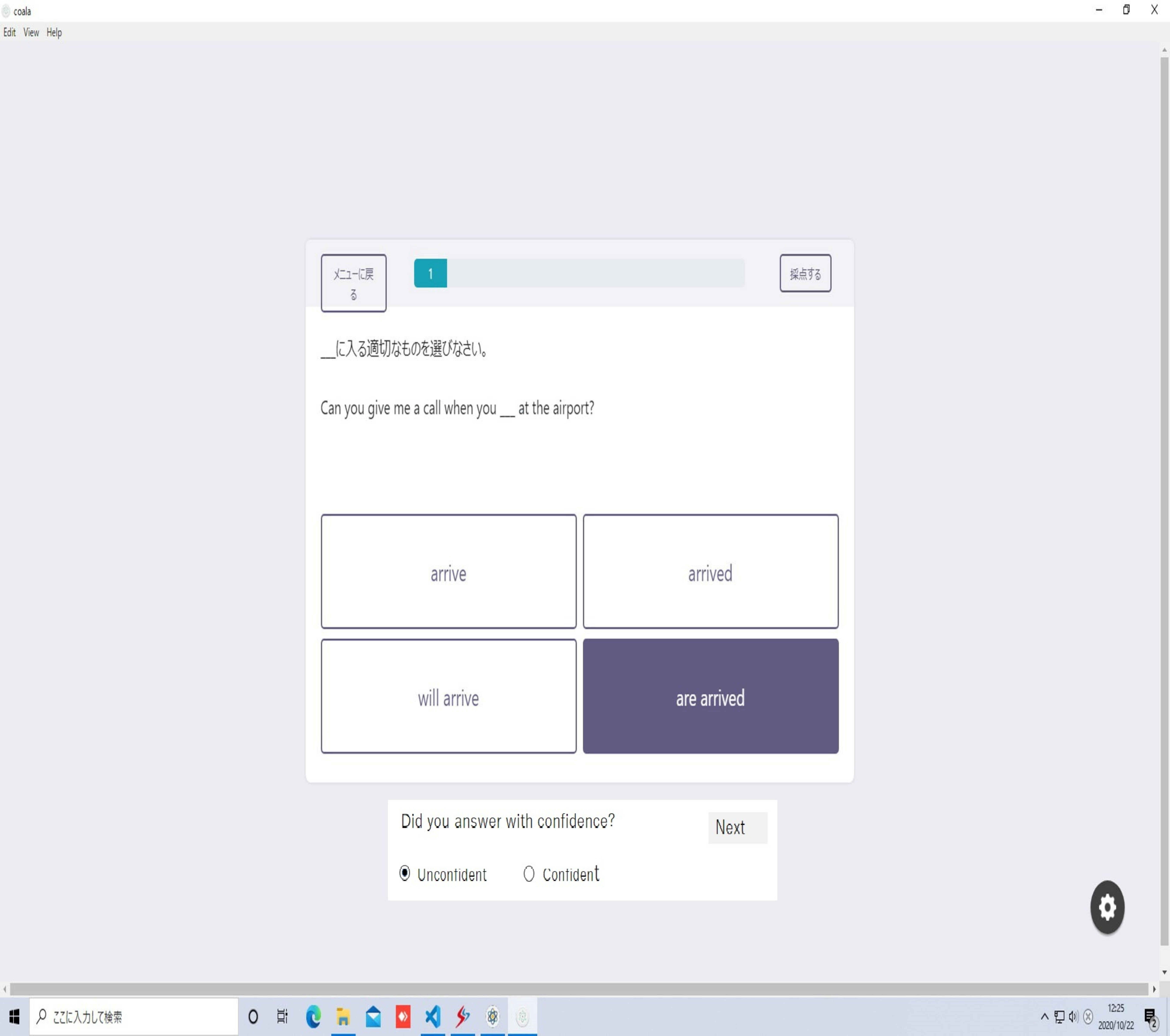}
        \subcaption{Screenshot of an MCQ with an inquiry about the confidence label}
        \label{fig:labeling_ce}
      \end{minipage}
      
    \caption{Data collection environment for confidence estimation.
    }
    \label{fig:ce_experiment}
  \end{figure}

We also use a labeled dataset and an unlabeled dataset for confidence estimation. 
We used the Tobii 4C pro-upgraded eye-tracker~\cite{tobii}, as shown in \autoref{fig:tobii}, a stationary eye-tracker whose sampling rate is 90 Hz.

\subsubsection{Labeled Dataset for Confidence Estimation}
We recruited 20 Japanese university students to generate the labeled dataset for confidence estimation. 
The data collection environement is shown in \autoref{fig:ce_experiment}. 
Each participant read and answered four-choice English grammatical questions on a computer screen. 
Right after answering each MCQ, participants were requested to assess the confidence behind
their answer, which then became a label for the data.
The summary of this dataset is shown in \autoref{tbl:data_ce}.
The comparison with the labeled dataset for reading detection shown in \autoref{tbl:rd-labeled-samples},
which is similar in size except NR case.
However, the labeled dataset includes a serious skew in the number of confident and unconfident answers.
This is because of the differences in English ability among the participants.
We did not impose any restrictions during the data collection, so that this dataset is also considered ``in-the-wild.''

\begin{table}[tb]
\centering
\caption{Summary of the labeled dataset for confidence estimation.}
\begin{tabular}{c|cccp{4mm}c|cccc}
    \cline{1-4}\cline{6-9}
    \multirow{2}{*}{Participant}  & \multicolumn{3}{c}{No. of MCQs answered} & & \multirow{2}{*}{Participant}  & \multicolumn{3}{c}{No. of MCQs answered}\\
    \cline{2-4} \cline{7-9}
     & Confident & Unconfident & Total & & & Confident & Unconfident & Total \\
    \cline{1-4}\cline{6-9}
    s1 & 361 & 108 & 469 && s11 & 159 & 271 & 430\\
    \cline{1-4}\cline{6-9}
    s2 & 398 & 55 & 453 &&  s12 & 296 & 243 & 539\\
    \cline{1-4}\cline{6-9}
    s3 & 415 & 103 & 518 &&  s13 & 325 & 140 & 465\\
   \cline{1-4}\cline{6-9}
    s4 & 420 & 14 & 434  && s14 & 263 & 253 & 516\\
   \cline{1-4}\cline{6-9}
    s5 & 390 & 175 & 565  && s15 & 174 & 2 & 176\\
    \cline{1-4}\cline{6-9}
    s6 & 68 & 458 & 526  && s16 & 556 & 109 &  665\\
    \cline{1-4}\cline{6-9}
    s7 & 222 & 253 &  475  &&  s17 & 202 & 354 & 556\\
    \cline{1-4}\cline{6-9}
    s8 & 263 & 272 &  535  && s18 & 316 & 260 &  576\\
    \cline{1-4}\cline{6-9}
    s9 & 348 & 87 &  435  && s19 & 306 & 140 & 446\\
    \cline{1-4}\cline{6-9}
    s10 & 210 & 180 & 390  && s20 & 304 & 135 &  439\\
    \cline{1-4}\cline{6-9}
\end{tabular}
\label{tbl:data_ce}
\end{table}

\subsubsection{Unlabeled Dataset for Confidence Estimation}
We recorded the unlabeled data for confidence estimation as described above except there was no inquiry about the confidence. We recruited 80 Japanese high school students, with each participant reading and answering four-choice English vocabulary questions.
The total number of samples in the unlabeled dataset is 57,460. 
Note that the age range of the participants and the contents of the four-choice questions are both different from that of the labeled dataset.

\section{Experimental Results and Discussion} \label{sec:resultsanddiscussion}
\subsection{Reading Detection}
\subsubsection{Experimental Conditions} 
The purpose of the experiment is to evaluate the effectiveness of the proposed self-supervised DL method
as compared to conventional methods. We selected the fully-supervised DL
and SVM for this purpose. 
We are interested in the change of effectiveness as a function of the number of labeled training samples.

For the fully-supervised DL, we simply use the proposed self-supervised DL method for the target task without the pre-training;
all the structure and parameters are identical to the proposed self-supervised DL method. Thus the results show the effectiveness
of pre-training by using an unlabeled dataset.
For SVM, the method described in \cite{rd_Ishimaru_13} is used. 
This allows us to evaluate the impact of DL methods on reading activity classification.
We used a total of ten features for SVM,  
the mean and variance of vertical and horizontal components of EOG data and the 
mean and variance of three axes components of ACC and GYRO data.

The training of each method was performed in the following way.
For the proposed self-supervised DL method, we first applied the pre-training by using the transformed sensor data. In data transformation, we defined the parameters $m$, $n$, and $d$ in \autoref{fig:transform_exp} as $5 \leq m \leq 10$, $n=2$, and $d=90$, respectively.
For each unlabeled segment, we selected a transformation value,  including no transformation, and applied it to the segment to produce pre-training data. Because only one transformation was applied to each segment,
the number of transformed unlabeled samples is equal to the number of original unlabeled samples. 
We applied each transformation equally so that 
the chance rates for the EOG data, ACC data, and GYRO data are 12.5\%, 11.1\%, and 11.1\%, respectively, where the first one is eight-class, and the latter two are nine-class classification.
After this, the target task training was applied. 
As shown in \autoref{tbl:rd-labeled-samples}, the number of available labeled samples is different for each class. 
We simply took all 5,340 samples (the smallest number) of  ``reading English'' and downsampled other classes to have them match in size.
Thus the chance rate for the target task is 25\%.
The same data were also employed for training the fully-supervised DL and SVM. We changed the number of labeled training samples per class for all three methods in the order of 10, 50, 100, 500, 1,000, and 5,340.

\begin{figure}[tb]
	\centering
	\includegraphics[width=0.5\hsize]{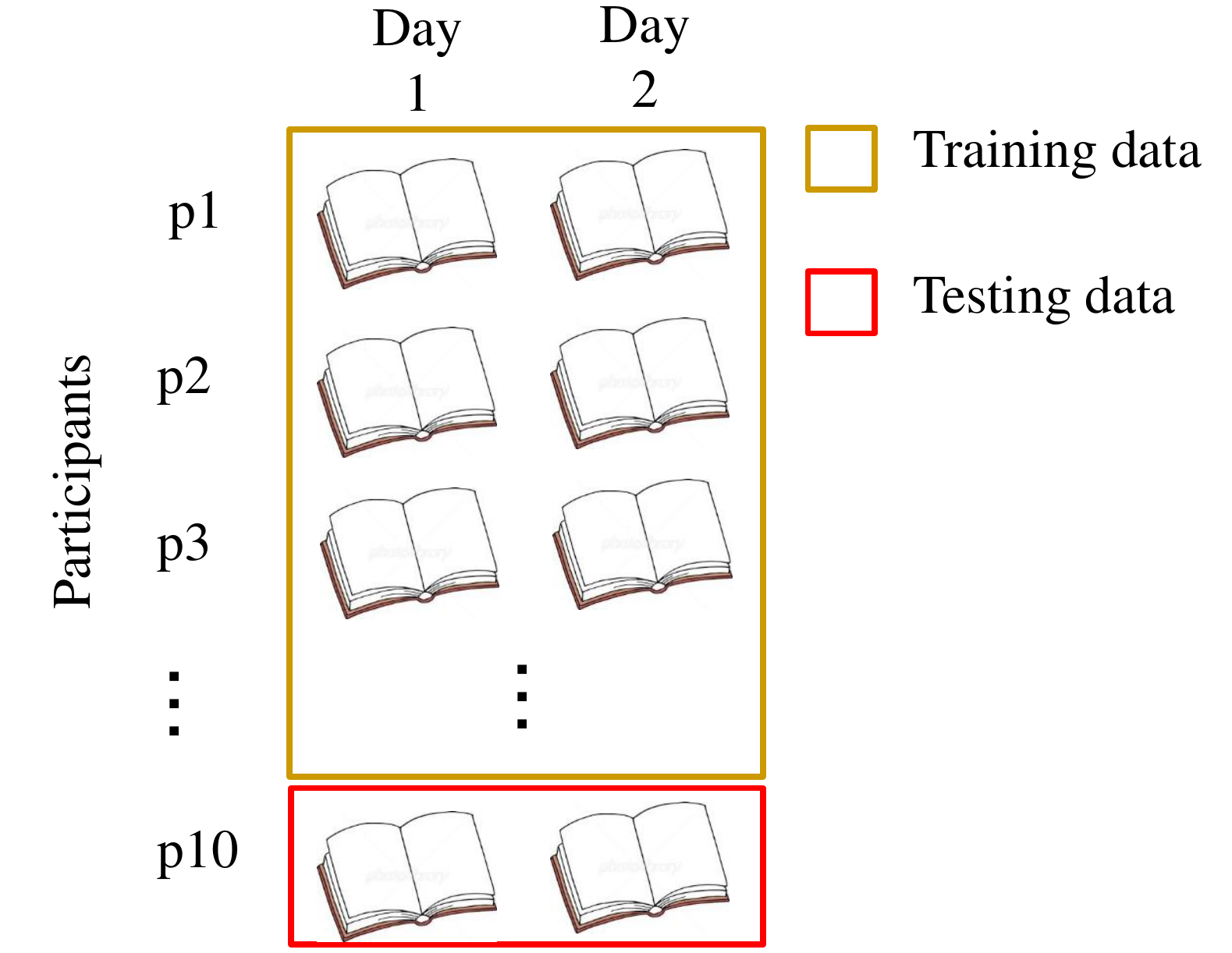}
	\caption{Leave-One-Participant-Out cross-validation for reading detection.}
	\label{fig:eval}
\end{figure}

All of the above methods were evaluated with the user independent Leave-One-Participant-Out
cross-validation approach as shown in \autoref{fig:eval}. 
In the target task training, 20\% of the training data were
randomly selected for validation, and the rest were employed for learning.

\subsubsection{Results of Pre-training}
\begin{figure}[tb]
      \begin{minipage}[b]{0.32\hsize}
        \centering
        \includegraphics[width=\hsize]{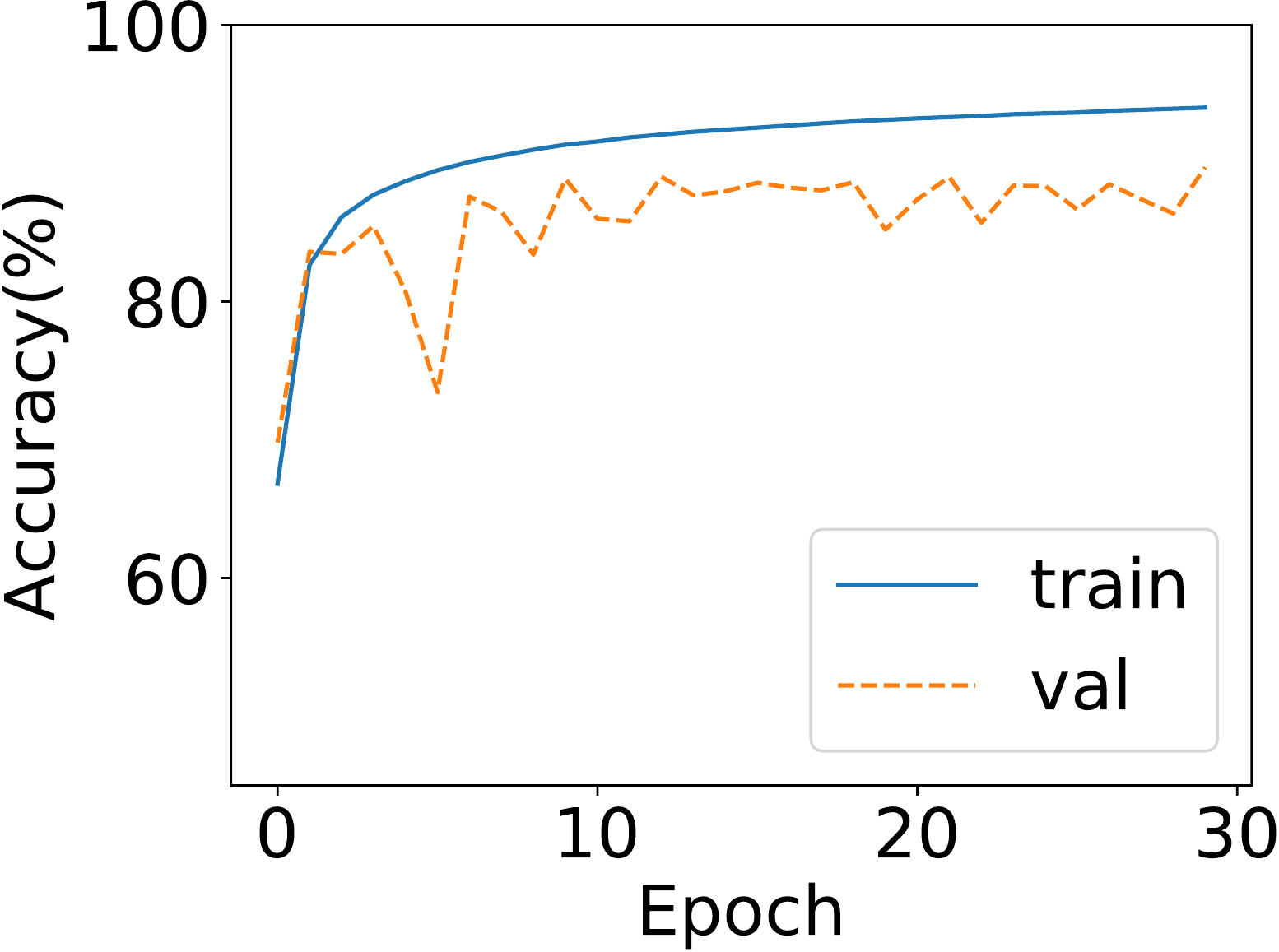}
        \subcaption{EOG}
        \label{fig:eog_acc}
      \end{minipage} 
      \hfill
      \begin{minipage}[b]{0.32\hsize}
        \centering
        \includegraphics[width=\hsize]{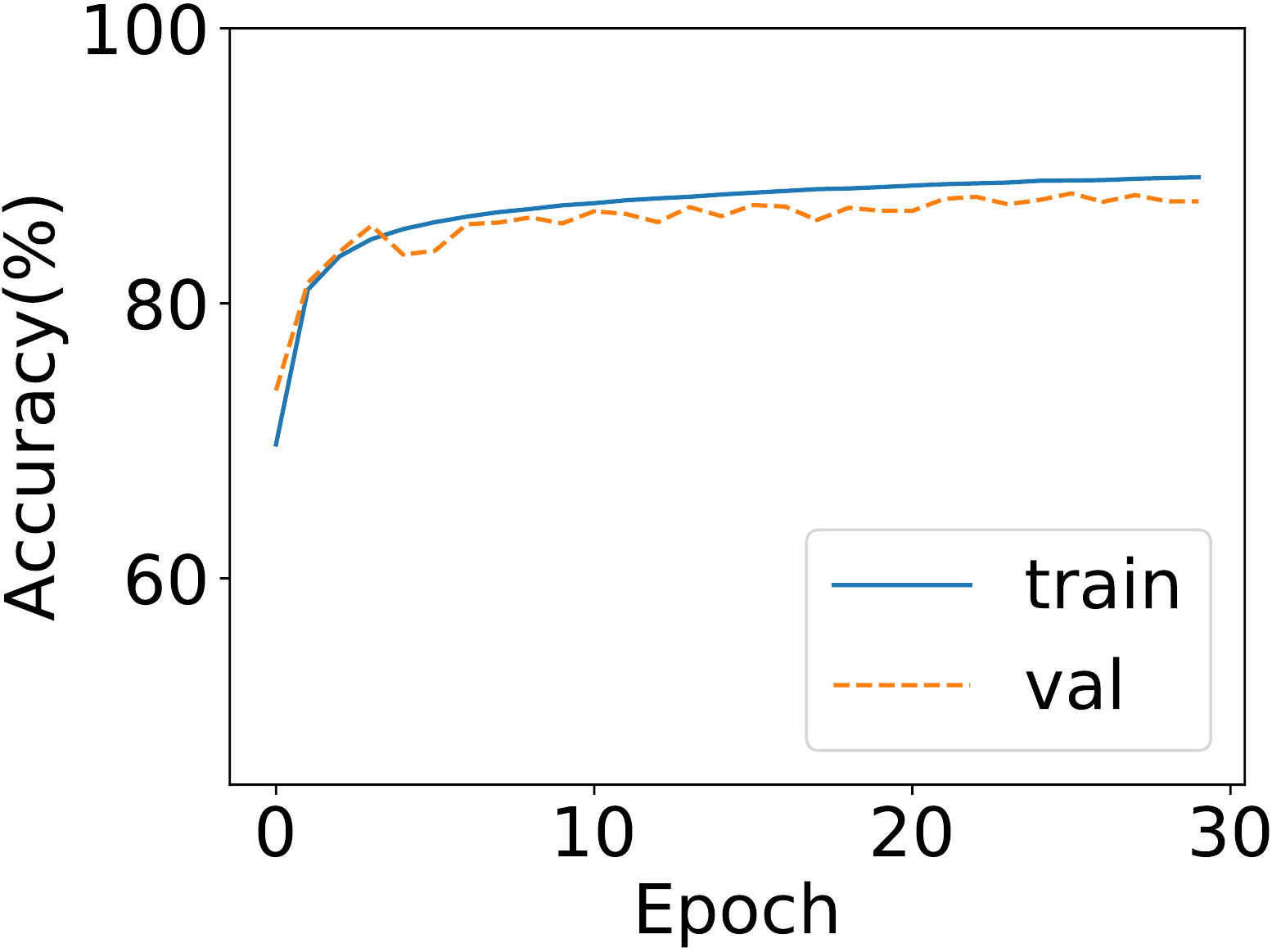}
        \subcaption{ACC}
        \label{fig:acc_acc}
      \end{minipage}
      \hfill
      \begin{minipage}[b]{0.32\hsize}
        \centering
        \includegraphics[width=\hsize]{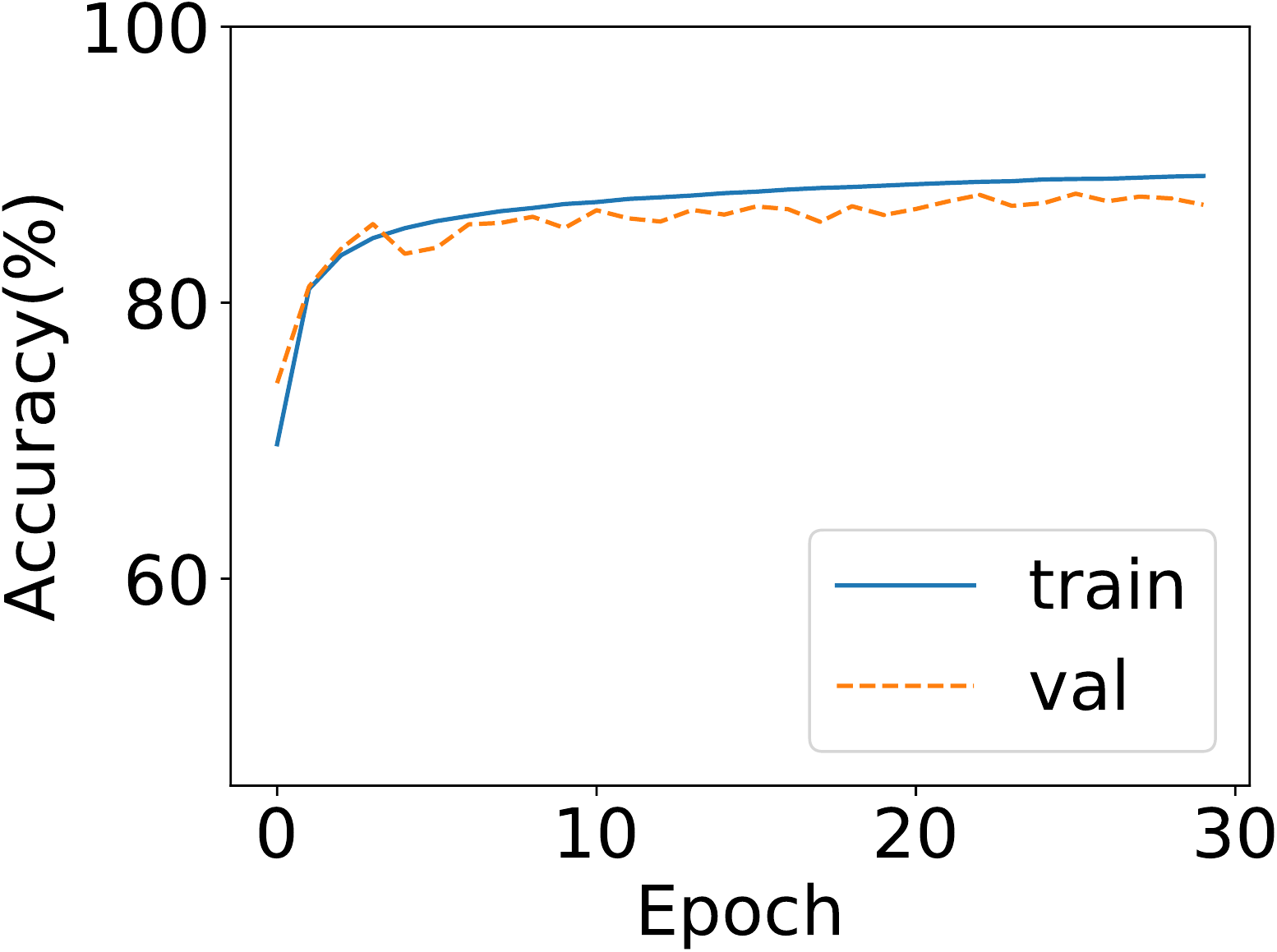}
        \subcaption{GYRO}
        \label{fig:gyro_acc}
      \end{minipage} 
    \caption{Training and validation accuracy curve of the self-supervised pre-training experiment for reading detection.}
    \label{fig:feature_extraction_acc}
  \end{figure}

Training and validation accuracy curves for EOG, ACC, and GYRO data are shown
in \autoref{fig:feature_extraction_acc}.
The average test accuracy of the ten participants is 93.2\% for EOG data, 95.5\% for ACC data, and 95.3\% for GYRO data. This high test accuracy primarily indicates that the network was trained well. 
The difference in EOG sensor data between the training and validation accuracy is slightly high, indicating the tendency of the network to overfit. Thus, the test accuracy for EOG data is also lower than that of the ACC and GYRO data.
This happened because we could not remove all the noisy segments from the EOG dataset. 

\subsubsection{Results of the Target Task}
\begin{figure}[tb]
	\centering
	\includegraphics[scale=.8]{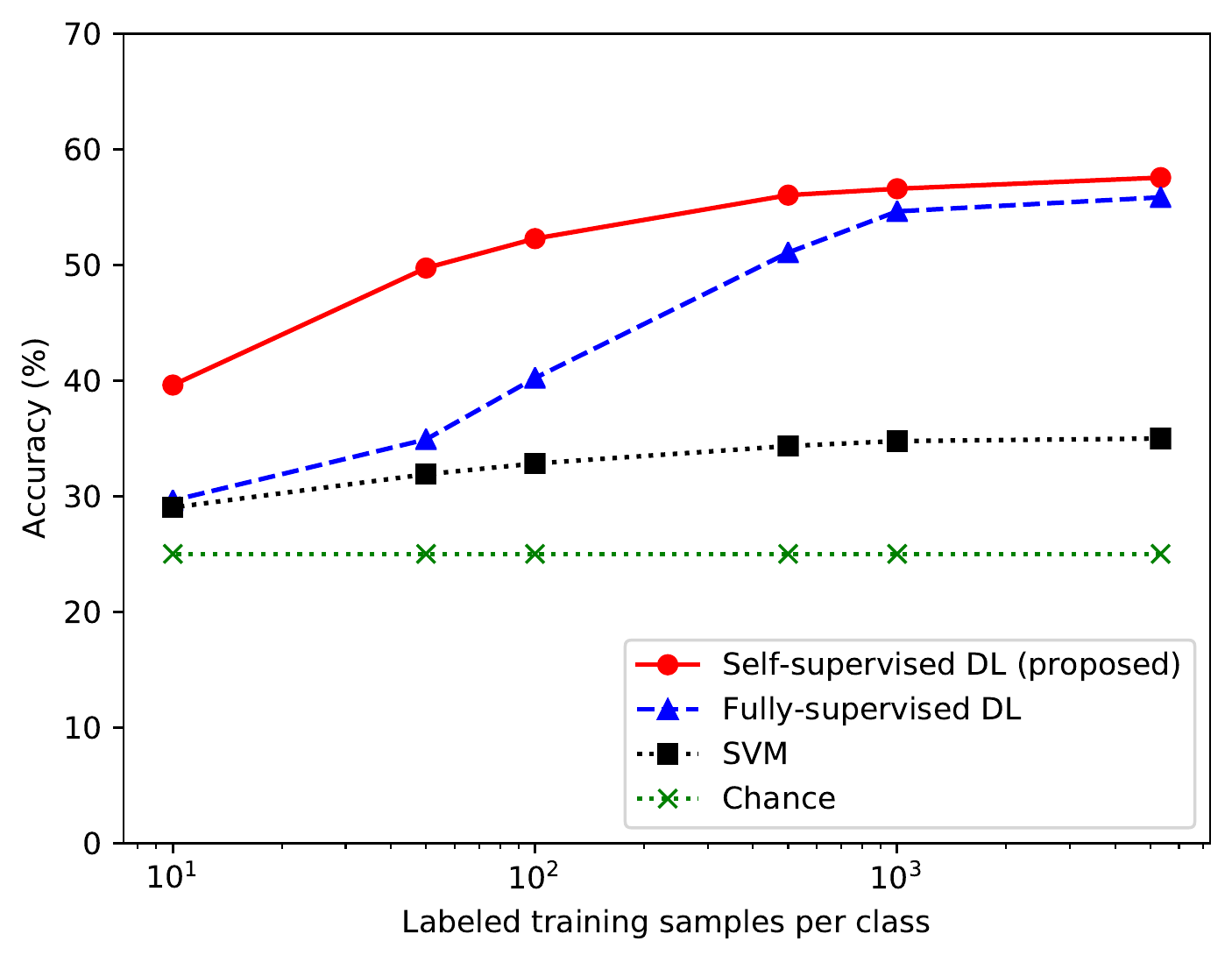}
	\caption{Dependency of test accuracy to the number of labeled training samples per class for reading detection.}
	\label{fig:result}
\end{figure}

\autoref{fig:result} shows the reading detection result. It describes the change of average test accuracy for the number of labeled training samples per class.
From this graph, we can observe the following.
First, the proposed self-supervised DL method performs best for all cases regardless of the number of training samples.
The proposed self-supervised DL method is more advantageous than the fully-supervised DL when the number of labeled training samples is smaller.
This indicates the effectiveness of the self-supervised DL.
As compared to SVM, the fully-supervised DL performs much better when the number of labeled training samples
is larger. However, this advantage disappears when the number of labeled training samples decreases.
This shows the limitation of the fully-supervised DL when a large enough number of training samples are not available.
On the other hand, the proposed self-supervised DL method is always much better than SVM and is never inferior to the fully-supervised DL.
In other words, we can always recommend to use the proposed self-supervised DL method.

\begin{figure}[tb]
	\centering
	\begin{minipage}[b]{0.49\hsize}
	\includegraphics[width=\hsize]{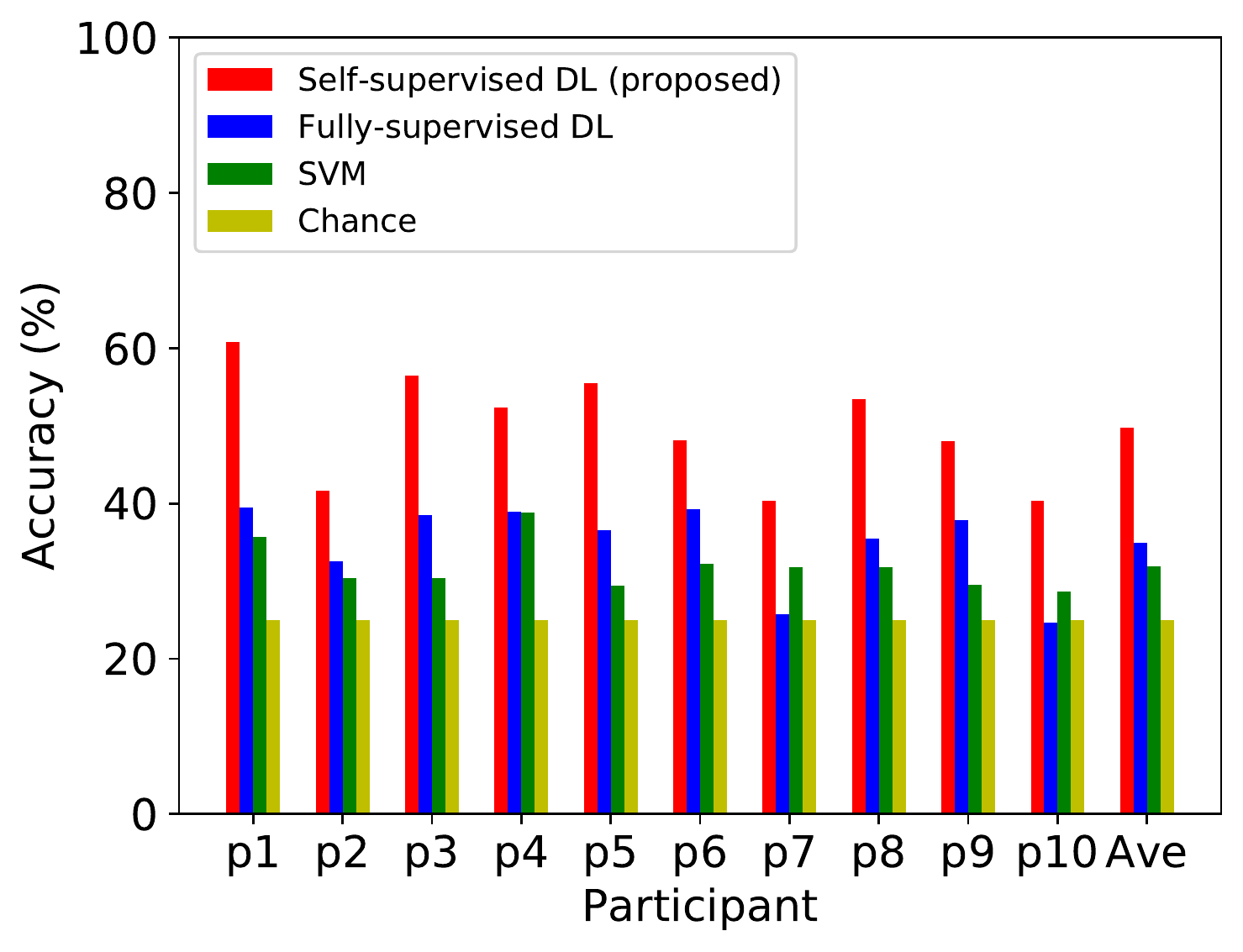}
	\subcaption{50 training samples per class}
	\label{fig:rd_50}
	\end{minipage}
	\hfill
	\begin{minipage}[b]{0.49\hsize}
	\includegraphics[width=\hsize]{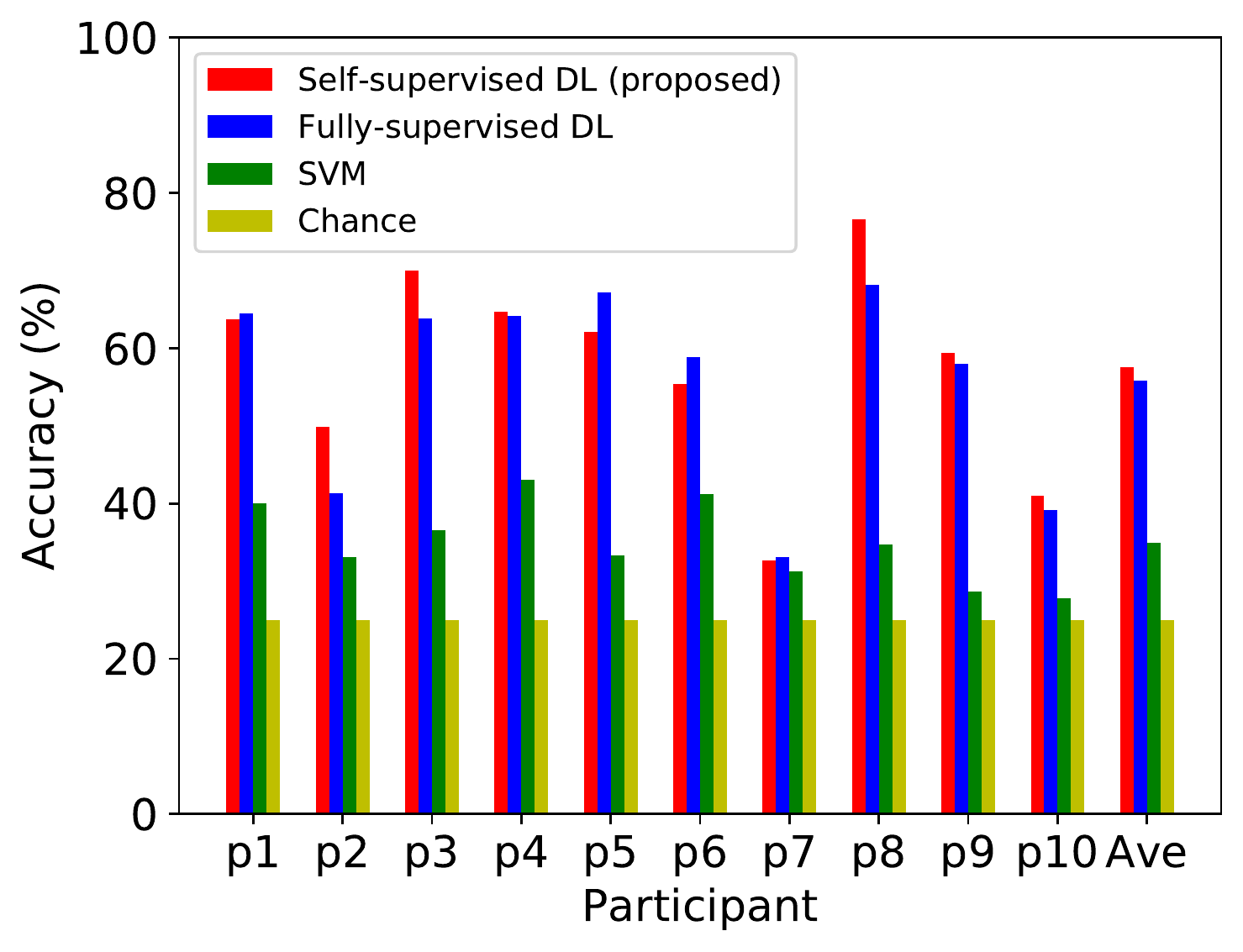}
	\subcaption{5,340 training samples per class}
	\label{fig:rd_5340}
	\end{minipage}
\caption{Test accuracy for each participant of reading detection.}
\label{fig:rd_each_paticipant}
\end{figure}

Let us take a closer look at the performance of cases when the number of labeled training samples per class
is 50 and 5,340.
\autoref{fig:rd_each_paticipant} displays the results for each participant.
For the case of 5,340 training samples, the advantage of the proposed self-supervised DL method over the fully-supervised DL depends on the participants.
On the other hand, for the case of 50 training samples, the proposed self-supervised DL method is better than the fully-supervised DL for all participants.

To make sure whether the difference is significant or not,
we applied the statistical analysis.
The one-way repeated measures analysis of variance (ANOVA) test was first applied to the results of
test accuracy. The null hypothesis is that the population means of all three methods are equal.
\autoref{tbl:ANOVA_rd} shows the results.
We have confirmed that the null hypothesis was rejected with the significance level $p<0.01$;
at least one population mean is different from the rest
for all cases of the number of training samples per class.
Then the post-hoc paired t-test was applied for the further analysis.
A null hypothesis here is that the population mean of one method is equal to 
that of another method. Because we have three methods, we conducted three t-test experiments
for all combinations. Multiple comparisons for the three methods was mitigated by a Bonferroni correction.
With the correction applied, significance is found if $p < 0.0033$.
Results are shown in \autoref{tbl:ttest_rd}.
For the pair of the proposed self-supervised DL method and the fully-supervised DL,
all but the case of 5,340 training samples, we have confirmed
that the proposed self-supervised DL method is statistically significantly better than the fully-supervised DL.
For the comparison with SVM, the advantage of the proposed self-supervised DL method is shown for all cases.
For the comparison between the fully-supervised DL and SVM,
we cannot reject the null hypothesis for the cases for smaller sample sizes (10 and 50).

\begin{table}[tb]
\caption{Repeated Measures ANOVA test result for reading detection.}
\label{tbl:ANOVA_rd}
\centering
   \begin{tabular}{c|llllll}
   \hline
    \multirow{2}{*}{Parameter} & \multicolumn{6}{c}{Training samples per class}\\
   \cline{2-7}
     & 10 & 50 & 100 & 500 & 1,000 & 5,340 \\
   \hline
   F value & 77.24 & 61.39 & 63.55 & 50.13 & 43.92 & 34.03 \\
   \hline
   p value & 0.00$^{*}$ & 0.00$^{*}$ & 0.00$^{*}$ & 0.00$^{*}$ & 0.00$^{*}$ & 0.00$^{*}$ \\
   \hline
   \multicolumn{7}{l}{ $^{*}$\footnotesize{p$<$0.01}} \\
   \end{tabular}
\end{table}

\begin{table}[tb]
\caption{Post-Hoc Paired t-test result for reading detection.}
\label{tbl:ttest_rd}
\centering
\scalebox{0.96}{
   \begin{tabular}{c|c|llllll}
   \hline
    \multirow{2}{*}{Pair}  & \multirow{2}{*}{Parameter} & \multicolumn{6}{c}{Training samples per class}\\
   \cline{3-8}
     &  & 10 & 50 & 100 & 500 & 1,000 & 5,340 \\
   \hline
   \multirow{2}{*}{ \makecell{Proposed self-supervised DL \\ and fully-supervised DL}} & t value & 11.30 & 10.69 & 11.87 & 4.79 & 3.02 & 1.15 \\
   \cline{2-8}
   & p value & 0.000$^{*}$ & 0.000$^{*}$ & 0.000$^{*}$ & 0.001$^{*}$ & 0.014 & 0.278 \\
   \hline
   \multirow{2}{*}{ \makecell{Proposed self-supervised DL \\ and SVM}} & t value & 9.88 & 8.49 & 8.65 & 7.60 & 6.68 & 6.09 \\
   \cline{2-8}
   & p value & 0.000$^{*}$ & 0.000$^{*}$ & 0.000$^{*}$ & 0.000$^{*}$ & 0.000$^{*}$ & 0.000$^{*}$ \\
   \hline
   \multirow{2}{*}{ \makecell{Fully-supervised DL and \\ SVM}} & t value & 0.68 & 1.88 & 4.26 & 6.70 & 6.69 & 6.04 \\
   \cline{2-8}
   & p value & 0.516 & 0.093 & 0.002$^{*}$ & 0.000$^{*}$ & 0.000$^{*}$ & 0.000$^{*}$ \\
   \hline
   \multicolumn{8}{l}{$^{*}$\footnotesize{p$<$0.0033}} \\
   \end{tabular}
   }
\end{table}

\begin{figure}[tb]
	\centering
	\begin{minipage}[b]{0.48\hsize}
	\includegraphics[width=\hsize]{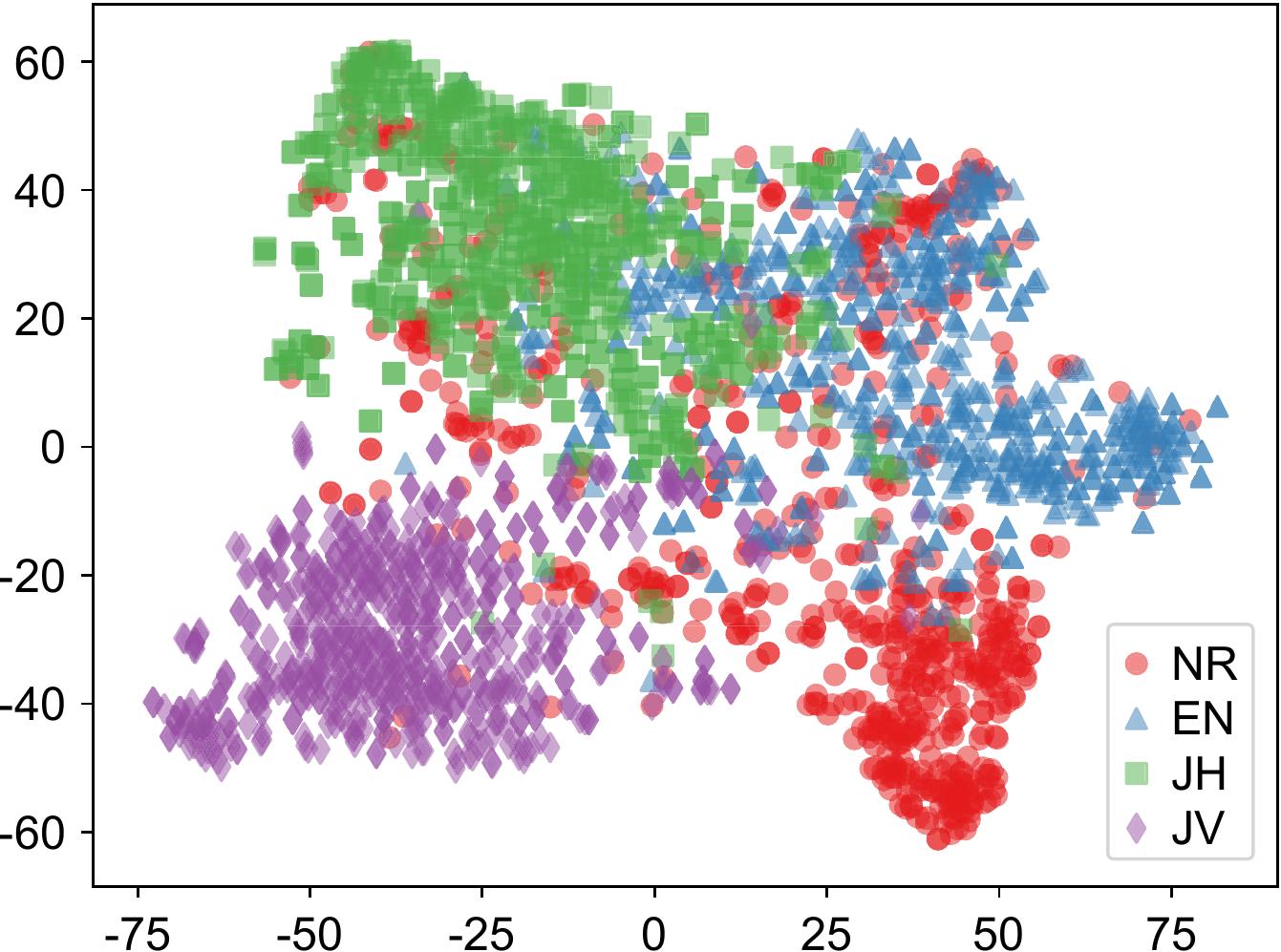}
	\subcaption{50 samples per class for the proposed self-supervised DL method}
	\label{fig:tsne_200_ssl_p1_rd}
	\end{minipage}
	\hfill
	\begin{minipage}[b]{0.48\hsize}
	\includegraphics[width=\hsize]{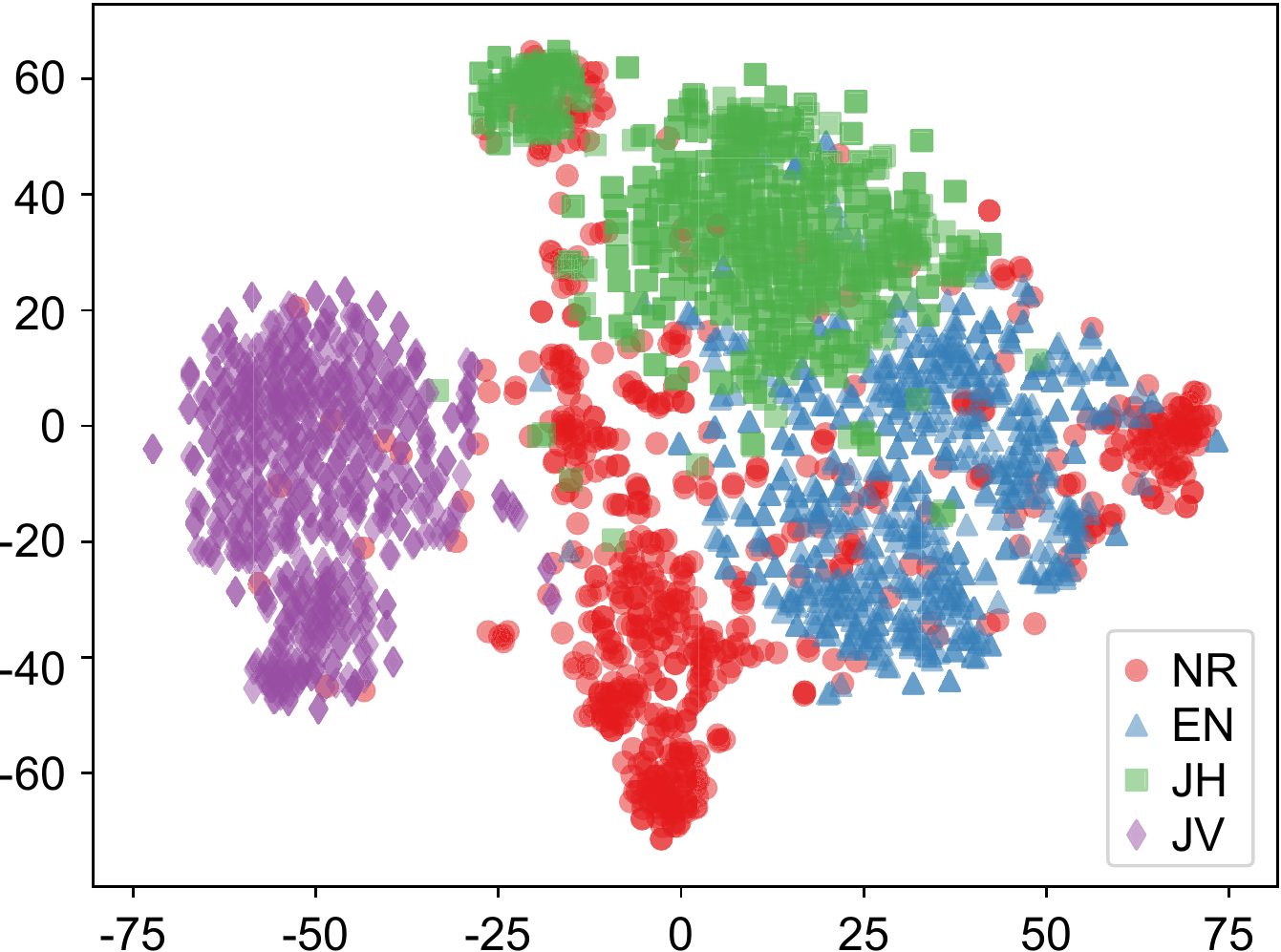}
	\subcaption{5,340 samples per class for the proposed self-supervised DL method}
	\label{fig:tsne_all_ssl_p1_rd}
	\end{minipage}
	
	\vspace*{\intextsep}

	\begin{minipage}[b]{0.48\hsize}
	\includegraphics[width=\hsize]{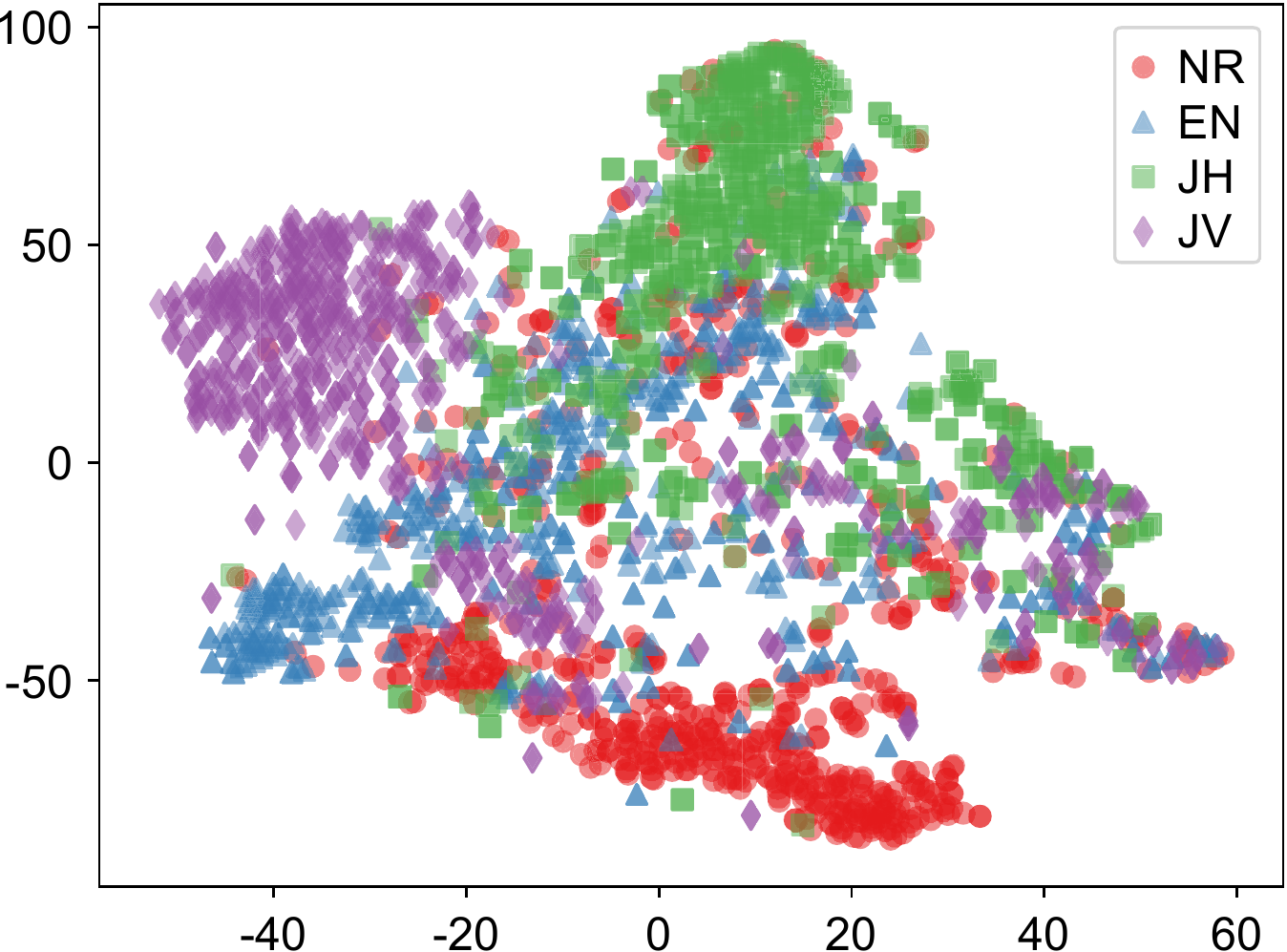}
	\subcaption{50 samples per class for the fully-supervised DL}
	\label{fig:tsne_100_supervised_p1_rd}
	\end{minipage}
	\hfill
	\begin{minipage}[b]{0.48\hsize}
	\includegraphics[width=\hsize]{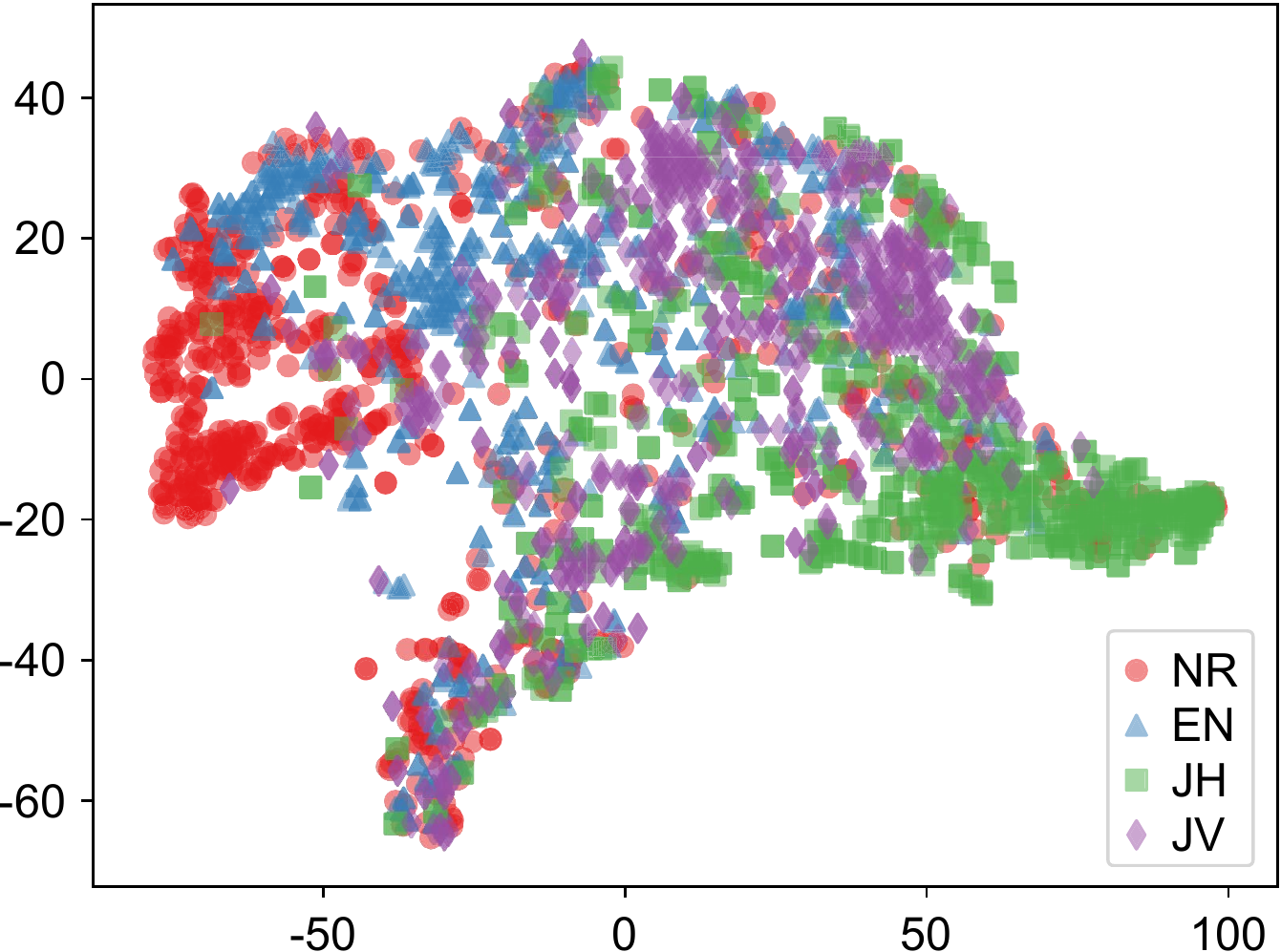}
	\subcaption{5,340 samples per class for the fully-supervised DL}
	\label{fig:tsne_all_supervised_p1_rd}
	\end{minipage}
\caption{Visualisation of features using t-SNE for reading detection.}
\label{fig:tsne_rd}
\end{figure}

We used a method called t-distributed Stochastic Neighbor Embedding (t-SNE)~\cite{tsne_maaten} to visualize the features learned by the proposed self-supervised DL method and the fully-supervised DL.
\autoref{fig:tsne_rd} is a visualization of the output of the global max-pooling layer in the network 
shown in the lower part of \autoref{fig:proposed_rd}. 
It is clear that the distributions for the 5,340 samples cases (\subref{fig:tsne_all_ssl_p1_rd} and \subref{fig:tsne_all_supervised_p1_rd})
are more clearly separated than those for the 50 samples cases (\subref{fig:tsne_200_ssl_p1_rd} and \subref{fig:tsne_100_supervised_p1_rd}).
As compared with the distributions by the fully-supervised DL,
the distributions by the proposed self-supervised DL method have clear separation.
Thus, more distinctive features were learned via the proposed self-supervised DL method, which generates higher accuracy as compared to the fully-supervised DL.

\autoref{tbl:confusion_rd} shows the confusion matrices for the proposed self-supervised DL method, the fully-supervised DL, and SVM.
\begin{table}[tb]
\caption{Confusion matrices for reading detection.}
\label{tbl:confusion_rd}
\centering
 \hfill
\begin{minipage}[t]{0.35\hsize}
\subcaption{50 samples per class for the proposed self-supervised DL method}
\label{tbl:50_SSDL}
\centering
  \begin{tabular}{|c|c|r|r|r|r|}
  \hline
  \multicolumn{2}{|c|}{} & \multicolumn{4}{c|}{Predicted} \\
  \cline{3-6}
  \multicolumn{2}{|c|}{} & NR & EN & JH & JV \\
  \hline
  \multirow{4}{*}{\rotatebox{90}{Actual}} & NR & \cellcolor{red!30}240 & 114 & 96 & 46 \\
  \cline{2-6}
  & EN & 100 & \cellcolor{mediumpersianblue!30}240 & 121 & 35 \\
  \cline{2-6}
  & JH & 117 & 185 & \cellcolor{mediumseagreen!40}166 & 28 \\
  \cline{2-6}
  & JV & 61 & 51 & 36 & \cellcolor{hanpurple!30}348 \\
  \hline
  \end{tabular}
 \end{minipage}
 \hfill
\begin{minipage}[t]{0.35\hsize}
\subcaption{5,340 samples per class for the proposed self-supervised DL method}
\label{tbl:5340_SSDL}
\centering
   \begin{tabular}{|c|c|r|r|r|r|}
   \hline
   \multicolumn{2}{|c|}{} & \multicolumn{4}{c|}{Predicted} \\
   \cline{3-6}
   \multicolumn{2}{|c|}{} & NR & EN & JH & JV \\
   \hline
   \multirow{4}{*}{\rotatebox{90}{Actual}} & NR & \cellcolor{red!30}345 & 67 & 54 & 30 \\
   \cline{2-6}
   & EN & 161 & \cellcolor{mediumpersianblue!30}237 & 84 & 14 \\
   \cline{2-6}
   & JH & 161 & 127 & \cellcolor{mediumseagreen!40}189 & 19 \\
   \cline{2-6}
   & JV & 76 & 24 & 15 & \cellcolor{hanpurple!30}381 \\
   \hline
   \end{tabular}
 \end{minipage}
\hfill
  \hspace{0mm} 

  \vspace*{\intextsep}

 \hfill
   \begin{minipage}[t]{0.35\hsize}
\subcaption{50 samples per class for the fully-supervised DL}
\label{tbl:50_FSDL}
\centering
   \begin{tabular}{|c|c|r|r|r|r|}
   \hline
   \multicolumn{2}{|c|}{} & \multicolumn{4}{c|}{Predicted} \\
   \cline{3-6}
   \multicolumn{2}{|c|}{} & NR & EN & JH & JV \\
   \hline
   \multirow{4}{*}{\rotatebox{90}{Actual}} & NR & \cellcolor{red!30}203 & 113 & 94 & 86 \\
   \cline{2-6}
   & EN & 113 & \cellcolor{mediumpersianblue!30}146 & 126 & 111\\
   \cline{2-6}
   & JH & 130 & 139 & \cellcolor{mediumseagreen!40}123 & 104 \\
   \cline{2-6}
   & JV & 87 & 99 & 83 & \cellcolor{hanpurple!30}227 \\
   \hline
   \end{tabular}
 \end{minipage}
 \hfill
\begin{minipage}[t]{0.35\hsize}
\subcaption{5,340 samples per class for the fully-supervised DL}
\label{tbl:5340_FSDL}
\centering
   \begin{tabular}{|c|c|r|r|r|r|}
   \hline
   \multicolumn{2}{|c|}{} & \multicolumn{4}{c|}{Predicted} \\
   \cline{3-6}
   \multicolumn{2}{|c|}{} & NR & EN & JH & JV \\
   \hline
   \multirow{4}{*}{\rotatebox{90}{Actual}} & NR & \cellcolor{red!30}338 & 56 & 71 & 31 \\
   \cline{2-6}
   & EN & 144 & \cellcolor{mediumpersianblue!30}196 & 141 & 15 \\
   \cline{2-6}
   & JH & 163 & 115 & \cellcolor{mediumseagreen!40}197 & 21 \\
   \cline{2-6}
   & JV & 76 & 13 & 19 & \cellcolor{hanpurple!30}388 \\
   \hline
   \end{tabular}
 \end{minipage}
 \hfill
  \hspace{0mm} 

  \vspace*{\intextsep}

 \hfill
\begin{minipage}[t]{0.35\hsize}
\subcaption{50 samples per class for the SVM}
\label{tbl:50_SVM}
\centering
   \begin{tabular}{|c|c|r|r|r|r|}
   \hline
   \multicolumn{2}{|c|}{} & \multicolumn{4}{c|}{Predicted} \\
   \cline{3-6}
   \multicolumn{2}{|c|}{} & NR & EN & JH & JV \\
   \hline
   \multirow{4}{*}{\rotatebox{90}{Actual}} & NR & \cellcolor{red!30}209 & 91 & 89 & 107 \\
   \cline{2-6}
   & EN & 92 & \cellcolor{mediumpersianblue!30}111 & 123 & 170\\
   \cline{2-6}
   & JH & 92 & 118 & \cellcolor{mediumseagreen!40}131 & 155 \\
   \cline{2-6}
   & JV & 68 & 117 & 130 & \cellcolor{hanpurple!30}181 \\
   \hline
   \end{tabular}
 \end{minipage}
 \hfill
\begin{minipage}[t]{0.35\hsize}
\subcaption{5,340 samples per class for the SVM}
\label{tbl:5340_SVM}
\centering
   \begin{tabular}{|c|c|r|r|r|r|}
   \hline
   \multicolumn{2}{|c|}{} & \multicolumn{4}{c|}{Predicted} \\
   \cline{3-6}
   \multicolumn{2}{|c|}{} & NR & EN & JH & JV \\
   \hline
   \multirow{4}{*}{\rotatebox{90}{Actual}} & NR & \cellcolor{red!30}238 & 65 & 146 & 47 \\
   \cline{2-6}
   & EN & 85 & \cellcolor{mediumpersianblue!30}73 & 239 & 99 \\
   \cline{2-6}
   & JH & 79 & 75 & \cellcolor{mediumseagreen!40}223 & 119 \\
   \cline{2-6}
   & JV & 51 & 61 & 219 & \cellcolor{hanpurple!30}165 \\
   \hline
   \end{tabular}
 \end{minipage}
  \hfill
  \hspace{0mm} 
\end{table}
We found the following tendencies: 
\begin{itemize}
\item ``Reading English'' and 
``Reading horizontally written Japanese''
tends to be confused, 
because both reading behaviors are dominated by horizontal eye movement. 
Among all the methods, the proposed self-supervised DL method
was most effective at distinguishing these classes
because it can learn slight differences in 
the reading behavior between these classes.
\item Some ``not reading'' behaviors are misclassified into reading, because they contain
similar behaviors by chance. Another reason is that participants sometimes read something unintentionally because we do not impose any restrictions on their behaviors. In other words, it was caused by the in-the-wild nature of the dataset.
\item SVM confuses reading horizontally written and vertically written Japanese because of the less distinctive features.
\end{itemize}

\subsection{Confidence Estimation}
\subsubsection{Experimental Conditions}

The purpose of the experiments is the same as the previous ones; we are interested in the effectiveness of the proposed self-supervised DL method as a function of the number of labeled training samples.
We employed the same methods for comparison as we did for reading detection.

For the self-supervised pre-training, we applied the transformations to the unlabeled data as follows.
For each image, we selected one of four transformations, including no transformation, as shown in \autoref{fig:transformation_ce} and applied it. Because each transformation was selected equally, the chance rate of the pre-training was 25\%.

After the pre-training, we created the target task network of confidence estimation by fine-tuning the pre-trained base network, as shown in the lower part of \autoref{fig:proposed_ce}. 
In this case, we used the labeled data. 
Although we applied Leave-One-Participant-Out cross-validation for reading detection,
it is not appropriate for confidence estimation 
due to the seriously skewed distribution of confident and unconfident labels
as shown in \autoref{tbl:data_ce}. We cannot balance the number of labels by simply applying over- nor under-sampling.
Thus we took a different approach for data preparation for training and testing.

\begin{figure}[tb]
    \centering
\includegraphics[scale=.45]{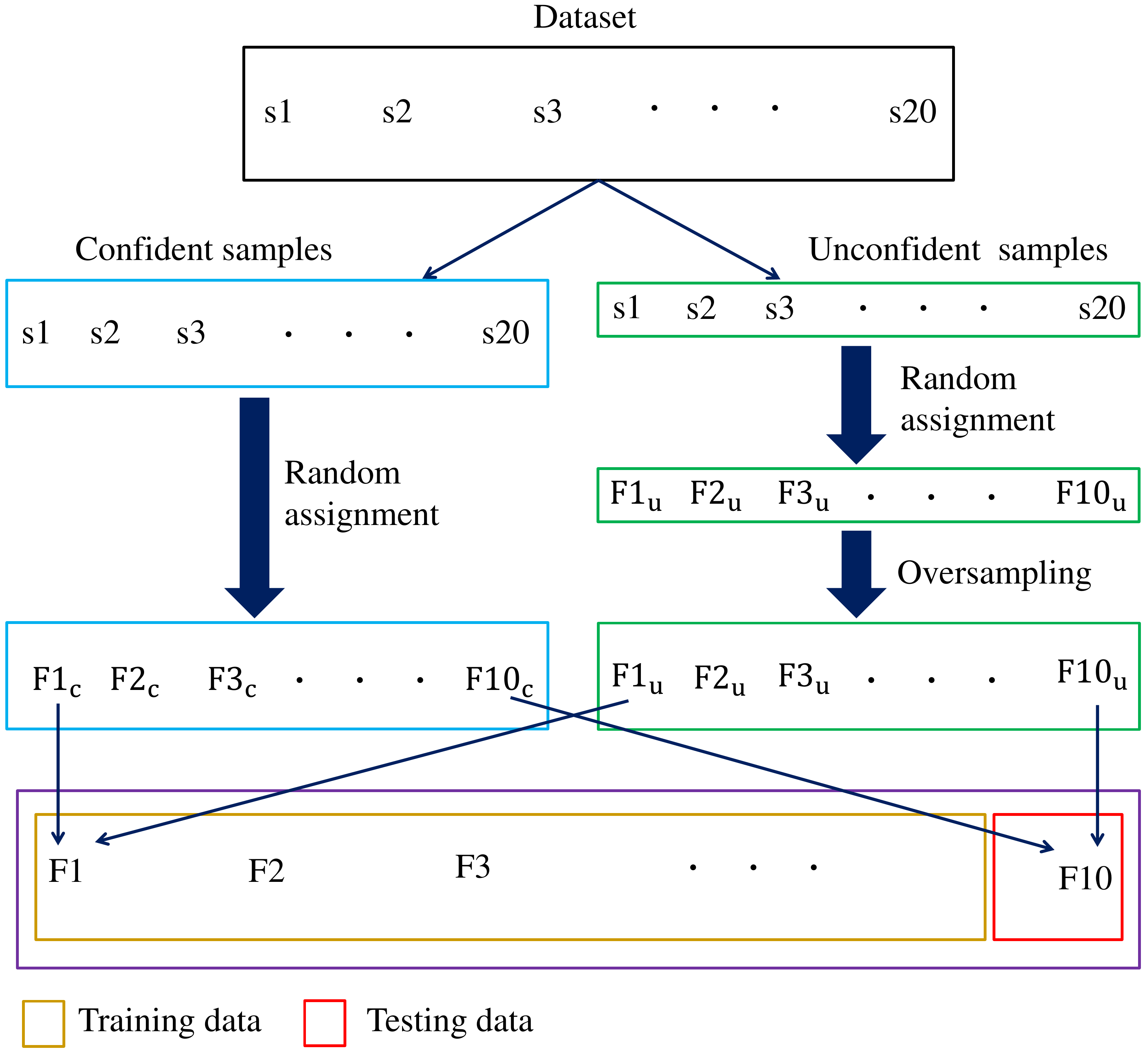}
    \caption{10-fold cross-validation for confidence estimation.}
    \label{fig:eval_ce}
\end{figure}

\autoref{fig:eval_ce} illustrates the procedure of data preparation.
First we separated the data into confident and unconfident samples.
Note that the number of samples available from each participant s$_i$
is different and the total number of confident samples is larger than
unconfident ones.
For each of the confident samples, we randomly assign a fold from F1$_c$ to F10$_c$
while keeping the number of samples in each fold is equal.
For the case of unconfident samples, we also assigned each sample to one of the 10 folds
F1$_u$--F10$_u$ in the same way. In order to make the number of samples in F$i_u$ 
equal to that in F$i_c$, we applied oversampling by using a 5-degree rotation to randomly
selected samples in each fold.
Finally we combined F$i_c$ and F$i_u$ to form a fold F$i$.
By using the 10 folds of data, we applied the 10-fold cross-validation.
In the training by using nine folds, we randomly selected 80\% for learning and the remaining 20\% for validation.
Because the number of training samples with the confident label is equal to that with the unconfident label
for all folds, the chance rate of the classification is 50\%.

In the case of the fully-supervised DL, we trained the network using only the labeled data.
We used SVM as a baseline method using basic statistical features such as mean and variance. We calculated and used four features from one sample (one MCQ); means and variances of the two axes of the eye gaze to classify data samples.
These two methods were also evaluated by the 10-fold cross-validation. We changed the number of labeled training samples per class for all three methods in the order of 10, 20, 50, 100, 150, 200, 300, 400, 500, 1,000, and 5,382.

\subsubsection{Results of Pre-training}
\begin{figure}[tb]
       \begin{minipage}[b]{\textwidth}
       \centering
          \includegraphics[scale=0.65]{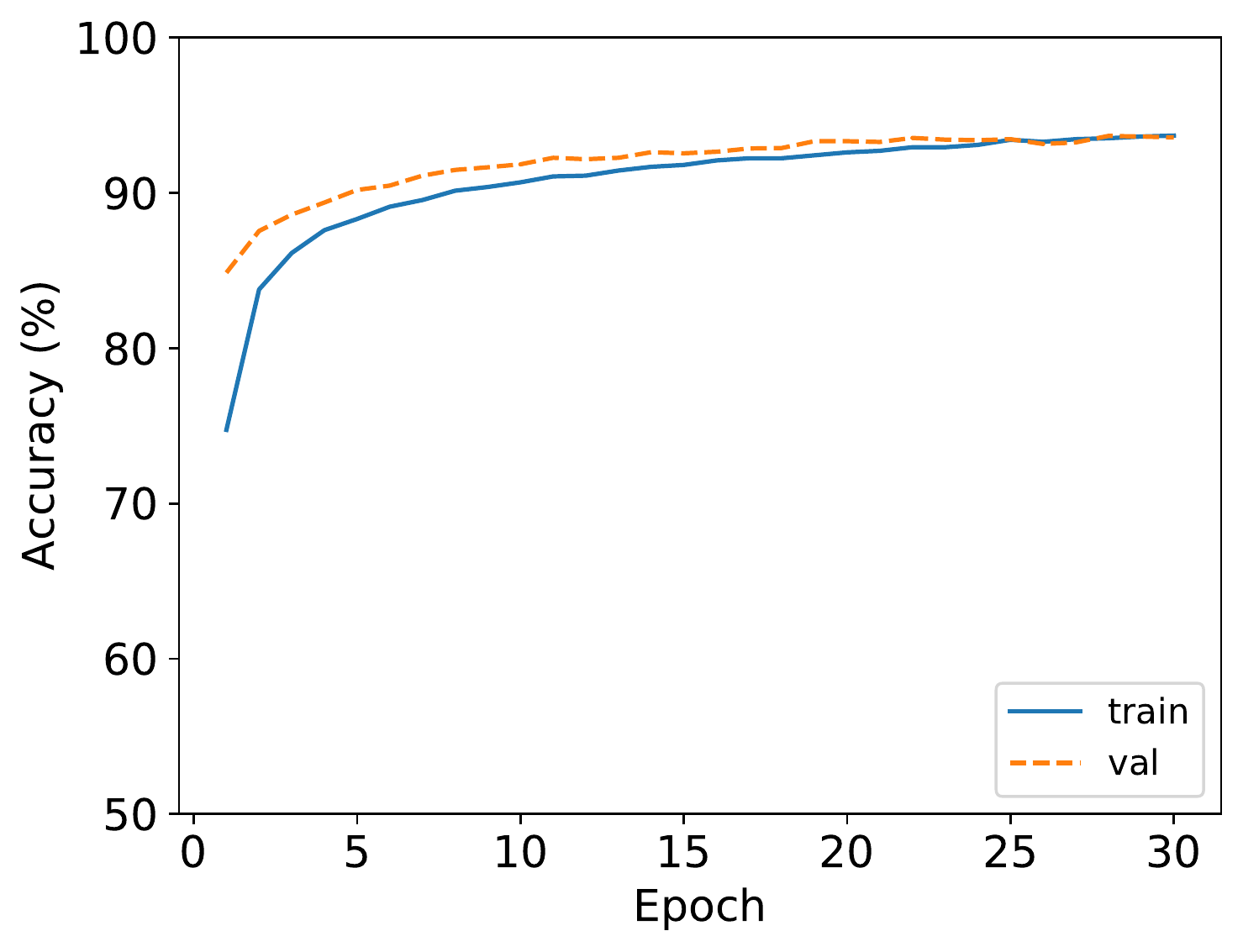}
        \caption{Training and validation accuracy curve of the self-supervised pre-training experiment for confidence estimation.}
        \label{fig:feature_extraction_ce}
       \end{minipage}\\
\vspace*{\intextsep}
       \begin{minipage}[b]{\textwidth}
          \centering
          \includegraphics[scale=0.85]{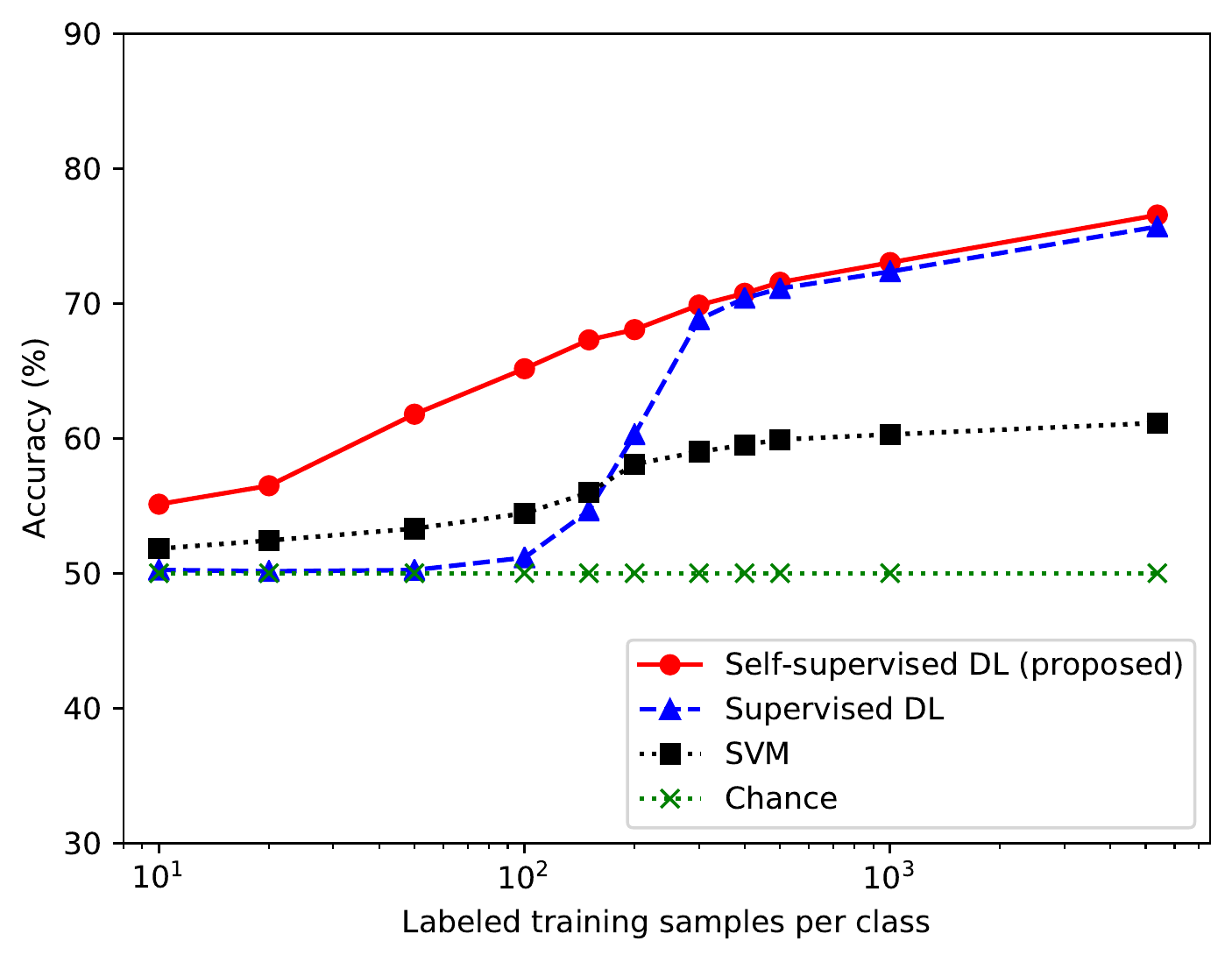}
          \caption{Dependency of test accuracy to the number of labeled training samples per class for confidence estimation.}
            \label{fig:result_ce}
       \end{minipage}
   \end{figure}

The average test accuracy of the pre-training experiment was 93.3\%; this high test accuracy indicates that the base network was trained well. 
\autoref{fig:feature_extraction_ce} illustrates
the training and validation accuracy curves for pre-training. 
The pre-trained network is good fit since the difference between the training and validation is almost zero. 
   
\subsubsection{Results of the Target Task} 
\autoref{fig:result_ce} shows the results of the target task.
Tendencies similar to the reading detection results were observed.
The proposed self-supervised DL method performed the best regardless of the number of labeled training samples.
The performance of the fully-supervised DL dropped when the number of labeled training samples
was insufficient.
With confidence estimation, SVM was not always worst, though performance
was limited even with a larger number of labeled samples.
From these results, we can always recommend to use the proposed self-supervised DL method in the case of reading detection.
\vspace{5mm}

As we did for the results of reading detection,
we also applied the statistical tests for the results of confidence estimation.
Tables~\ref{tbl:ANOVA_ce} and \ref{tbl:ttest_ce} show the results.
\begin{table}[tb]
\caption{Repeated Measures ANOVA test result for confidence estimation.}
\label{tbl:ANOVA_ce}
\centering
   \begin{tabular}{c|lllllllllll}
   \hline
    \multirow{2}{*}{Parameter} & \multicolumn{11}{c}{Training samples per class}\\
   \cline{2-12}
      & 10 & 20 & 50 & 100 & 150 & 200 & 300 & 400 & 500 & 1,000 & 5,382 \\
   \hline
   F value & 617.2 & 341.2 & 780.7 & 1472.1 & 453.4 & 492.7 & 486.4 & 537.9 & 510.5 & 615.6 & 876.3\\
   \hline
   p value & 
   0.00$^{*}$ & 0.00$^{*}$ & 0.00$^{*}$ & 0.00$^{*}$ & 0.00$^{*}$ & 0.00$^{*}$ & 0.00$^{*}$ & 0.00$^{*}$ & 0.00$^{*}$ & 0.00$^{*}$ & 0.00$^{*}$\\
   \hline
   \multicolumn{12}{l}{$^{{*}}$\footnotesize{p$<$0.01}} \\
   \end{tabular}
\end{table}

\begin{table}[tb]
\caption{Post-Hoc Paired t-test result for confidence estimation.}
\label{tbl:ttest_ce}
\centering
 \scalebox{0.75}{
   \begin{tabular}{c|c|lllllllllll}
   \hline
    \multirow{2}{*}{Pair} & \multirow{2}{*}{Param.} & \multicolumn{11}{c}{Training samples per class}\\
   \cline{3-13}
     &  & 10 & 20 & 50 & 100 & 150 & 200 & 300 & 400 & 500 & 1,000 & 5,382 \\
   \hline
   \multirow{2}{*}{ \makecell{Proposed self-supervised DL\\ and fully-supervised DL}} & t value & 44.96 & 23.51 & 35.14 & 41.33 & 34.91 & 25.38 & 7.21 & 1.89 & 2.21 & 2.78 & 4.28 \\
   \cline{2-13}
   & p value & 0.000$^{*}$ & 0.000$^{*}$ & 0.000$^{*}$ & 0.000$^{*}$ & 0.000$^{*}$ & 0.000$^{*}$ & 0.000$^{*}$ & 0.092 & 0.054 & 0.022 & 0.002$^{*}$\\
   \hline
   \multirow{2}{*}{ \makecell{Proposed self-supervised DL \\ and SVM}} & t value & 20.51 & 12.95 & 28.24 & 39.69 & 21.43 & 25.03 & 26.19 & 23.15 & 22.11 & 33.28 & 33.12 \\
   \cline{2-13}
   & p value & 0.000$^{*}$ & 0.000$^{*}$ & 0.000$^{*}$ & 0.000$^{*}$ & 0.000$^{*}$ & 0.000$^{*}$ & 0.000$^{*}$ & 0.000$^{*}$ & 0.000$^{*}$ & 0.000$^{*}$ & 0.000$^{*}$\\
   \hline
   \multirow{2}{*}{ \makecell{Fully-supervised DL and \\ SVM}} & t value & -10.54 & -22.24 & -11.10 & -18.75 & -2.80 & 7.67 & 19.66 & 25.41 & 25.48 & 22.20 & 28.72 \\
   \cline{2-13}
   & p value & 0.000$^{*}$ & 0.000$^{*}$ & 0.000$^{*}$ & 0.000$^{*}$ & 0.021 & 0.000$^{*}$ & 0.000$^{*}$ & 0.000$^{*}$ & 0.000$^{*}$ & 0.000$^{*}$ & 0.000$^{*}$ \\
   \hline
   \multicolumn{13}{l}{$^{{*}}$\footnotesize{p$<$0.0033}} \\
   \end{tabular}
   }
\end{table}
From the one-way repeated measures ANOVA test, we have confirmed that
there exists at least one population mean is different from the rest.
From the results of the post-hoc paired t-test, we have confirmed the following:
For the comparison between the proposed self-supervised DL method and the fully-supervised DL,
we found significant differences except for the cases of 400, 500, and 1,000 training samples.
For the comparison between the proposed self-supervised DL method and SVM, all cases show a significant difference.
For the comparison between the fully-supervised DL and SVM,
we could not reject the null hypothesis for the case of 150 training samples per class,
but it was rejected for all other cases. For the cases with the larger number of training 
samples, the fully-supervised DL worked better than SVM, and for the case with the smaller
number of training samples, the opposite conclusion is held.
From these results of statistical analysis, 
we can also confirm that our discussion of the performance
comparison we made above has been supported.
\begin{figure}[tb]
	\centering
	\begin{minipage}[b]{0.48\hsize}
	\includegraphics[width=\hsize]{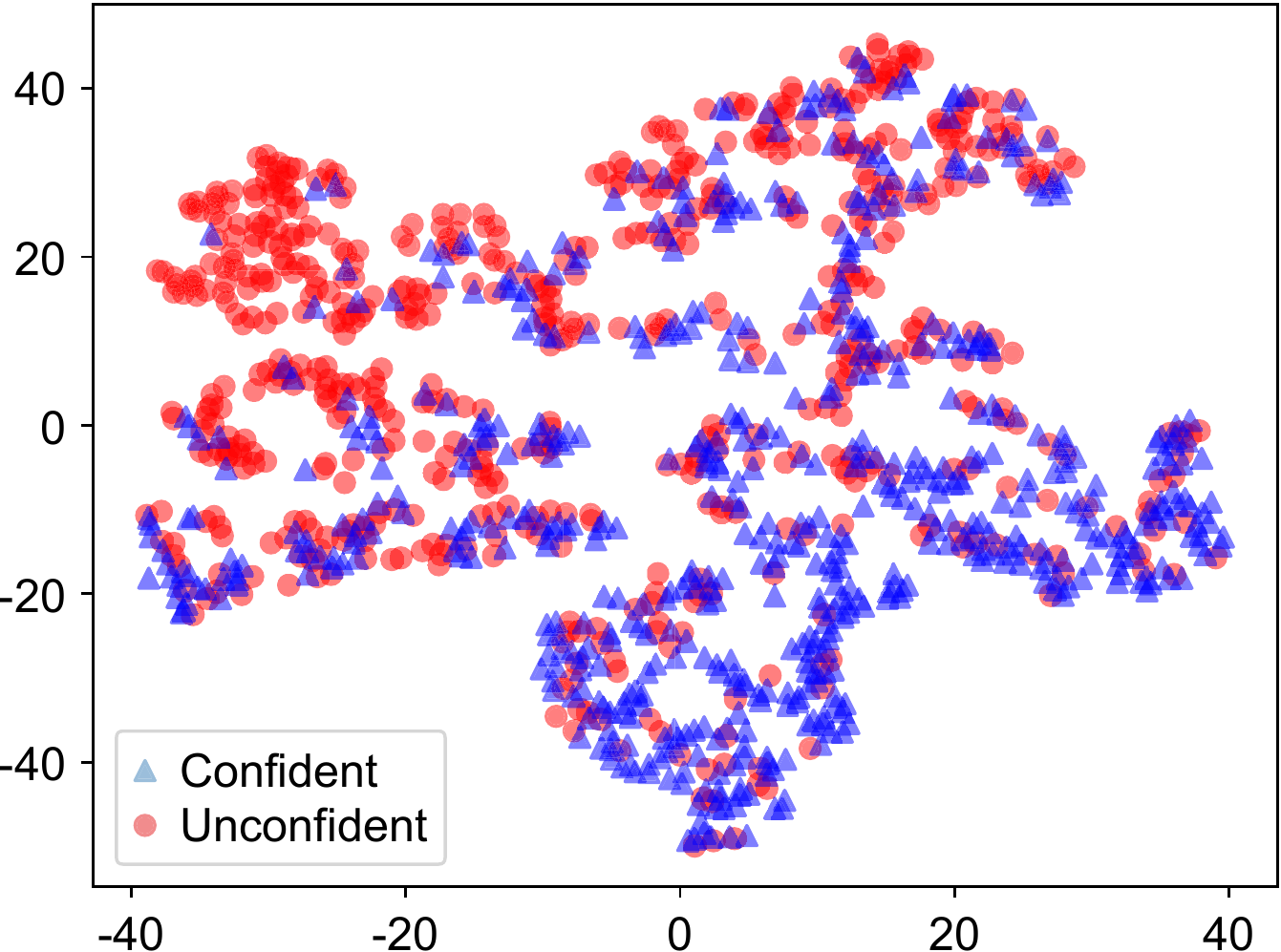}
	\subcaption{200 samples per class for the proposed self-supervised DL method}
	\label{fig:tsne_200_ssl_F1_ce}
	\end{minipage}
	\hfill
	\begin{minipage}[b]{0.48\hsize}
	\includegraphics[width=\hsize]{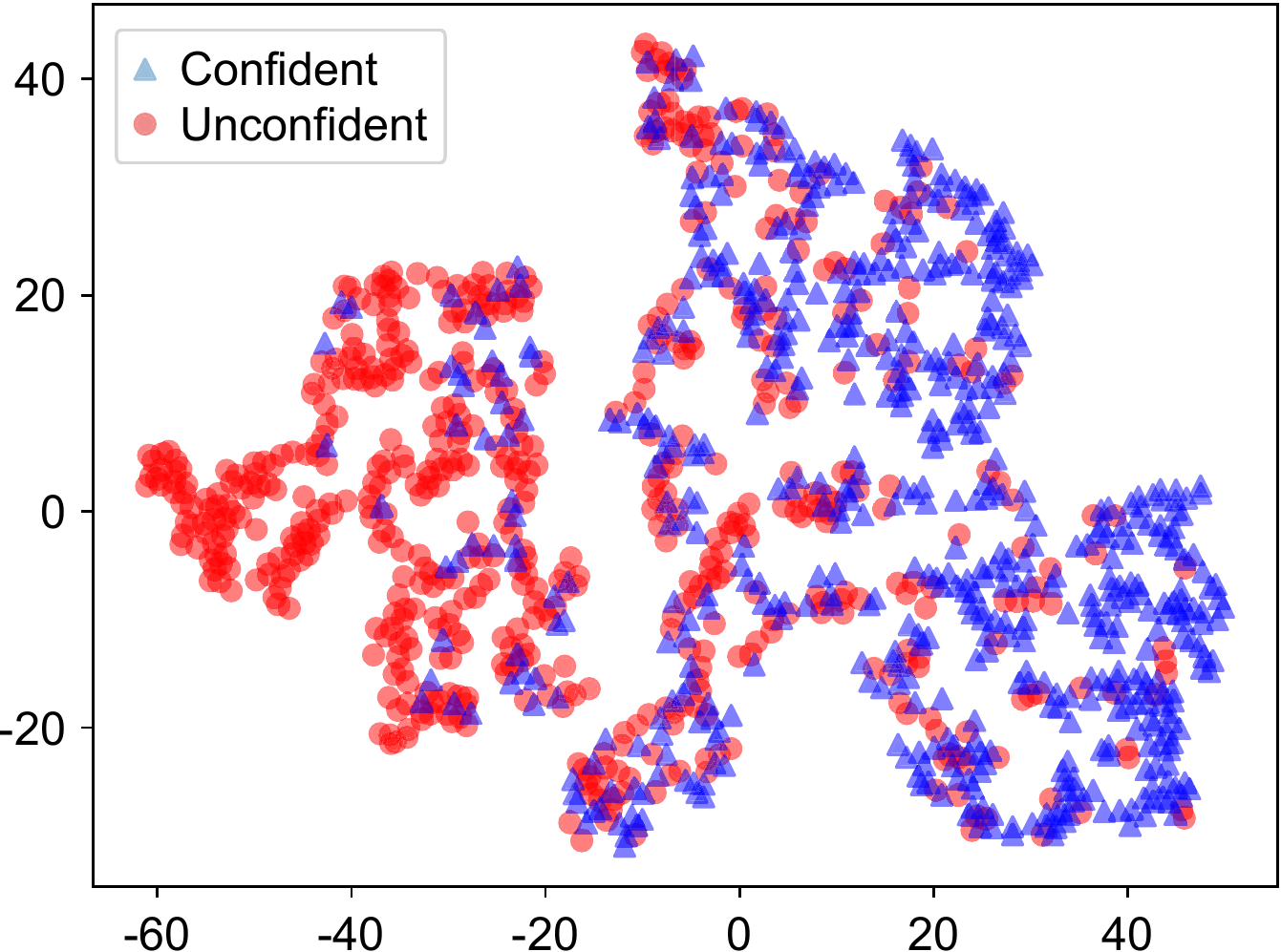}
	\subcaption{5,382 samples per class for the proposed self-supervised DL method}
	\label{fig:tsne_all_ssl_F1_ce}
	\end{minipage}
	
	  \vspace*{\intextsep}

	\begin{minipage}[b]{0.48\hsize}
	\includegraphics[width=\hsize]{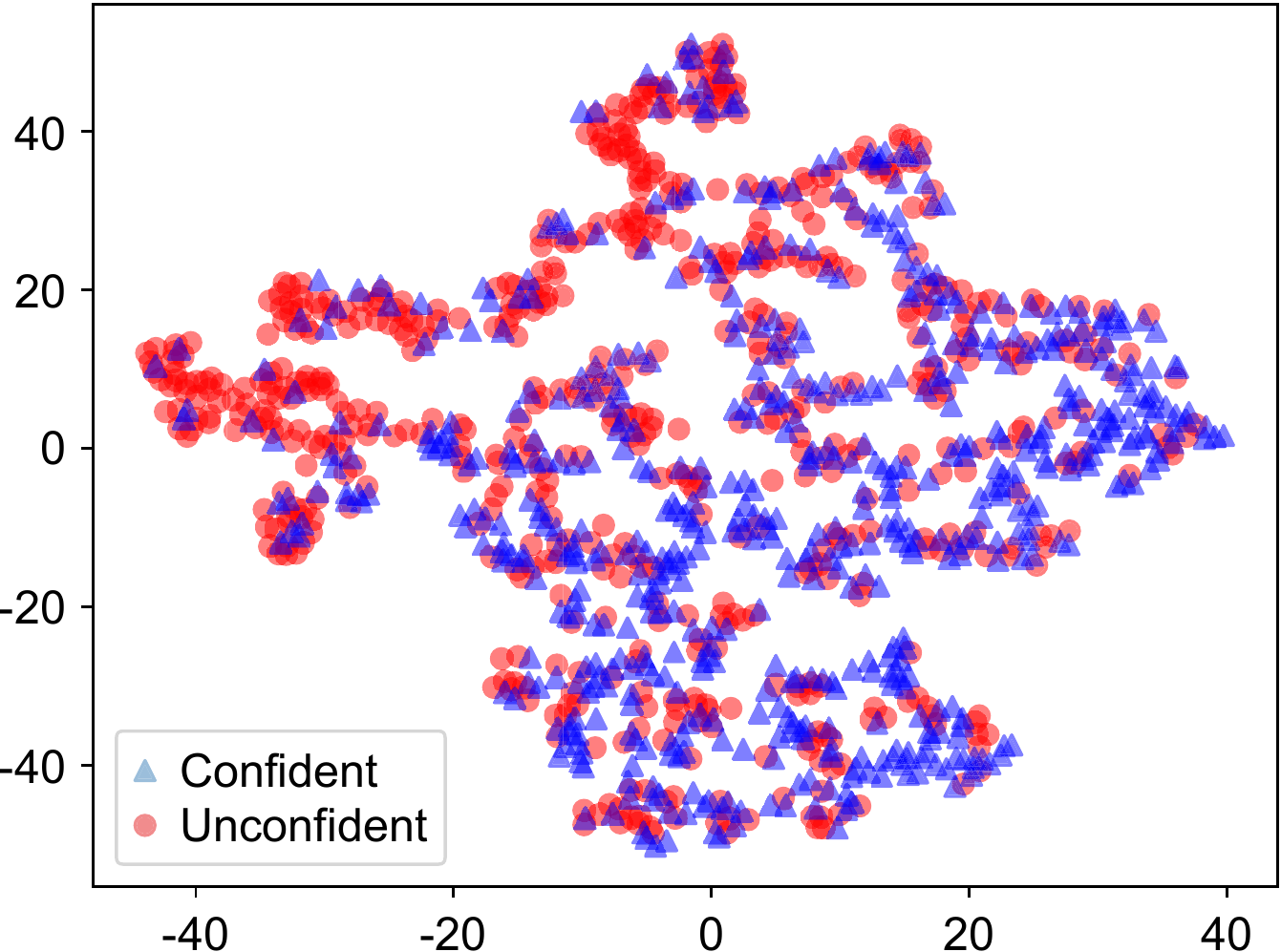}
	\subcaption{200 samples per class for the fully-supervised DL}
	\label{fig:tsne_200_supervised_F1_ce}
	\end{minipage}
	\hfill
	\begin{minipage}[b]{0.48\hsize}
	\includegraphics[width=\hsize]{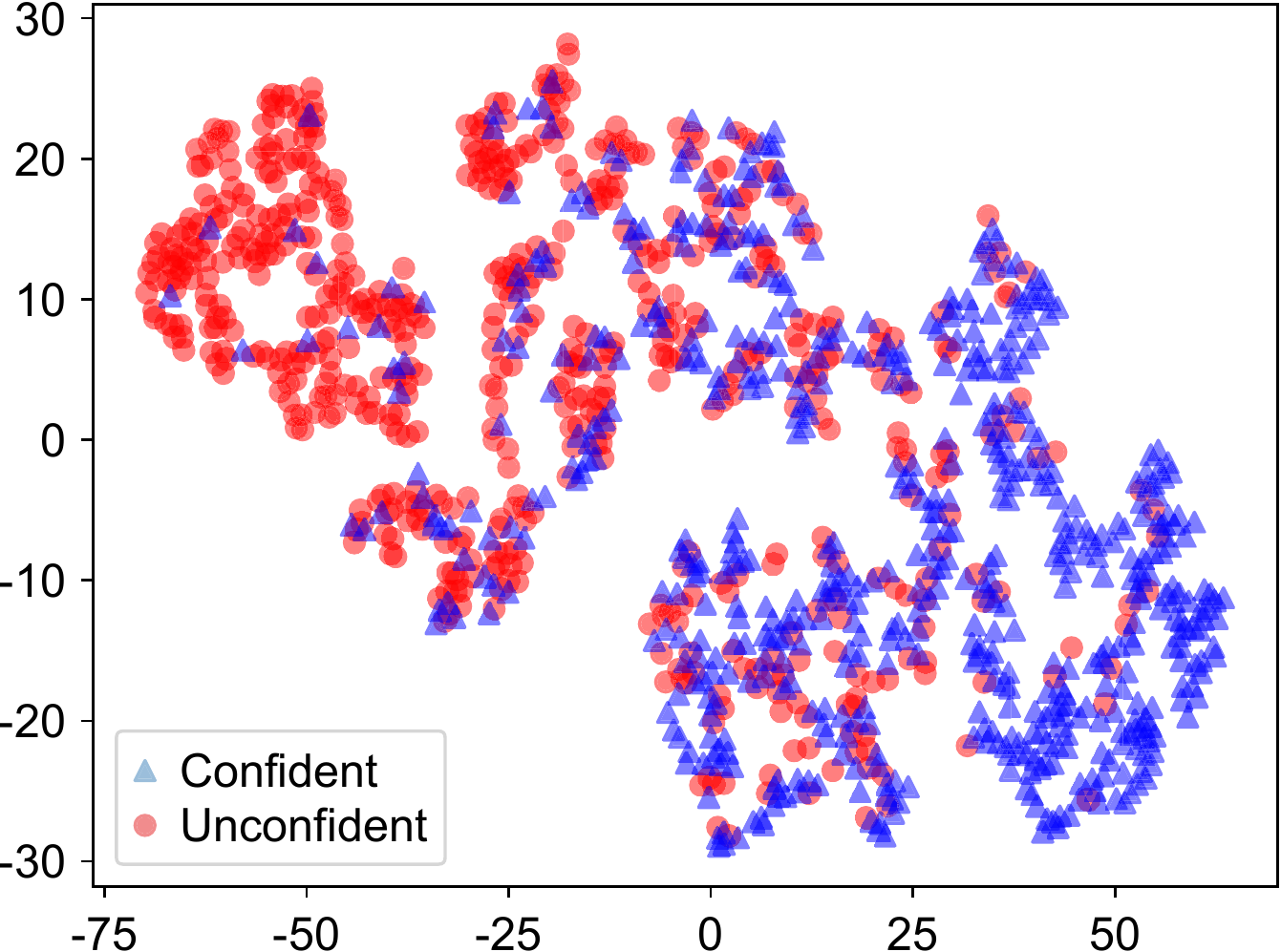}
	\subcaption{5,382 samples per class for the fully-supervised DL}
	\label{fig:tsne_all_supervised_F1_ce}
	\end{minipage}
\caption{Visualisation of features using t-SNE for confidence estimation.}
\label{fig:tsne_ce}
\end{figure}

Next, we show in \autoref{fig:tsne_ce} the t-SNE results for 200 samples per class, and 5,382 samples per class 
for both the proposed self-supervised DL method and the fully-supervised DL.
As compared to the results for reading detection, we have less separated distributions.
However, we can still see that the distributions with 5,382 samples are better than those with 200 samples.

\autoref{tbl:confusion_ce} shows the confusion matrices for 200 samples per class, and 5,382 samples per class cases for all three methods.
From this table, we can find that for all the methods predictions were not biased with the larger number of 
training samples, but biased with the smaller number of training samples except for the case of the proposed self-supervised DL method.

\begin{table}[tb]
\caption{Confusion matrices for confidence estimation.}
\label{tbl:confusion_ce}
\centering
\hfill
\begin{minipage}[t]{0.46\hsize}
\subcaption{200 samples per class for the proposed self-supervised DL method}
\centering
   \begin{tabular}{|c|c|c|c|}
   \hline
   \multicolumn{2}{|c|}{} & \multicolumn{2}{c|}{Predicted} \\
   \cline{3-4}
   \multicolumn{2}{|c|}{} & Confident & Unconfident\\
   \hline
   \multirow{2}{*}{Actual} & Confident & \cellcolor{blue!30}402 & 197\\
   \cline{2-4}
   & Unconfident & 186 & \cellcolor{red!30}412\\
   \hline
   \end{tabular}
   \label{tbl:CM_200_proposed self-supervised DL _ce}
 \end{minipage}
\hfill
\begin{minipage}[t]{0.46\hsize}
\subcaption{5,382 samples per class for the proposed self-supervised DL method}
\centering
   \begin{tabular}{|c|c|c|c|}
   \hline
   \multicolumn{2}{|c|}{} & \multicolumn{2}{c|}{Predicted} \\
   \cline{3-4}
   \multicolumn{2}{|c|}{} & Confident & Unconfident\\
   \hline
   \multirow{2}{*}{Actual} & Confident & \cellcolor{blue!30}468 & 131\\
   \cline{2-4}
   & Unconfident & 150 & \cellcolor{red!30}448\\
   \hline
   \end{tabular}
   \label{tbl:CM_all_proposed_ce}
 \end{minipage}
\hfill
   \hspace{0mm} 

	  \vspace*{\intextsep}

\hfill
\begin{minipage}[t]{0.46\hsize}
\subcaption{200 samples per class for the fully-supervised DL}
\centering
   \begin{tabular}{|c|c|c|c|}
   \hline
   \multicolumn{2}{|c|}{} & \multicolumn{2}{c|}{Predicted} \\
   \cline{3-4}
   \multicolumn{2}{|c|}{} & Confident & Unconfident\\
   \hline
   \multirow{2}{*}{Actual} & Confident & \cellcolor{blue!30}276 & 323\\
   \cline{2-4}
   & Unconfident & 152 & \cellcolor{red!30}446\\
   \hline
   \end{tabular}
   \label{tbl:CM_200_supervised_ce}
 \end{minipage}
 \hfill
\begin{minipage}[t]{0.46\hsize}
\subcaption{5,382 samples per class for the fully-supervised DL}
\centering
   \begin{tabular}{|c|c|c|c|}
   \hline
   \multicolumn{2}{|c|}{} & \multicolumn{2}{c|}{Predicted} \\
   \cline{3-4}
   \multicolumn{2}{|c|}{} & Confident & Unconfident\\
   \hline
   \multirow{2}{*}{Actual} & Confident & \cellcolor{blue!30}468 & 131\\
   \cline{2-4}
   & Unconfident & 161 & \cellcolor{red!30}437\\
   \hline
   \end{tabular}
   \label{tbl:CM_all_supervised_ce}
 \end{minipage}
\hfill
   \hspace{0mm} 
 
	  \vspace*{\intextsep}

\hfill
\begin{minipage}[t]{0.46\hsize}
\subcaption{200 samples per class for the SVM}
\centering
   \begin{tabular}{|c|c|c|c|}
   \hline
   \multicolumn{2}{|c|}{} & \multicolumn{2}{c|}{Predicted} \\
   \cline{3-4}
   \multicolumn{2}{|c|}{} & Confident & Unconfident\\
   \hline
   \multirow{2}{*}{Actual} & Confident & \cellcolor{blue!30}426 & 173\\
   \cline{2-4}
   & Unconfident & 328 & \cellcolor{red!30}270\\
   \hline
   \end{tabular}
   \label{tbl:CM_200_svm_ce}
 \end{minipage}
 \hfill
\begin{minipage}[t]{0.46\hsize}
\subcaption{5,382 samples per class for the SVM}
\centering
   \begin{tabular}{|c|c|c|c|}
   \hline
   \multicolumn{2}{|c|}{} & \multicolumn{2}{c|}{Predicted} \\
   \cline{3-4}
   \multicolumn{2}{|c|}{} & Confident & Unconfident\\
   \hline
   \multirow{2}{*}{Actual} & Confident & \cellcolor{blue!30}390 & 209\\
   \cline{2-4}
   & Unconfident & 257 & \cellcolor{red!30}341\\
   \hline
   \end{tabular}
   \label{tbl:CM_all_svm_ce}
 \end{minipage}
 \hfill
   \hspace{0mm} 
 \end{table}

\begin{figure}[tb]
    \begin{minipage}[b]{0.24\hsize}
    \centering
    \includegraphics[width=\hsize]{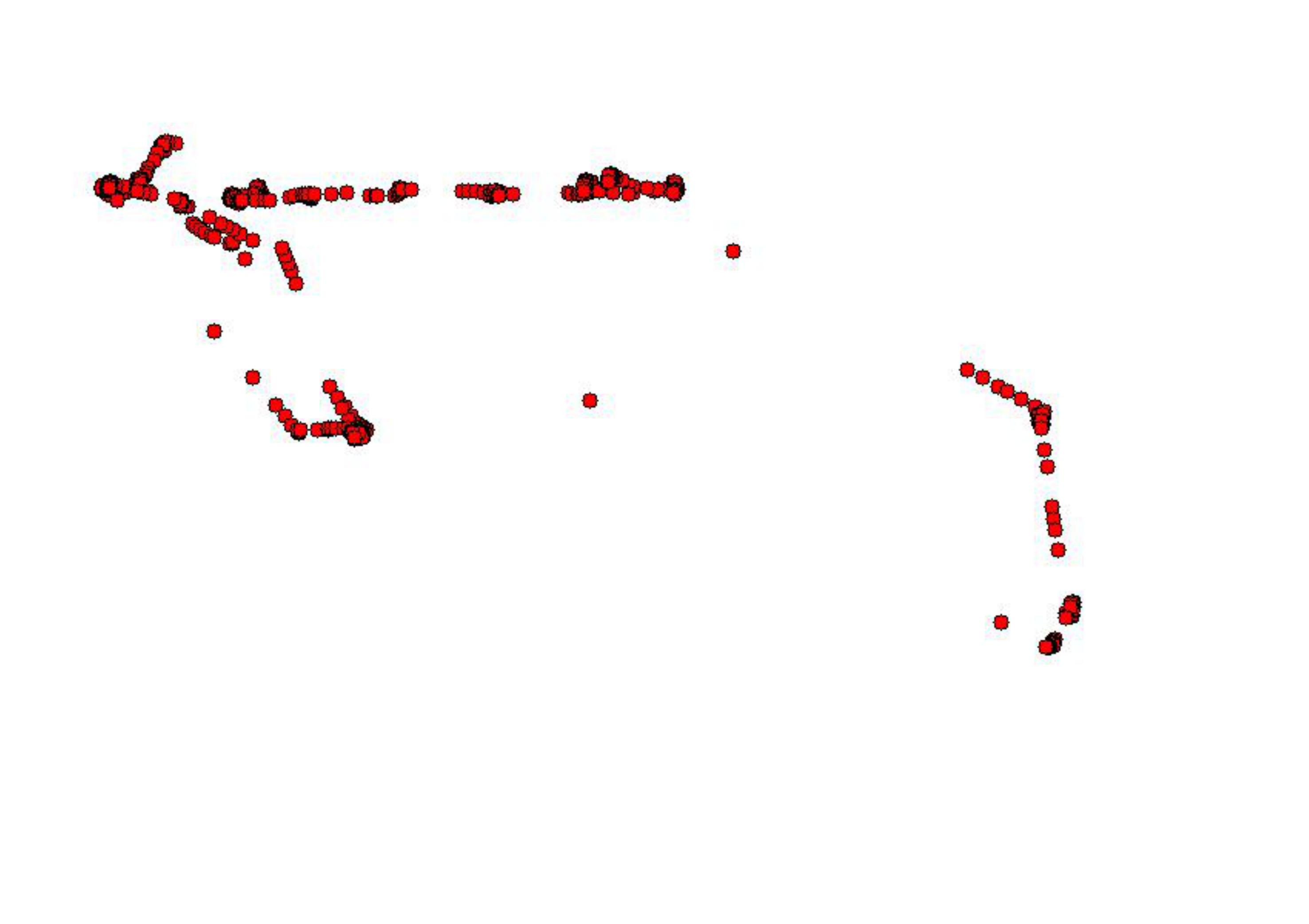}
    \subcaption{}
    \label{fig:correct1}
    \end{minipage}
    \hfill
    \begin{minipage}[b]{0.24\hsize}
    \centering
    \includegraphics[width=\hsize]{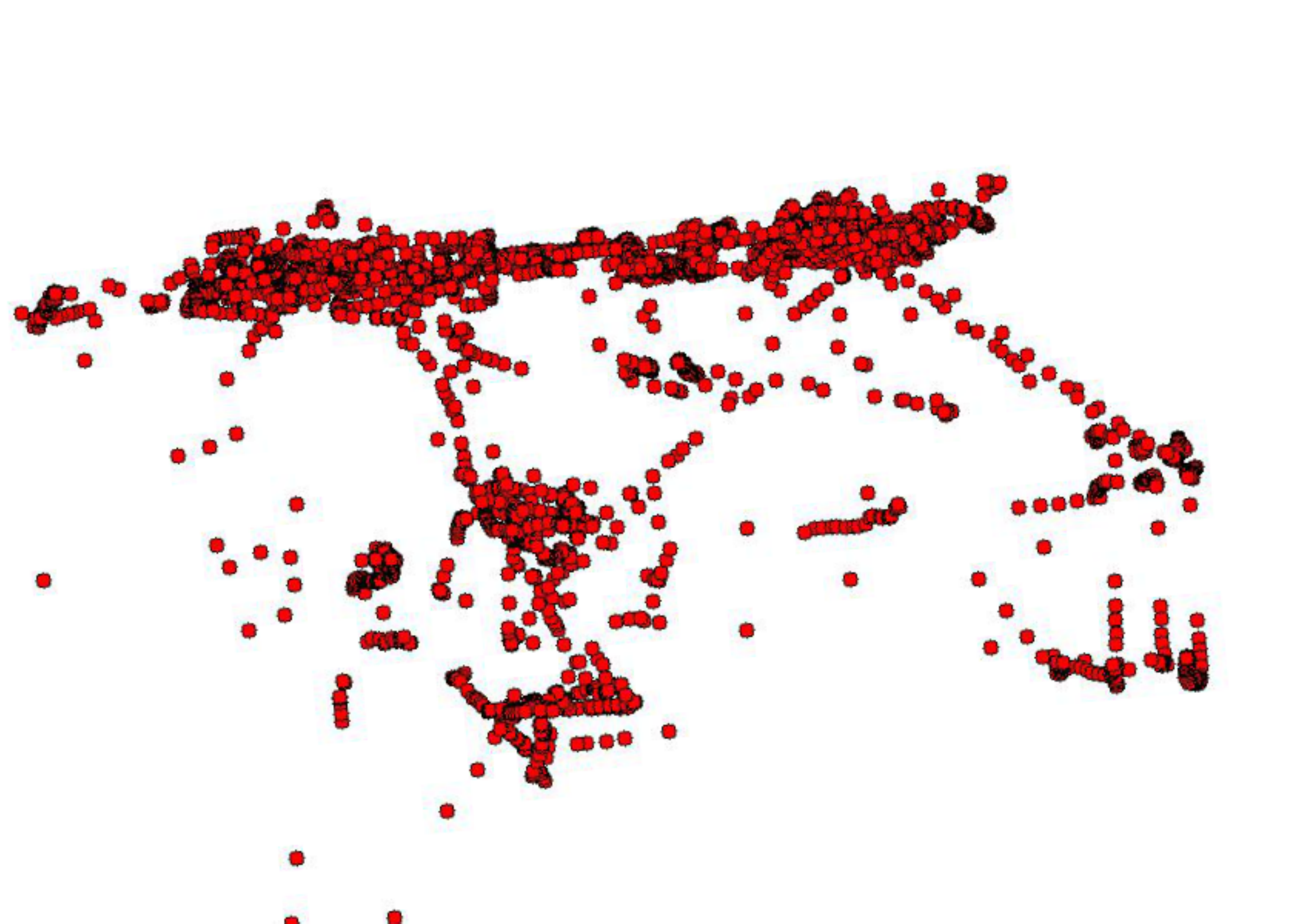}
    \subcaption{}
    \label{fig:correct2}
    \end{minipage}
    \hfill
      \begin{minipage}[b]{0.24\hsize}
        \centering
        \includegraphics[width=\hsize]{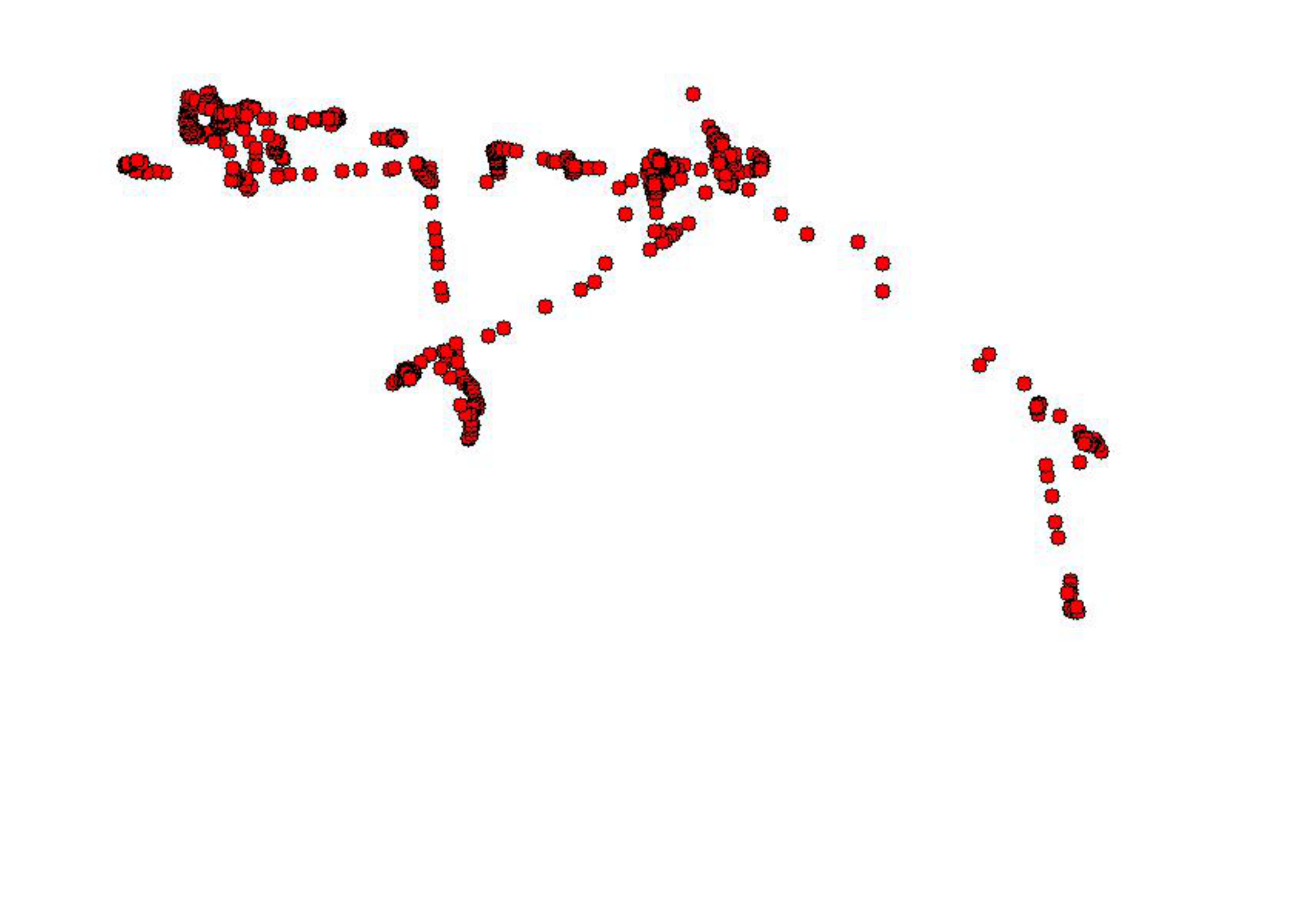}
        \subcaption{}
        \label{fig:false1}
      \end{minipage}
    \hfill
      \begin{minipage}[b]{0.24\hsize}
        \centering
        \includegraphics[width=\hsize]{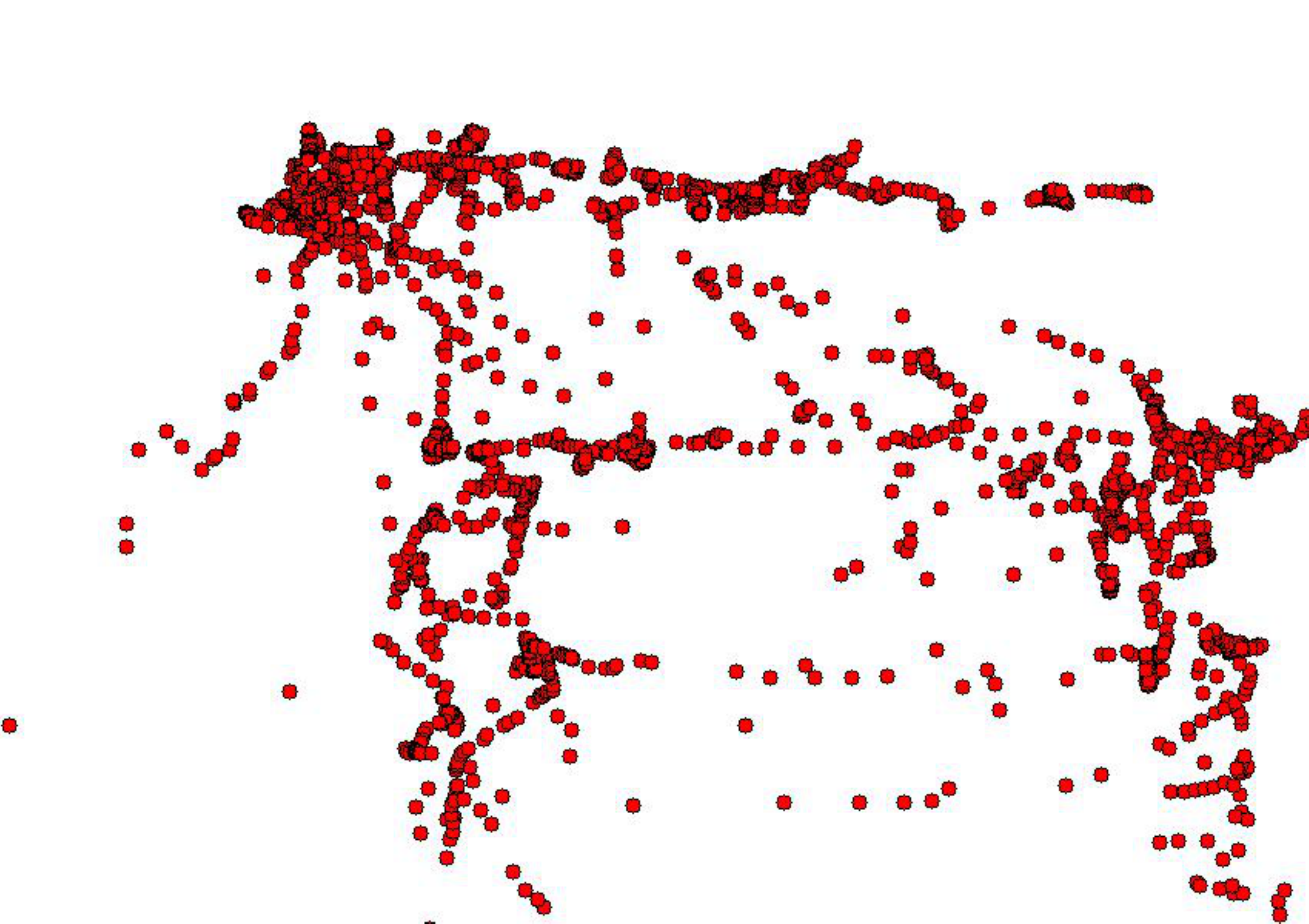}
        \subcaption{}
        \label{fig:false2}
      \end{minipage}
\caption{Case study for confidence estimation; (a) data sample corresponds to confident class and correctly predicted and (b) data sample corresponds to unconfident class and correctly predicted, (c) data sample corresponds to unconfident class and incorrectly predicted, and (d) data sample corresponds to confident class and incorrectly predicted.}
\label{fig:data_example_ce}
\end{figure}

\autoref{fig:data_example_ce} shows some
typical cases of correct and incorrect classification.
In many cases, the user's eye gaze was distributed like
either \autoref{fig:correct1} or \subref{fig:correct2}.
The former is a typical confident behavior;
the user spends less time and goes immediately to an answer.
The latter is also a typical unconfident behavior;
the user spends more time and looks at choices more carefully.
However, we also observed some difficult cases
as shown in Figures~\ref{fig:false1} and \subref{fig:false2},
which are unconfident and confident cases, respectively.
The former is the case of an early surrender;
the user quickly realizes that they do not know the answer.
In this case, the user randomly selects a choice.
The latter represents finding an answer
after deliberation. A careful examination of the
choices allows the user to select an answer with confidence.
In general, these two cases are difficult to distinguish
even by the proposed self-supervised DL method. 
To solve this issue, we need to introduce other sensors.
For example, a camera to capture facial expression may be 
of help to obtain good features for classification.

\section{Conclusion} \label{sec:conclusion}
Automatic recording and reading behavior analysis allow users to examine their reading habits which can help with the development of reading strategies. 
Methods that use classical machine learning approaches and handcrafted features may achieve good results in laboratory settings, but may not obtain satisfactory results in-the-wild. DL methods that can solve this issue that requires a large-sized labeled dataset to extract useful features. However, a large-sized labeled data collections are difficult to obtain. As a step towards tackling providing robust and feasible reading analysis, we have proposed a  self-supervised DL method.
We evaluated the effectiveness of the proposed self-supervised DL method by selecting two reading activities that explore physical reading and confidence, respectively. We evaluated both tasks with the proposed self-supervised DL method, the fully-supervised DL, and SVM. 

The proposed self-supervised DL method for reading detection consists of two stages. In the first stage, we trained the network by solving pretext tasks automatically applied to the unlabeled data for representation learning. In the second stage, we created the reading detection target task network by fine-tuning the pre-trained base network using labeled data. 
Confidence estimation followed a similar process. 

From the experimental results, we have confirmed that 
the proposed self-supervised DL method performs the best for both reading activity classification tasks 
compared to the fully-supervised DL method and SVM for all cases of the numbers of training samples.
Therefore we can always recommend to use the proposed self-supervised DL method regardless of the available number of training samples.

Future work includes further improvement in accuracy of the proposed self-supervised DL method by introducing other sensors, as well as its application for other reading activity classification tasks using various sensors.

\section*{Acknowledgments}
This work was supported in part by the JST CREST (Grant No. JPMJCR16E1), JSPS Grant-in-Aid for Scientific Research (20H04213), Grand challenge of the Initiative for Life Design Innovation (iLDi), and OPU Keyproject.
\printbibliography

@book{rayner_psychology_2012,
    author = {Rayner, Keith and Pollatsek, Alexander and Ashby, Jane and Jr, Charles Clifton},
	title = {Psychology of Reading},
	edition   = {2nd},
	publisher = {Psychology Press},
	year = {2012},
	address   = {New York, NY, USA},
}

@inproceedings{rd_Kelton_2,
author = {Kelton, Conor and Wei, Zijun and Ahn, Seoyoung and Balasubramanian, Aruna and Das, Samir R. and Samaras, Dimitris and Zelinsky, Gregory},
title = {Reading Detection in Real-Time},
year = {2019},
isbn = {9781450367097},
publisher = {ACM},
address = {New York, NY, USA},
url = {https://doi.org/10.1145/3314111.3319916},
doi = {10.1145/3314111.3319916},
booktitle = {Proceedings of the 11th ACM Symposium on Eye Tracking Research \& Applications},
articleno = {43},
numpages = {5},
location = {Denver, Colorado},
series = {ETRA '19}
}

@inproceedings{rd_campbell_4,
author = {Campbell, Christopher S. and Maglio, Paul P.},
title = {A Robust Algorithm for Reading Detection},
year = {2001},
isbn = {9781450374736},
publisher = {ACM},
address = {New York, NY, USA},
url = {https://doi.org/10.1145/971478.971503},
doi = {10.1145/971478.971503},
booktitle = {Proceedings of the 2001 Workshop on Perceptive User Interfaces},
pages = {1–7},
numpages = {7},
location = {Orlando, Florida, USA},
series = {PUI '01}
}

@inproceedings{rd_keat_7,
title = {Eye gaze based reading detection},
author = {{Foo Tun Keat} and Ranganath, Surendra and Venkatesh, Y.V.},
year = {2003},
booktitle = {{TENCON} 2003. {Conference} on {Convergent} {Technologies} for {Asia}-{Pacific} {Region}},
publisher = {IEEE},
location = {Bangalore, India},
volume={2},
number={},
numpages = {4},
pages={825-828},
isbn = {978-0-7803-8162-9},
url = {http://ieeexplore.ieee.org/document/1273294/},
doi={10.1109/TENCON.2003.1273294},
urldate = {2020-10-02}
}

@inproceedings{rd_Ishimaru_8,
author = {Ishimaru, Shoya and Kunze, Kai and Tanaka, Katsuma and Uema, Yuji and Kise, Koichi and Inami, Masahiko},
title = {Smart Eyewear for Interaction and Activity Recognition},
year = {2015},
isbn = {9781450331463},
publisher = {ACM},
address = {New York, NY, USA},
url = {https://doi.org/10.1145/2702613.2725449},
doi = {10.1145/2702613.2725449},
booktitle = {Proceedings of the 33rd Annual ACM Conference Extended Abstracts on Human Factors in Computing Systems},
pages = {307–310},
numpages = {4},
location = {Seoul, Republic of Korea},
series = {CHI EA '15}
}

@inproceedings{rd_Ishimaru_9,
author = {Ishimaru, Shoya and Hoshika, Kensuke and Kunze, Kai and Kise, Koichi and Dengel, Andreas},
title = {Towards Reading Trackers in the Wild: Detecting Reading Activities by EOG Glasses and Deep Neural Networks},
year = {2017},
isbn = {9781450351904},
publisher = {ACM},
address = {New York, NY, USA},
url = {https://doi.org/10.1145/3123024.3129271},
doi = {10.1145/3123024.3129271},
booktitle = {Proceedings of the 2017 ACM International Joint Conference on Pervasive and Ubiquitous Computing and Proceedings of the 2017 ACM International Symposium on Wearable Computers},
pages = {704–711},
numpages = {8},
location = {Maui, Hawaii},
series = {UbiComp '17}
}

@inproceedings{rd_landsmann_10,
author = {Landsmann, Manuel and Augereau, Olivier and Kise, Koichi},
title = {Classification of Reading and Not Reading Behavior Based on Eye Movement Analysis},
year = {2019},
publisher = {ACM},
address = {New York, NY, USA},
url = {https://doi.org/10.1145/3341162.3343811},
doi = {10.1145/3341162.3343811},
booktitle = {Adjunct Proceedings of the 2019 ACM International Joint Conference on Pervasive and Ubiquitous Computing and Proceedings of the 2019 ACM International Symposium on Wearable Computers},
pages = {109–112},
numpages = {4},
location = {London, UK},
series = {UbiComp/ISWC '19 Adjunct}
}

@article{rd_Bulling_12,
author = {Bulling, Andreas and Ward, Jamie A. and Gellersen, Hans and Tröster, Gerhard},
title = {Eye Movement Analysis for Activity Recognition Using Electrooculography},
year = {2011},
issue_date = {April 2011},
publisher = {IEEE Computer Society},
address = {USA},
volume = {33},
number = {4},
issn = {0162-8828},
url = {https://doi.org/10.1109/TPAMI.2010.86},
doi = {10.1109/TPAMI.2010.86},
journal = {IEEE Trans. Pattern Anal. Mach. Intell.},
month = apr,
pages = {741–753},
numpages = {13}
}

@inproceedings{rd_Ishimaru_13,
author = {Ishimaru, Shoya and Maruichi, Takanori and Landsmann, Manuel and Kise, Koichi and Dengel, Andreas},
title = {Electrooculography Dataset for Reading Detection in the Wild},
year = {2019},
isbn = {9781450368698},
publisher = {ACM},
address = {New York, NY, USA},
url = {https://doi.org/10.1145/3341162.3343812},
doi = {10.1145/3341162.3343812},
booktitle = {Adjunct Proceedings of the 2019 ACM International Joint Conference on Pervasive and Ubiquitous Computing and Proceedings of the 2019 ACM International Symposium on Wearable Computers},
pages = {85–88},
numpages = {4},
location = {London, UK},
series = {UbiComp/ISWC '19 Adjunct}
}

@inproceedings{rd_Steil_14,
author = {Steil, Julian and Bulling, Andreas},
title = {Discovery of Everyday Human Activities from Long-Term Visual Behaviour Using Topic Models},
year = {2015},
isbn = {9781450335744},
publisher = {ACM},
address = {New York, NY, USA},
url = {https://doi.org/10.1145/2750858.2807520},
doi = {10.1145/2750858.2807520},
booktitle = {Proceedings of the 2015 ACM International Joint Conference on Pervasive and Ubiquitous Computing},
pages = {75–85},
numpages = {11},
location = {Osaka, Japan},
series = {UbiComp '15}
}

@inproceedings{rd_Ishimaru_16,
author = {Ishimaru, Shoya and Kunze, Kai and Kise, Koichi and Weppner, Jens and Dengel, Andreas and Lukowicz, Paul and Bulling, Andreas},
title = {In the Blink of an Eye: Combining Head Motion and Eye Blink Frequency for Activity Recognition with Google Glass},
year = {2014},
isbn = {9781450327619},
publisher = {ACM},
address = {New York, NY, USA},
url = {https://doi.org/10.1145/2582051.2582066},
doi = {10.1145/2582051.2582066},
booktitle = {Proceedings of the 5th Augmented Human International Conference},
articleno = {15},
numpages = {4},
location = {Kobe, Japan},
series = {AH '14}
}

@inproceedings{rd_kunze_17,
author = {Kunze, Kai and Utsumi, Yuzuko and Shiga, Yuki and Kise, Koichi and Bulling, Andreas},
title = {I Know What You Are Reading: Recognition of Document Types Using Mobile Eye Tracking},
year = {2013},
isbn = {9781450321273},
publisher = {ACM},
address = {New York, NY, USA},
url = {https://doi.org/10.1145/2493988.2494354},
doi = {10.1145/2493988.2494354},
booktitle = {Proceedings of the 2013 International Symposium on Wearable Computers},
pages = {113–116},
numpages = {4},
location = {Zurich, Switzerland},
series = {ISWC '13}
}

@article{rd_Bulling_20,
author = {Bulling, Andreas and Ward, Jamie A. and Gellersen, Hans},
title = {Multimodal Recognition of Reading Activity in Transit Using Body-Worn Sensors},
year = {2012},
issue_date = {March 2012},
publisher = {ACM},
address = {New York, NY, USA},
volume = {9},
number = {1},
issn = {1544-3558},
url = {https://doi.org/10.1145/2134203.2134205},
doi = {10.1145/2134203.2134205},
journal = {ACM Trans. Appl. Percept.},
month = mar,
articleno = {2},
numpages = {21}
}

@article{mcq_Tsai_1,
author = {Tsai, Meng-Jung and Hou, Huei-Tse and Lai, Meng-Lung and Liu, Wan-Yi and Yang, Fang-Ying},
title = {Visual Attention for Solving Multiple-Choice Science Problem: An Eye-Tracking Analysis},
year = {2012},
issue_date = {January, 2012},
publisher = {Elsevier Science Ltd.},
address = {GBR},
volume = {58},
number = {1},
issn = {0360-1315},
url = {https://doi.org/10.1016/j.compedu.2011.07.012},
doi = {10.1016/j.compedu.2011.07.012},
journal = {Computers \& Education},
month = jan,
pages = {375–385},
numpages = {11}
}

@inproceedings{mcq_Yamada_2,
author = {Yamada, Kento and Kise, Koichi and Augereau, Olivier},
title = {Estimation of Confidence Based on Eye Gaze: An Application to Multiple-Choice Questions},
year = {2017},
isbn = {9781450351904},
publisher = {ACM},
address = {New York, NY, USA},
url = {https://doi.org/10.1145/3123024.3123138},
doi = {10.1145/3123024.3123138},
booktitle = {Proceedings of the 2017 ACM International Joint Conference on Pervasive and Ubiquitous Computing and Proceedings of the 2017 ACM International Symposium on Wearable Computers},
pages = {217–220},
numpages = {4},
location = {Maui, Hawaii},
series = {UbiComp '17}
}

@inproceedings{ubiDL_Hammerla_1,
author = {Hammerla, Nils Y. and Halloran, Shane and Pl\"{o}tz, Thomas},
title = {Deep, Convolutional, and Recurrent Models for Human Activity Recognition Using Wearables},
year = {2016},
isbn = {9781577357704},
publisher = {AAAI Press},
booktitle = {Proceedings of the Twenty-Fifth International Joint Conference on Artificial Intelligence},
pages = {1533–1540},
numpages = {8},
location = {New York, NY, USA},
series = {IJCAI'16},
url = {https://www.ijcai.org/Proceedings/16/Papers/220.pdf}
}

@article{ubiDL_Ordez_2,
author = {Ordóñez, Francisco Javier and Roggen, Daniel},
title = {Deep {Convolutional} and {LSTM} {Recurrent} {Neural} {Networks} for {Multimodal} {Wearable} {Activity} {Recognition}},
journal = {Sensors},
volume = {16},
number = {1},
month = jan,
year = {2016},
Publisher= {Multidisciplinary Digital Publishing Institute},
pages = {115},
url = {https://www.mdpi.com/1424-8220/16/1/115},
doi = {10.3390/s16010115},
}

@article{ubiDL_sheng_3,
author = {Sheng, Taoran and Huber, Manfred},
title = {Weakly Supervised Multi-Task Representation Learning for Human Activity Analysis Using Wearables},
year = {2020},
issue_date = {June 2020},
publisher = {ACM},
address = {New York, NY, USA},
volume = {4},
number = {2},
url = {https://doi.org/10.1145/3397330},
doi = {10.1145/3397330},
journal = {Proc. ACM Interact. Mob. Wearable Ubiquitous Technol.},
month = jun,
articleno = {57},
numpages = {18}
}

@article{ubiDL_Chen_4,
author = {Chen, Ling and Zhang, Yi and Peng, Liangying},
title = {METIER: A Deep Multi-Task Learning Based Activity and User Recognition Model Using Wearable Sensors},
year = {2020},
issue_date = {March 2020},
publisher = {ACM},
address = {New York, NY, USA},
volume = {4},
number = {1},
url = {https://doi.org/10.1145/3381012},
doi = {10.1145/3381012},
journal = {Proc. ACM Interact. Mob. Wearable Ubiquitous Technol.},
month = mar,
articleno = {5},
numpages = {18}
}

@article{perveDL_Sarhan_1,
	title = {Multipose {Face} {Recognition}-{Based} {Combined} {Adaptive} {Deep} {Learning} {Vector} {Quantization}},
	author = {Sarhan, Shahenda and Nasr, Aida A. and Shams, Mahmoud Y.},
	journal = {Computational Intelligence and Neuroscience},
	month = sep,
	year = {2020},
	ISSN = {1687-5265},
    numpages= {11},
    Publisher= {Hindawi},
    Volume={2020},
	url = {https://www.hindawi.com/journals/cin/2020/8821868/},
	doi = {10.1155/2020/8821868},
}

@article{perveDL_Yao_2,
author = {Yao, Shuochao and Zhao, Yiran and Shao, Huajie and Zhang, Chao and Zhang, Aston and Hu, Shaohan and Liu, Dongxin and Liu, Shengzhong and Su, Lu and Abdelzaher, Tarek},
title = {SenseGAN: Enabling Deep Learning for Internet of Things with a Semi-Supervised Framework},
year = {2018},
issue_date = {September 2018},
publisher = {ACM},
address = {New York, NY, USA},
volume = {2},
number = {3},
url = {https://doi.org/10.1145/3264954},
doi = {10.1145/3264954},
journal = {Proc. ACM Interact. Mob. Wearable Ubiquitous Technol.},
month = sep,
articleno = {144},
numpages = {21}
}

@article{DLhealth_nonaka_1,
	title = {Data {Augmentation} for {Electrocardiogram} {Classification} with {Deep} {Neural} {Network}},
	url = {http://arxiv.org/abs/2009.04398},
	urldate = {2020-10-02},
	journal = {arXiv:2009.04398 [eess]},
	author = {Nonaka, Naoki and Seita, Jun},
	month = sep,
	year = {2020}
}

@article{DLhealth_Hannun_2,
	title = {Cardiologist-level arrhythmia detection and classification in ambulatory electrocardiograms using a deep neural network},
	author = {Hannun, Awni Y. and Rajpurkar, Pranav and Haghpanahi, Masoumeh and Tison, Geoffrey H. and Bourn, Codie and Turakhia, Mintu P. and Ng, Andrew Y.},
	volume = {25},
	issn = {1546-170X},
	number = {1},
	journal = {Nature Medicine},
	month = jan,
	year = {2019},
    Publisher = {Nature Publishing Group},
	pages = {65-69},
	url = {https://www.nature.com/articles/s41591-018-0268-3},
	doi = {10.1038/s41591-018-0268-3},
}

@article{DLhealth_Ji_3,
	title = {Electrocardiogram {Classification} {Based} on {Faster} {Regions} with {Convolutional} {Neural} {Network}},
	author = {Ji, Yisheng and Zhang, Sen and Xiao, Wendong},
	journal = {Sensors},
	month = jun,
	year = {2019},
	volume = {19},
	number = {11},
	publisher = {MDPI},
	url = {https://pubmed.ncbi.nlm.nih.gov/31195603/},
	doi = {10.3390/s19112558},
	ISSN = {1424-8220}
}

@article{DLwellbeing_Saeed_1,
  author    = {Aaqib Saeed and Stojan Trajanovski},
  title     = {Personalized Driver Stress Detection with Multi-task Neural Networks using Physiological Signals},
  journal   = {Computing Research Repository},
  volume    = {abs/1711.06116},
  year      = {2017},
  url       = {http://arxiv.org/abs/1711.06116},
  archivePrefix = {arXiv},
  eprint    = {1711.06116},
  bibsource = {dblp computer science bibliography, https://dblp.org}
}

@inproceedings{DLwellbeing_Liu_2,
author = {Liu, Chang and Cao, Yu and Luo, Yan and Chen, Guanling and Vokkarane, Vinod and Ma, Yunsheng},
title = {DeepFood: Deep Learning-Based Food Image Recognition for Computer-Aided Dietary Assessment},
year = {2016},
isbn = {9783319396002},
publisher = {Springer-Verlag},
address = {Berlin, Heidelberg},
url = {https://doi.org/10.1007/978-3-319-39601-9_4},
doi = {10.1007/978-3-319-39601-9_4},
booktitle = {Proceedings of the 14th International Conference on Inclusive Smart Cities and Digital Health},
pages = {37–48},
numpages = {12},
location = {Wuhan, China},
series = {ICOST 2016}
}

@inproceedings{SSL_agrawal_1,
	title = {Learning to {See} by {Moving}},
	author = {Agrawal, Pulkit and Carreira, João and Malik, Jitendra},
	month = dec,
	year = {2015},
	publisher = {IEEE},
	booktitle = {2015 IEEE International Conference on Computer Vision (ICCV)},
	ISSN = {2380-7504},
	pages = {37-45},
	url = {https://doi.org/10.1109/ICCV.2015.13},
	doi={10.1109/ICCV.2015.13}
}

@article{SSL_imagevision_gidaris_4,
author    = {Spyros Gidaris and Praveer Singh and Nikos Komodakis},
title     = {Unsupervised Representation Learning by Predicting Image Rotations},
journal   = {CoRR},
volume    = {abs/1803.07728},
year      = {2018},
url       = {http://arxiv.org/abs/1803.07728},
archivePrefix = {arXiv},
eprint    = {1803.07728},
bibsource = {dblp computer science bibliography, https://dblp.org}
}

@inproceedings{SSL_imagevision_lee_8,
	title = {Unsupervised {Representation} {Learning} by {Sorting} {Sequences}},
	booktitle = {2017 {IEEE} {International} {Conference} on {Computer} {Vision} ({ICCV})},
	author = {Lee, Hsin-Ying and Huang, Jia-Bin and Singh, Maneesh and Yang, Ming-Hsuan},
	month = oct,
	year = {2017},
	ISSN = {2380-7504},
	publisher = {IEEE},
	pages = {667--676},
	url = {https://doi.org/10.1109/ICCV.2017.79},
}

@inproceedings{SSL_imagevision_owens_12,
title = {Ambient Sound Provides Supervision for Visual Learning},
author = {Owens, Andrew and Wu, Jiajun and McDermott, Josh H. and Freeman, William T. and Torralba, Antonio},
booktitle = {Computer {Vision} – {ECCV} 2016},
isbn = {978-3-319-46448-0},
volume = {9905},
publisher = {Springer},
address = {Cham},
year = {2016},
pages = {801--816},
url = {https://doi.org/10.1007/978-3-319-46448-0_48},
series = {Lecture {Notes} in {Computer} {Science}}
}

@inproceedings{SSL_imagevision_pathak_13,
title = {Curiosity-Driven Exploration by Self-Supervised Prediction},
author = {Deepak Pathak and Pulkit Agrawal and Alexei A. Efros and Trevor Darrell},
booktitle={2017 IEEE Conference on Computer Vision and Pattern Recognition Workshops (CVPRW)},
publisher = {IEEE},
month = jul,
year = {2017},
ISSN = {2160-7516},
url = {https://doi.org/10.1109/CVPRW.2017.70},
pages = {488--489},
}

@article{emDL_copeland_2,
	title = {Predicting reading comprehension scores from eye movements using artificial neural networks and fuzzy output error},
	author = {Copeland, Leana and Gedeon, Tom and Mendis, Sumudu},
	journal = {Artificial Intelligence Research},
	volume = {3},
	number = {3},
	issn = {1927-6982},
	doi = {10.5430/air.v3n3p35},
	month = aug,
	year = {2014},
	pages = {35},
	url = {https://doi.org/10.5430/air.v3n3p35},
}

@article{tsne_maaten,
	title = {Visualizing Data using t-SNE},
	author = {Maaten, Laurens van der and Hinton, Geoffrey},
	journal = {Journal of Machine Learning Research},
	volume = {9},
	number = {86},
	year = {2008},
	pages = {2579-2605},
	url = {http://jmlr.org/papers/v9/vandermaaten08a.html},
}

@article{SSL_saeed,
author = {Saeed, Aaqib and Ozcelebi, Tanir and Lukkien, Johan},
title = {Multi-Task Self-Supervised Learning for Human Activity Detection},
year = {2019},
issue_date = {June 2019},
publisher = {ACM},
address = {New York, NY, USA},
volume = {3},
number = {2},
url = {https://doi.org/10.1145/3328932},
doi = {10.1145/3328932},
journal = {Proc. ACM Interact. Mob. Wearable Ubiquitous Technol.},
month = jun,
articleno = {61},
numpages = {30},
}

@book{MCQ_haladyna,
title = {Developing and Validating Multiple-choice Test Items},
author = {Haladyna, Thomas M.},
edition = {3rd Edition},
publisher = {Routledge},
address = {New York, NY, USA},
month = nov,
year = {2015},
isbn = {978-1-138-96747-2},
language = {English},
}

@article{MCQ_lau,
	title = {Guessing, Partial Knowledge, and Misconceptions in Multiple-Choice Tests.},
	author = {Lau, Paul and Lau, Sie and Hong, Kian and Usop, Hasbee},
	journal = {Educational Technology \& Society},
	volume = {14},
	month = jan,
	year = {2011},
	pages = {99-110},
	url = {https://eric.ed.gov/?id=EJ963283}
}

@article{MCQ_mckenna,
	title = {Multiple choice questions: answering correctly and knowing the answer},
	volume = {16},
	issn = {1741-5659},
	url = {https://doi.org/10.1108/ITSE-09-2018-0071},
	doi = {10.1108/ITSE-09-2018-0071},
	number = {1},
	journal = {Interactive Technology and Smart Education},
	author = {McKenna, Peter},
	month = jan,
	year = {2019},
	Publisher = {Emerald Publishing Limited},
	pages = {59-73}
}

@article{MCQ_melovitz,
	title = {Analysis of testing with multiple choice versus open-ended questions: {Outcome}-based observations in an anatomy course},
	author = {Melovitz Vasan, Cheryl A. and DeFouw, David O. and Holland, Bart K. and Vasan, Nagaswami S.},
	volume = {11},
	number = {3},
	issn = {1935-9780},
	journal = {Anatomical Sciences Education},
	month = may,
	year = {2018},
	pmid = {28941215},
	pages = {254--261},
	url = {https://doi.org/10.1002/ase.1739},
	doi = {10.1002/ase.1739},
}

@article{MCQ_nehm,
	title = {Biology Majors Knowledge and Misconceptions of Natural Selection},
	volume = {57},
	issn = {0006-3568},
	number = {3},
	journal = {BioScience},
	author = {Nehm, Ross H. and Reilly, Leah},
	month = mar,
	year = {2007},
	Publisher = {Oxford Academic},
	pages = {263-272},
	url = {https://doi.org/10.1641/B570311},
	doi = {10.1641/B570311}
}

@article{MCQ_brassil,
	title = {Multiple-true-false questions reveal more thoroughly the complexity of student thinking than multiple-choice questions: a {Bayesian} item response model comparison},
	author = {Brassil, Chad E. and Couch, Brian A.},
	volume = {6},
	issn = {2196-7822},
	number = {1},
	journal = {International Journal of STEM Education},
	month = may,
	year = {2019},
	pages = {16},
	url = {https://doi.org/10.1186/s40594-019-0169-0},
	doi = {10.1186/s40594-019-0169-0},
}

@article{MCQ_sam,
	title = {Validity of very short answer versus single best answer questions for undergraduate assessment},
	author = {Sam, Amir H. and Hameed, Saira and Harris, Joanne and Meeran, Karim},
	volume = {16},
	issn = {1472-6920},
	number = {1},
	journal = {BMC Medical Education},
	month = oct,
	year = {2016},
	pages = {266},
	url = {https://doi.org/10.1186/s12909-016-0793-z},
	doi = {10.1186/s12909-016-0793-z},
}

@article{MCQ_chan,
	title = {Are Multiple-Choice Exams Easier for Economics Students? A Comparison of Multiple-Choice and "Equivalent" Constructed-Response Exam Questions},
	author = {Chan, Nixon and Kennedy, Peter E.},
	volume = {68},
	issn = {0038-4038},
	number = {4},
	urldate = {2020-10-15},
	journal = {Southern Economic Journal},
	year = {2002},
	Publisher = {Southern Economic Association},
	pages = {957--971},
	url = {https://doi.org/10.2307/1061503},
	doi = {10.2307/1061503},
}

@article{ImageDL_rawat,
	title = {Deep Convolutional Neural Networks for Image Classification: A Comprehensive Review},
	author = {Rawat, Waseem and Wang, Zenghui},
	volume = {29},
	issn = {0899-7667, 1530-888X},
	number = {9},
	journal = {Neural Computation},
	month = sep,
	year = {2017},
	pages = {2352-2449},
	url = {https://doi.org/10.1162/neco_a_00990},
	doi = {10.1162/neco_a_00990},
}

@article{imageDL_imagenet,
title = {ImageNet Classification with Deep Convolutional Neural Networks},
author = {Krizhevsky, Alex and Sutskever, Ilya and Hinton, Geoffrey E.},
year = {2017},
issue_date = {June 2017},
publisher = {ACM},
address = {New York, NY, USA},
volume = {60},
number = {6},
issn = {0001-0782},
url = {https://doi.org/10.1145/3065386},
doi = {10.1145/3065386},
journal = {Commun. ACM},
month = may,
pages = {84–90},
numpages = {7}
}

@article{imageDL_chan,
	title = {PCANet: A Simple Deep Learning Baseline for Image Classification.},
	author = {Chan, Tsung-Han and Jia, Kui and Gao, Shenghua and Lu, Jiwen and Zeng, Zinan and Ma, Yi},
	volume = {24},
	issn = {1057-7149, 1941-0042},
	url = {https://doi.org/10.1109/TIP.2015.2475625},
	doi = {10.1109/TIP.2015.2475625},
	number = {12},
	journal = {IEEE Transactions on Image Processing},
	month = dec,
	year = {2015},
	pages = {5017-5032},
}

@article{DLHAR_ronao,
	title = {Human activity recognition with smartphone sensors using deep learning neural networks},
	author = {Ronao, Charissa Ann and Cho, Sung-Bae},
	volume = {59},
	issn = {09574174},
	urldate = {2020-10-16},
	journal = {Expert Systems with Applications},
	month = oct,
	year = {2016},
	pages = {235-244},
	url = {https://doi.org/10.1016/j.eswa.2016.04.032},
	doi = {10.1016/j.eswa.2016.04.032},
}

@article{DLHAR_jobanputra,
	title = {Human Activity Recognition: A Survey},
	author = {Jobanputra, Charmi and Bavishi, Jatna and Doshi, Nishant},
	volume = {155},
	issn = {18770509},
	url = {https://doi.org/10.1016/j.procs.2019.08.100},
	doi = {10.1016/j.procs.2019.08.100},
	journal = {Procedia Computer Science},
	year = {2019},
	pages = {698-703}
}

@inproceedings{SSL_Harish,
author = {Haresamudram, Harish and Beedu, Apoorva and Agrawal, Varun and Grady, Patrick L. and Essa, Irfan and Hoffman, Judy and Pl\"{o}tz, Thomas},
title = {Masked Reconstruction Based Self-Supervision for Human Activity Recognition},
year = {2020},
isbn = {9781450380775},
publisher = {ACM},
address = {New York, NY, USA},
url = {https://doi.org/10.1145/3410531.3414306},
doi = {10.1145/3410531.3414306},
booktitle = {Proceedings of the 2020 International Symposium on Wearable Computers},
pages = {45–49},
numpages = {5},
location = {Virtual Event, Mexico},
series = {ISWC '20}
}

@inproceedings{speech_graves,
	title = {Speech recognition with deep recurrent neural networks},
	author = {Graves, Alex and Mohamed, Abdel-rahman and Hinton, Geoffrey},
	isbn = {978-1-4799-0356-6},
	booktitle = {2013 {IEEE} {International} {Conference} on {Acoustics}, {Speech} and {Signal} {Processing}},
	publisher = {IEEE},
	month = may,
	year = {2013},
	pages = {6645-6649},
	location = {Vancouver, BC, Canada},
	url = {https://doi.org/10.1109/ICASSP.2013.6638947},
	doi = {10.1109/ICASSP.2013.6638947}
}

@inproceedings{NLP_stanford,
	title = {The Stanford CoreNLP Natural Language Processing Toolkit},
	author = {Manning, Christopher and Surdeanu, Mihai and Bauer, John and Finkel, Jenny and Bethard, Steven and McClosky, David},
	url = {https://doi.org/10.3115/v1/P14-5010},
	doi = {10.3115/v1/P14-5010},
	urldate = {2020-10-16},
	booktitle = {Proceedings of 52nd {Annual} {Meeting} of the {Association} for {Computational} {Linguistics}: {System} {Demonstrations}},
	publisher = {Association for Computational Linguistics},
	month = jun,
	year = {2014},
	pages = {55-60},
	location = {Baltimore, Maryland},
}

@techreport{MCQ_siren,
	title = {The Best of both Worlds: Expanding the Depth and Breadth of Multiple-Choice Questions},
	author = {Siren, Kathleen},
	url = {https://papers.ssrn.com/abstract=3660997},
	number = {ID 3660997},
	institution = {Social Science Research Network},
	address = {Rochester, NY},
	month = may,
	year = {2020},
}

@inproceedings{rd_Biedert,
author = {Biedert, Ralf and Hees, J\"{o}rn and Dengel, Andreas and Buscher, Georg},
title = {A Robust Realtime Reading-Skimming Classifier},
year = {2012},
isbn = {9781450312219},
publisher = {ACM},
address = {New York, NY, USA},
url = {https://doi.org/10.1145/2168556.2168575},
doi = {10.1145/2168556.2168575},
booktitle = {Proceedings of the Symposium on Eye Tracking Research and Applications},
pages = {123–130},
numpages = {8},
location = {Santa Barbara, California},
series = {ETRA '12}
}

@article{rd_strukelj,
	title = {One page of text: Eye movements during regular and thorough reading, skimming, and spell checking.},
	author = {Strukelj, Alexander and Niehorster, Diederick C.},
	volume = {11},
	issn = {1995-8692},
	_url = {https://bop.unibe.ch/JEMR/article/view/Strukelj},
	url = {https://doi.org/10.16910/jemr.11.1.1},
	doi = {10.16910/jemr.11.1.1},
	number = {1},
	journal = {Journal of Eye Movement Research},
	month = feb,
	year = {2018},
}

@article{rd_srivastava,
author = {Srivastava, Namrata and Newn, Joshua and Velloso, Eduardo},
title = {Combining Low and Mid-Level Gaze Features for Desktop Activity Recognition},
year = {2018},
issue_date = {December 2018},
publisher = {ACM},
address = {New York, NY, USA},
volume = {2},
number = {4},
url = {https://doi.org/10.1145/3287067},
doi = {10.1145/3287067},
journal = {Proc. ACM Interact. Mob. Wearable Ubiquitous Technol.},
month = dec,
articleno = {189},
numpages = {27}
}

@inproceedings{timeseriestoimage_wang,
author = {Wang, Zhiguang and Oates, Tim},
title = {Imaging Time-Series to Improve Classification and Imputation},
year = {2015},
isbn = {9781577357384},
publisher = {AAAI Press},
booktitle = {Proceedings of the 24th International Conference on Artificial Intelligence},
pages = {3939–3945},
numpages = {7},
location = {Buenos Aires, Argentina},
series = {IJCAI'15},
url = {https://www.ijcai.org/Proceedings/15/Papers/553.pdf}
}

@inproceedings{timeseriestoimage_wu,
author = {Wu, Wei and Zhang, Yuan},
month = jul,
booktitle={2019 Chinese Control Conference (CCC)}, 
title={Activity Recognition from Mobile Phone using Deep CNN}, 
year={2019},
volume={},
number={},
pages={7786-7790},
url = {https://doi.org/10.23919/ChiCC.2019.8865142},
doi={10.23919/ChiCC.2019.8865142}
}

@inproceedings{timeseriestoimage_Jiang,
author = {Jiang, Wenchao and Yin, Zhaozheng},
title = {Human Activity Recognition Using Wearable Sensors by Deep Convolutional Neural Networks},
year = {2015},
isbn = {9781450334594},
publisher = {ACM},
address = {New York, NY, USA},
url = {https://doi.org/10.1145/2733373.2806333},
doi = {10.1145/2733373.2806333},
booktitle = {Proceedings of the 23rd ACM International Conference on Multimedia},
pages = {1307–1310},
numpages = {4},
location = {Brisbane, Australia},
series = {MM '15}
}

@article{timeseriestoimage_hatami,
author    = {Nima Hatami and Yann Gavet and Johan Debayle},
  title     = {Classification of Time-Series Images Using Deep Convolutional Neural Networks},
  journal   = {CoRR},
  volume    = {abs/1710.00886},
  year      = {2017},
  url       = {http://arxiv.org/abs/1710.00886},
  archivePrefix = {arXiv},
  eprint    = {1710.00886},
  bibsource = {dblp computer science bibliography, https://dblp.org}
}

@article{timeseriestoimage_chen,
author    = {Yun{-}Cheng Tsai and Jun Hao Chen and Chun{-}Chieh Wang},
 title     = {Encoding Candlesticks as Images for Patterns Classification Using
 Convolutional Neural Networks},
  journal   = {CoRR},
  volume    = {abs/1901.05237},
  year      = {2019},
  url       = {http://arxiv.org/abs/1901.05237},
  archivePrefix = {arXiv},
  eprint    = {1901.05237},
  bibsource = {dblp computer science bibliography, https://dblp.org}
}

@inproceedings{CV_toshev,
	title = {DeepPose: Human Pose Estimation via Deep Neural Networks},
	booktitle = {2014 {IEEE} {Conference} on {Computer} {Vision} and {Pattern} {Recognition}},
	author = {Toshev, Alexander and Szegedy, Christian},
	publisher = {IEEE},
	month = jun,
	year = {2014},
	ISSN = {1063-6919},
	pages = {1653-1660},
	doi = {10.1109/CVPR.2014.214},
	url = {https://doi.org/10.1109/CVPR.2014.214},
}

@article{CV_moschoglou,
	title = {3DFaceGAN: Adversarial Nets for 3D Face Representation, Generation, and Translation},
	author = {Moschoglou, Stylianos and Ploumpis, Stylianos and Nicolaou, Mihalis A. and Papaioannou, Athanasios and Zafeiriou, Stefanos},
	volume = {128},
	issn = {1573-1405},
	number = {10},
	journal = {International Journal of Computer Vision},
	month = nov,
	year = {2020},
	pages = {2534-2551},
	url = {https://doi.org/10.1007/s11263-020-01329-8},
	doi = {10.1007/s11263-020-01329-8},
}

@misc{japanese_h_v,
	author = {Wikipedia},
	title = {Horizontal and vertical writing in East Asian scripts},
	howpublished = "\url{https://en.wikipedia.org/w/index.php?title=Horizontal_and_vertical_writing_in_East_Asian_scripts&oldid=984358336}",
	year = {2020},
	note = "Accessed: Oct 29, 2020."
}

@misc{jins_meme,
	author = {JINS},
	title = {JINS MEME electrooculography glasses},
	howpublished = "\url{https://jins-meme.com/en/}",
	year = {2020},
	note = "Accessed: Nov 10, 2020."
}

@misc{tobii,
	author = {Tobii},
	title = {Tobii eye-tracker},
	howpublished = "\url{https://gaming.tobii.com/getstarted/}",
	year = {2020},
	note = "Accessed: Nov 10, 2020."
}

@article{speech_Park_2019,
   title={SpecAugment: A Simple Data Augmentation Method for Automatic Speech Recognition},
   url={http://dx.doi.org/10.21437/Interspeech.2019-2680},
   DOI={10.21437/interspeech.2019-2680},
   journal={Interspeech 2019},
   publisher={ISCA},
   author={Park, Daniel S. and Chan, William and Zhang, Yu and Chiu, Chung-Cheng and Zoph, Barret and Cubuk, Ekin D. and Le, Quoc V.},
   year={2019},
   month=Sep
}

@inproceedings{NLP_ruder,
    title = "Transfer Learning in Natural Language Processing",
    author = "Ruder, Sebastian  and
      Peters, Matthew E.  and
      Swayamdipta, Swabha  and
      Wolf, Thomas",
    booktitle = "Proceedings of the 2019 Conference of the North {A}merican Chapter of the Association for Computational Linguistics: Tutorials",
    month = jun,
    year = {2019},
    address = "Minneapolis, Minnesota",
    publisher = "Association for Computational Linguistics",
    doi = "10.18653/v1/N19-5004",
    pages = {15-18},
    url={http://dx.doi.org/10.18653/v1/N19-5004},
}

@article{data_Roh,
  author    = {Yuji Roh and Geon Heo and Steven Euijong Whang},
  title     = {A Survey on Data Collection for Machine Learning: a Big Data - {AI} Integration Perspective},
  journal   = {CoRR},
  volume    = {abs/1811.03402},
  year      = {2018},
  url       = {http://arxiv.org/abs/1811.03402},
}

@article{SSL_amis,
	title = {Self-supervised {ARTMAP}},
	author = {Amis, Gregory P. and Carpenter, Gail A.},
	volume = {23},
	issn = {0893-6080},
	url = {https://doi.org/10.1016/j.neunet.2009.07.026},
	doi = {10.1016/j.neunet.2009.07.026},
	number = {2},
	journal = {Neural Networks},
	month = mar,
	year = {2010},
}

@ARTICLE{SSL_Liu,
  author={X. {Liu} and J. v. d. {Weijer} and A. D. {Bagdanov}},
  journal={IEEE Transactions on Pattern Analysis and Machine Intelligence}, 
  title={Exploiting Unlabeled Data in CNNs by Self-Supervised Learning to Rank}, 
  year={2019},
  volume={41},
  number={8},
  pages={1862-1878},
  doi={10.1109/TPAMI.2019.2899857},
  url = {https://doi.org/10.1109/TPAMI.2019.2899857},
  }

@article{SSL_Lan,
  author    = {Zhenzhong Lan and Mingda Chen and Sebastian Goodman and Kevin Gimpel and Piyush Sharma and Radu Soricut},
  title = {{ALBERT:} {A} Lite {BERT} for Self-supervised Learning of Language Representations},
  journal   = {Computing Research Repository},
  volume    = {abs/1909.11942},
  year      = {2019},
  url       = {http://arxiv.org/abs/1909.11942},
}

@InProceedings{Nair_ICML2010,
  author    = {Vinod Nair and Geoffrey E. Hinton},
  booktitle = {Proceedings of the 27th International Conference on International Conference on Machine Learning},
  title     = {Rectified linear units improve restricted boltzmann machines},
  pages     = {807--14},
  url       = {https://dl.acm.org/doi/10.5555/3104322.3104425},
  month     = jun,
  year      = {2010},
}

@InProceedings{Kingma_ICLR2015,
  author    = {Diederik P. Kingma and Jimmy Ba},
  booktitle = {Proceedings of the 3rd International Conference for Learning Representations},
  title     = {Adam: A Method for Stochastic Optimization},
  year      = {2015},
}

@article{HAR_SSL_Zaki,
  title={Self-Supervised Human Activity Recognition by Augmenting Generative Adversarial Networks},
  author={Mohammad Zaki Zadeh and A. R. Babu and Ashish Jaiswal and F. Makedon},
  journal={ArXiv},
  year={2020},
  volume={abs/2008.11755},
  url = {https://arxiv.org/abs/2008.11755},
}

@inproceedings{cs_Pascual_10,
   author = {Mart\'{\i}nez-G\'{o}mez, Pascual and Aizawa, Akiko},
   title = {Recognition of Understanding Level and Language Skill Using Measurements of Reading Behavior},
   year = {2014},
   isbn = {9781450321846},
   publisher = {ACM},
   address = {New York, NY, USA},
   url = {https://doi.org/10.1145/2557500.2557546},
   doi = {10.1145/2557500.2557546},
   booktitle = {Proceedings of the 19th International Conference on Intelligent User Interfaces},
   pages = {95–104},
   numpages = {10},
   location = {Haifa, Israel},
   series = {IUI '14}
 }

@inproceedings{cs_Okoso_11,
 title = "Towards extraction of subjective reading incomprehension: Analysis of eye gaze features",
 author = "Ayano Okoso and Takumi Toyama and Kunze, {Kai Steven} and Joachim Folz and Marcus Liwicki and Koichi Kise",
 year = "2015",
 month = apr,
 day = "18",
 doi = "10.1145/2702613.2732896",
 language = "English",
 volume = "18",
 pages = "1325--1330",
 booktitle = "CHI 2015 - Extended Abstracts Publication of the 33rd Annual CHI Conference on Human Factors in Computing Systems: Crossings",
 publisher = "ACM"
 }

@inproceedings{cs_Cheng_12,
 author = {Cheng, Shiwei and Sun, Zhiqiang and Sun, Lingyun and Yee, Kirsten and Dey, Anind K.},
 title = {Gaze-Based Annotations for Reading Comprehension},
 year = {2015},
 isbn = {9781450331456},
 publisher = {ACM},
 address = {New York, NY, USA},
 url = {https://doi.org/10.1145/2702123.2702271},
 doi = {10.1145/2702123.2702271},
 booktitle = {Proceedings of the 33rd Annual ACM Conference on Human Factors in Computing Systems},
 pages = {1569–1572},
 numpages = {4},
 location = {Seoul, Republic of Korea},
 series = {CHI '15}
 }
\end{document}